\documentclass[preprint]{aastex631}

\newcommand{\rom}[1]{\MakeUppercase{\romannumeral#1}}
\usepackage{comment}
\usepackage{hyperref}
\usepackage{amsmath}
\usepackage{cleveref}
\usepackage{enumitem}
\usepackage{graphicx}
\usepackage{lmodern}
\usepackage{CJK}

\newcommand{\Mnor}{\hbox{$\mathcal{M}^{\rm N}_\odot$}}
\newcommand{\Lnor}{\hbox{$\mathcal{L}^{\rm N}_\odot$}}
\newcommand{\Rnor}{\hbox{$\mathcal{R}^{\rm N}_\odot$}}

\turnoffediting

\received{November 12, 2022}
\revised{December 27, 2022}
\accepted{March 21, 2023}

\shorttitle{{\rm Be} Binary Systems}
\graphicspath{{./}{figures/}}

\begin{document}
\begin{CJK*}{UTF8}{gbsn}
\title{The orbital and physical properties of five southern Be+sdO binary systems}

\correspondingauthor{Luqian Wang}  
\email{wangluqian@ynao.ac.cn, dgies@gsu.edu, gpeters@usc.edu, zhanwenhan@ynao.ac.cn}

\author[0000-0003-4511-6800]{Luqian Wang (王璐茜)}
\altaffiliation{Visiting astronomer, Cerro Tololo Inter-American Observatory at NSF’s NOIRLab, which is managed by the Association of Universities for Research in Astronomy (AURA) under a cooperative agreement with the National Science Foundation.}
\affiliation{Yunnan Observatories, Chinese Academy of Sciences (CAS), Kunming 650216, Yunnan, China}
\affiliation{Center for High Angular Resolution Astronomy and Department 
of Physics and Astronomy, Georgia State University, P.O. Box 5060, Atlanta,
GA 30302-5060, USA}

\author[0000-0001-8537-3583]{Douglas R. Gies}
\altaffiliation{Visiting astronomer, Cerro Tololo Inter-American Observatory at NSF’s NOIRLab, which is managed by the Association of Universities for Research in Astronomy (AURA) under a cooperative agreement with the National Science Foundation.}
\affiliation{Center for High Angular Resolution Astronomy and Department 
of Physics and Astronomy, Georgia State University, P.O. Box 5060, Atlanta,
GA 30302-5060, USA}

\author[0000-0003-4202-269X]{Geraldine J. Peters}
\affiliation{Department of Physics and Astronomy,
University of Southern California, Los Angeles, CA 90089-0484, USA}

\author[0000-0001-9204-7778]{Zhanwen Han}
\affiliation{Yunnan Observatories, Chinese Academy of Sciences (CAS), Kunming 650216, Yunnan, China}



\begin{abstract} 
Close binary interactions may play a critical role in the formation of the rapidly rotating Be stars.  Mass transfer can result in a mass gainer star spun up by the accretion of mass and angular momentum, while the mass donor is stripped of its envelope to form a hot and faint helium star. FUV spectroscopy has led to the detection of about 20 such binary Be+sdO systems. Here we report on a three-year program of high quality spectroscopy designed to determine the orbital periods and physical properties of five Be binary systems. These binaries are long orbital period systems with $P =$ {95 to 237} days and with small semi-amplitude $K_1<11$ km s$^{-1}$. We combined the Be star velocities with prior sdO measurements to obtain mass ratios. A Doppler tomography algorithm shows the presence of the \ion{He}{2} $\lambda 4686$ line in the faint spectrum of the hot companion in four of the targets.  We discuss the observed line variability and show evidence of phased-locked variations in the emission profiles of HD 157832, suggesting a possible disk spiral density wave due to the presence of the companion star. The stripped companions in HD 113120 and HD 137387 may have a mass larger than the 1.4 $M_\odot$ indicating that they could be progenitors of Type Ib and Ic supernovae. 
\end{abstract}



\section{Introduction} \label{sec:intro}
Be stars are B-type main-sequence stars whose spectra show or have shown Balmer emission lines. Such emission features probably originate in a circumstellar decretion disk around the central star. The exact formation mechanism of the disk is not clear, but it is widely accepted that the disk is geometrically thin and moving in Keplerian motion governed by viscosity \citep{Quirrenbach1997,Meilland2007,Rivinius2013}. Be stars often display photometric and spectroscopic line variations, and such variability spans timescales from days to months \citep{Hanuschik1996,Arcos2018}. Be stars are rapid rotators, and their projected rotational velocities can reach up to $\sim80\%$ of their critical rotational velocities \citep{Rivinius2013}.      

Some theoretical studies suggest that single Be stars may spin up due to  physical changes at certain evolutionary stages. These processes include the transportation of angular momentum from the central contracting core to the outer envelope \citep{Meynet2007}, the effects of large initial rotational velocity and non-solar metallicity \citep{Ekstrom2008}, and the influence of equatorial mass-loss for the near-critical rotation of Be stars \citep{Granada2013}. 

Alternatively, close binary interactions offer another way to understand the rapid rotation of the Be stars and, consequently, the Be phenomenon. \citet{Pols1991} and \citet{vanBever1997} performed calculations for case B close binary evolution and concluded that a non-negligible fraction of Be population was formed through close binary interactions. In such a binary scenario, the initially more massive star loses its outer hydrogen envelope and transfers mass to the gainer star through Roche-lobe overflow. The mass and angular momentum transfer process may rejuvenate and spin up the gainer star, which now appears as a rapidly rotating Be star.  The former donor star becomes an evolved remnant, such as a black hole, neutron star, or helium subdwarf star (sdO). Recent binary population synthesis simulations from \citet{Shao2014,Shao2021} indicate that the Be+sdO binary systems are abundant in the Milky Way with an estimated number of order $10^4$. Using various assumptions for the mass ratio distribution and the mass transfer efficiency, these authors suggest that Be+sdO binaries likely have orbital periods from 80 to 200 days, helium companion masses from 0.3 to 0.6 \Mnor ~(nominal solar value\footnote{The superscript N indicates the nominal values of the solar units that are the standard values set forth by the International Astronomical Union at the \rom{29}th General Assembly \citep{Prsa2016}. They are recommended for usage in stellar and planetary publications for a homogeneous unit system to avoid inconsistent conversion constants appearing in the SI unit transformation.}), and Be star semi-amplitudes of $\sim8$ km s$^{-1}$. 

Despite the large number of Be+sdO binaries predicted from theoretical simulations, only a handful of such systems are found observationally. The first known Be+sdO binary system, $\phi$ Per, was detected through optical spectroscopy of the anti-phase motion of the \ion{He}{2} $\lambda$4686 emission line formed in hot gas near the sdO star \citep{Poeckert1981}. The spectral signature of the sdO star itself was subsequently confirmed from the International Ultraviolet Explorer (IUE) FUV spectral analysis by \citet{Thaller1995}. The physical and orbital properties of this system were later determined by \cite{Gies1998} using observations from HST FUV spectroscopy and by \citet{Mourard2015} using near-IR interferometry from the CHARA Array. Because the sdO stars are hotter than their Be primaries, identifying their spectral features is best accomplished in the FUV since they contribute relatively more flux towards the short wavelength part of the spectrum. Studies from IUE FUV spectroscopy led to detection of sdO companions in Be binaries FY CMa \citep{Peters2008}, 59 Cyg \citep{Peters2013}, HR 2142 \citep{Peters2016}, and 60 Cyg \citep{Wang2017}. Later assessment of available FUV observations for a sample of 264 stars in the IUE database by \citet{Wang2018} led to the detection of an additional 12 Be+sdO candidate binary systems, and the spectral signature of the sdO stars in nine of them was later confirmed by HST FUV spectroscopy \citep{Wang2021}. The subdwarf stars in these Be+sdO binary systems have effective temperatures ($T_{\rm eff}$) in the range from 34 to 53 kK, small projected rotational velocities $V\sin{i} < 36$ km s$^{-1}$ (with the exception of $V\sin{i} = 102 \pm 4$ km s$^{-1}$ in case of HD 51354; \citealt{Wang2021}), flux ratios $f_\mathrm{sdO}/f_\mathrm{Be}$ between $1$ and about $10\%$, small radii $R_\mathrm{sdO} < 0.6$ \Rnor, low luminosity $\log{L_\mathrm{sdO}} < 4.1$ \Lnor, orbital periods of 28 to 126 days, and estimated masses of 0.1 to 1.0 \Mnor. However, the masses and the orbital parameters of these hot subdwarf stars remain poorly constrained due to the limited number of FUV spectra.      

Although it is difficult to identify the spectral features of the sdO star from optical spectroscopy due to their faintness and small size compared to the Be star, indirect evidence indicating the detection of sdO stars is reported in several works. \citet{Chojnowski2018} utilized high-quality optical observations to trace the signature of the sdO companion in the HD 55606 binary system from the \ion{He}{2} $\lambda4686$ profile appearing in the spectra. Studies of the \ion{He}{1} $\lambda6678$ emission profile revealed possible sdO companions in the cases of $o$ Pup \citep{Koubsky2012}, HD 161306 \citep{Koubsky2014}, and 7 Vul \citep{Harmanec2020}. Furthermore, periodic $V/R$ variations appearing in the violet and red emission peaks of the \ion{He}{1} $\lambda6678$ profile in $o$ Pup may also indicate its binary nature \citep{Rivinius2012}. Recently, \citet{Naze2022} conducted a radial velocity monitoring survey for 16 $\gamma$~Cas-type stars with strong X-ray emission, and they reported that these systems likely harbor a small mass companion ({$0.6-1.0$} \Mnor) in orbits with long periods ({$80-120$} days). \citet{Harmanec2022} performed a comprehensive study to investigate the spectral, photometric, and color variations of V1294 Aql using observations spanning about 70 years, and they concluded that this is likely another Be+sdO binary candidate system. An extensive analysis of archival optical spectra and new CHARA Array interferometry of $\kappa$ Dra by \citet{Klement2022b} suggests the existence of a cooler subdwarf companion (sdB) in this Be binary.         

Long-term monitoring programs of the Be+sdO binary systems to determine their orbits and stellar parameters are vital to tracing their evolutionary history. Direct comparison of the measured quantities to models will provide constraints on the nature of these binary systems, such as evaluating the dynamic stability of the binaries based upon the critical mass ratio of the system, estimating the mass-loss or accretion state of the component stars, investigating the mass stripping process, and tracing their formation efficiency \citep{Shao2014,Shao2021}. Investigating the physical properties of the Be binaries provides insights into understanding the formation and rapid rotation features of massive stars \citep{deMink2013}. Stripped hot stars are also important contributors to the ionizing flux of stellar populations \citep{Gotberg2019}. Massive subdwarf stars may explode as SN~\rom{1}b and SN~\rom{1}c \citep{Laplace2021,Aguilera-Dena2022} and leave compact neutron star \citep{Reig2011} or black hole \citep{Schneider2021} remnants. 

Motivated by the recent detection of hot subdwarf stars in Be+sdO binary systems from FUV spectroscopy by \citet{Wang2021}, we conducted a long-term optical spectroscopic monitoring program to measure radial velocities and determine the orbital and physical parameters. Here we report on the first part of the work that is focused on five Be+sdO binary systems visible in the southern sky. In Section~\ref{sec:obser} we describe the observations. Section~\ref{sec:RVs} presents the line profiles identified in the optical spectra of each target star, the radial velocity (RV) measurements made from strong lines, and the derivation of the orbital elements. We document the line profile variations and the emission line equivalent widths in Section~\ref{sec:Variability}. 
Section \ref{sec:4686} summarizes our use of Doppler tomography to detect the \ion{He}{2} $\lambda4686$ absorption profile from the sdO companion stars. 
We present in Section~\ref{sec:dis} a comparison of the physical and orbital properties with those of known Be+sdO (sdB) binary systems, and we give a brief discussion of theoretical aspects of the formation of such Be binaries. We offer our conclusions in Section~\ref{sec:conc}.   

\newpage

\section{Spectroscopy Program} \label{sec:obser}

The spectra of five southern Be targets were obtained with the high resolving power, echelle spectrometer CHIRON installed on the Small and Moderate Aperture Research Telescope System (SMARTS) 1.5-m telescope at the Cerro Tololo Inter-American Observatory (CTIO) in Chile. The spectra were collected between 2018 Apr 23 and 2021 Jul 23 through the queue observing mode using the image slicer with a spectral resolving power $R=80000$.  The spectra are recorded in 59 echelle orders covering a wavelength range from 4150 to 8800 \AA\ \citep{Tokovinin2013}. Table~\ref{tab:list} lists the HD number, star name, Hipparcos HIP number, $V$-band magnitude, spectral classification and the associated reference, number of observations, and UT date range for each target star. These stars were identified as Be+sdO systems through FUV spectroscopic detection of the sdO line patterns \citep{Wang2018,Wang2021}. {One of the stars, HD 137387, was identified as a spectroscopic binary system based upon the analysis from \citet{Jilinski2010}, but the rest of the stars in the list were not previously known binaries. }  Appendix D provides a summary of recent studies of these stars. 

The spectra were reduced following the standard reduction procedures, including bias subtraction, flat-field correction, removal of cosmic rays, echelle order extraction, and wavelength calibration through the usage of Th-Ar lamps \citep{Paredes2021}. The spectra are dominated by broad Balmer lines associated with the Be star disk, and these often span more than one echelle order.  Consequently, we adopted the technique from \cite{Kolbas2015} to use the flux in adjoining orders to remove the blaze function variation in orders recording the broad features. {Because the atmospheric telluric lines were not removed from the observed spectra, we thus adopted the atmospheric transmission spectral library by \citet{Hinkle2003} to correct for these absorption features. We applied a fitting scheme to identify the telluric absorption lines by matching the observations with the atmospheric transmission profiles and subsequently removed these features from the observed spectra.}  We then transformed the spectra onto a uniform wavelength grid on a $\log{\lambda}$ scale with a step size of 2.26 km s$^{-1}$ per pixel, and we rectified the spectra to a unit continuum from selected continuum points in relatively line-free regions. The final working products are matrix arrays of normalized flux as a function of heliocentric wavelength and heliocentric corrected Julian date (HJD) for each echelle order. 

\begin{deluxetable*}{lclllccc}
\tablecaption{Spectroscopic Program Targets \label{tab:list}}
\tablewidth{0pt}
\tablehead{
\colhead{HD} & \colhead{Star} & \colhead{HIP} & \colhead{$V$} &  \colhead{Spectral} & \colhead{Ref.\tablenotemark{a}} &\colhead{Number of } &\colhead{UT Date} \\
\colhead{Number} & \colhead{Name} & \colhead{Number} & \colhead{(mag)} &\colhead{Classification} & \colhead{} & \colhead{Spectra} & \colhead{Range}
}
\startdata
113120  &	LS Mus		& 63688	& 6.03 & B2 \rom{4}ne		& 1	& 45	& 	2018 Apr 23 -- 2021 Apr 18\\   
137387	&	$\kappa$ Aps	& 76013	& 5.49 & B2 \rom{5}npe	& 1	& 46	&	2019 Jan 12 -- 2021 Jul 13\\	
152478	&	V846 Ara		& 82868	& 6.33  & B3 \rom{5}npe	& 1	& 49	&	2019 Feb 21 -- 2021 Jul 20\\
157042	&	$\iota$ Ara  	& 85079	& 5.25  & B2.5 \rom{4}e	& 2	& 47	&	2019 Feb 23 -- 2021 Jul 18\\
157832	& 	V750 Ara		& 85467	& 6.66 &B1.5 \rom{5}e	& 3	& 47    &   2019 Feb 26  -- 2021 Jul 23\\
\enddata
\tablenotetext{a}{(1) \citet{Levenhagen2006}, (2) \citet{Slettebak1982}, (3) \citet{Lopes2011}. }
\end{deluxetable*}

\section{Radial Velocities and Orbital Elements}
\label{sec:RVs}
In order to determine the orbital solutions of these Be+sdO binary systems, we first inspected the spectra to identify suitable line profiles for measurement of the radial velocities (RVs) of the Be component. Below we describe the selected spectral features and the detailed procedures for radial velocity measurements of each star.

\subsection{HD 113120}\label{subsec:RVHD113120}
\begin{figure*}[h]
\gridline{\fig{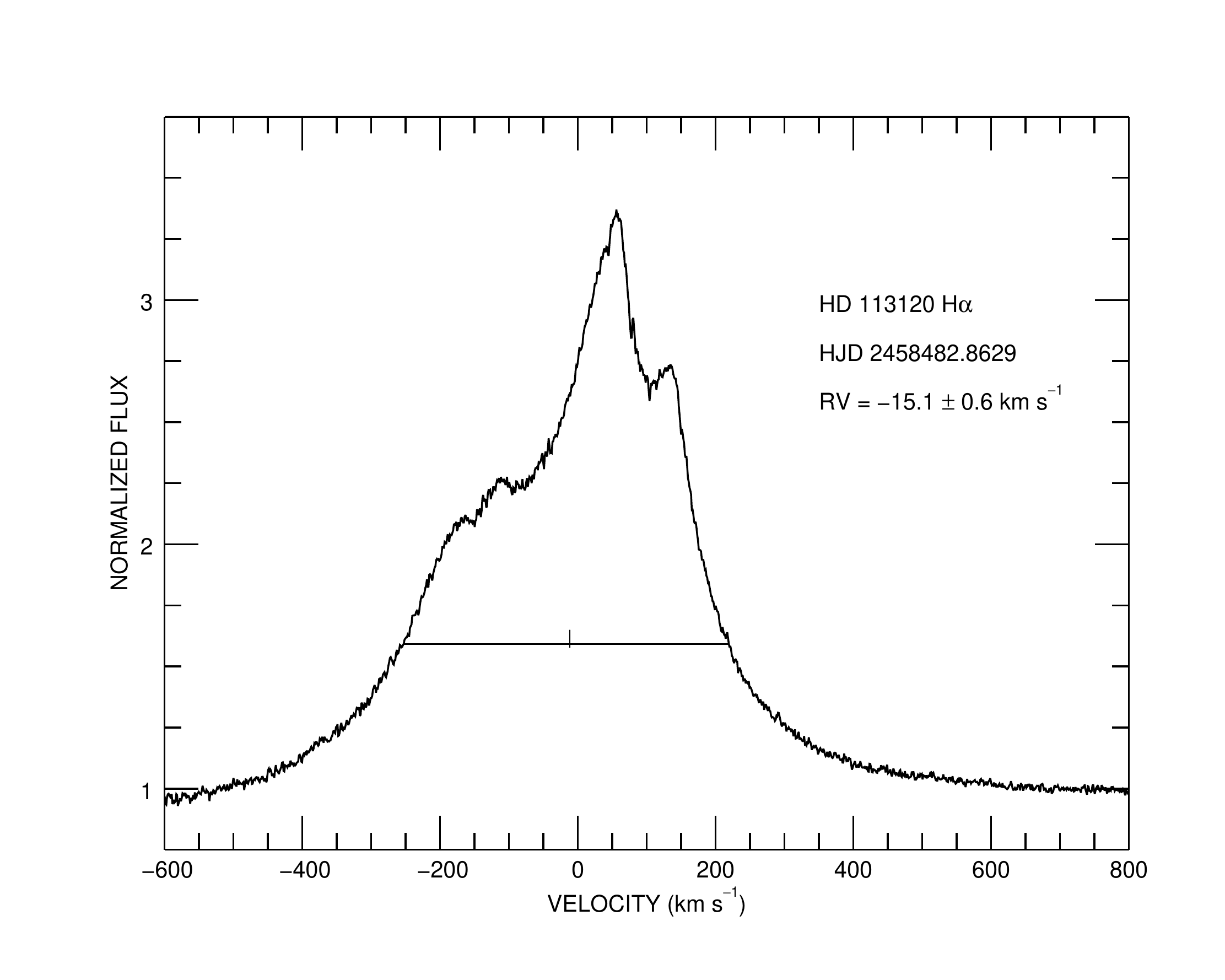}{\textwidth}{} }
\caption{An example of the H$\alpha$ profile observed on the night of HJD 2458482 for HD 113120. The horizontal line indicates 25\% of the peak height of the profile, and two oppositely signed Gaussian functions were located at the intersection points to sample the wings of the profile and to determine the bisector position of the Doppler shift (shown as the tick mark). }
\label{fig:spec_HD113120}
\end{figure*}

The spectra of HD 113120 contain emission-line profiles of the Balmer series, including H$\alpha$ and H$\beta$, \ion{He}{1} lines, and metallic lines (including \ion{Fe}{2}, \ion{Si}{2}, and the \ion{Ca}{2} triplet series). The metallic line profiles, such as \ion{Si}{2} and the \ion{Ca}{2} triplet, as well as \ion{He}{1} profiles, are generally too weak to make reliable radial velocity measurements. The \ion{He}{1} $\lambda\lambda 5875, 6678, 7065$ and H$\beta$ profiles display line variations that affect the line wings (see Section~\ref{subsec:VarHD113120}). Thus, we restrict the Be component's RV measurements to the echelle spectral orders containing the H$\alpha$, H$\beta$, and \ion{Fe}{2} profiles only. The H$\alpha$ profile is very broad and displays emission peak variations (see Fig.~\ref{fig:vari_HD113120} below). \citet{Porter2003} suggest that the Be disks rotate with Keplerian motion, so the high-speed wings are likely formed close to the Be stars. Thus, measurements of the Doppler shifts of the line wings provide a good proxy for the orbital motion of the Be star.  We adopted the wing bisector technique from \citet{Shafter1986} to use two oppositely signed Gaussian functions to sample the wings of the profile to determine the bisector position and measure the velocity of the Be star. We adopted the derivations from \citet{Grundstrom2007} (in her Appendix A.\ 2) to estimate the error of the RV measurements from samplings of the spectral flux, separation of the two Gaussian functions, and the uncertainties in the continuum. An example of the H$\alpha$ profile observed on the night of HJD 2458482 with the associated RV measurement is shown in Figure~\ref{fig:spec_HD113120}. Alternatively, we utilized a calculation of the cross-correlation function (CCF) of the observed spectra with the co-added mean spectrum to measure the relative RVs for the H$\beta$ and \ion{Fe}{2} profiles ($\lambda\lambda 5276, 5316, 5362, 8451$). These ancillary RV measurements are included in Table~\ref{tab:RV_HD113120_other} of Appendix~\ref{sec:RV_other}.    

Similar to other known Be binary systems (e.g., \citealt{Harmanec2000, Harmanec2002}), the RVs measured from the H$\alpha$ wings of HD 113120 display long-term variations on a timescale of about a year that are most likely due to the variations in the emission profiles caused by structural changes in the disk. We thus applied corrections to the measured Doppler shifts to remove such long-term variations. The detailed procedures are discussed in Appendix~\ref{sec:RVcorrect}.  Table~\ref{tab:RV_table} lists the star name, observational date, derived orbital phase from the SB2 solution, corrected radial velocity and the associated error, derived residual velocity, and the line profile used to measure the Doppler shift for each observation of the five target stars. RV measurements and orbital solutions for the other four target stars are discussed in the following few subsections.  

We applied a global minimization convergence method developed by \citet{Iglesias-Marzoa2015} to obtain the orbital solution using the RV measurements made from the H$\alpha$ profile. {These include the orbital period, the epoch when the Be star reaches superior conjunction in its orbit, the semi-amplitude of the velocity curves and the systemic velocity of each component star, eccentricity, and the longitude of periastron of the fitting orbit. } Equal weight was assigned to each velocity measurement for the orbital fitting, except in the case of the spectrum obtained on the night of HJD 2458232, in which a zero weight was given because the measured RV deviated greatly from the derived orbital solution. This spectrum showed a steep drop in intensity, and the profile displayed a redshift compared to the adjacent spectrum in the dataset observed about 250 days later. We found no improvements in the fit residuals by adopting an eccentric orbital solution, and thus, we report the derived single-lined, circular orbital elements in column two of Table~\ref{tab:orbit_HD113120}.

\begin{figure*}[h]
\gridline{
        \fig{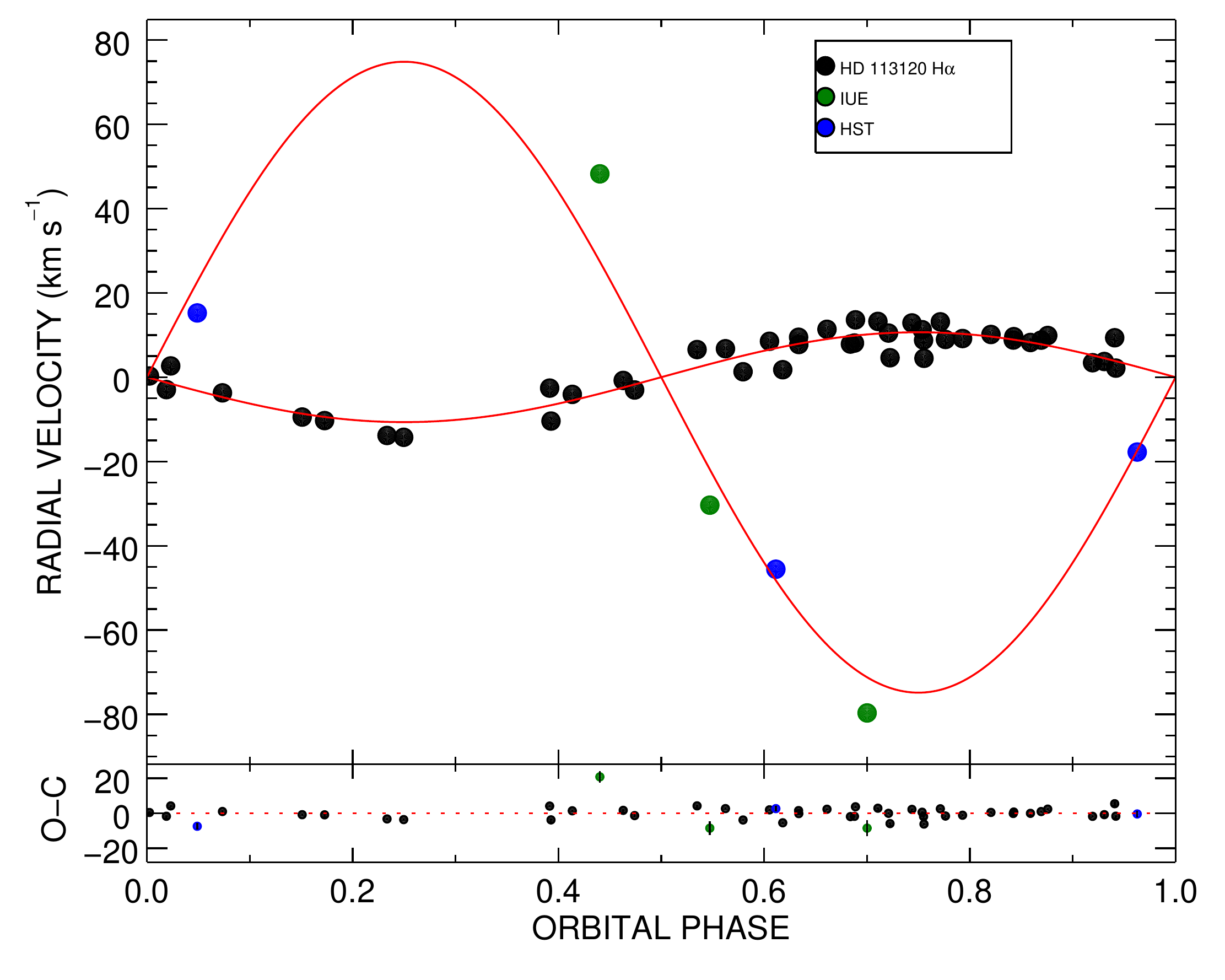}{1.0\textwidth}{}
        }
\caption{Top panel: The relative radial velocity curves (red) for the Be+sdO binary system HD 113120. The zero orbital phase is defined as the epoch when the Be star reaches superior conjunction in its orbit. RVs of the Be star measured from the CHIRON H$\alpha$ profiles are labeled in black dots, RVs of the sdO companion star obtained from the prior studies using FUV spectroscopy from IUE \citep{Wang2018} and HST \citep{Wang2021} are shown in green and blue, respectively. Systemic velocities for each stellar component obtained from the orbital fits were subtracted from individual measurements for the purposes of this plot. Bottom panel: The measured radial velocities minus the calculated values obtained from the orbital fits. The zero residuals are shown as the red dotted line. The typical error of the measured Be RVs is $\sim0.5$ km s$^{-1}$, which is too small to be shown in the figure. }
\label{fig:RV_HD113120}
\end{figure*}

\citet{Wang2018,Wang2021} performed an FUV spectral analysis of this star and reported RV measurements for the sdO companion obtained from IUE and HST observations. We utilized these measurements to derive the orbital solutions for both stellar components, and the best-fit radial velocity curves are shown in red in Figure~\ref{fig:RV_HD113120}. RV measurements made from the high-resolution HST spectra ($R = 45800$, marked in blue) achieve a better velocity fit compared to those of lower resolution IUE observations ($R = 13000$, in green). We give the derived circular orbital solutions from both components in column three of Table~\ref{tab:orbit_HD113120}. Note that the uncertainties quoted for the different parameters are the internal errors from the fitting routine \citep{Iglesias-Marzoa2015}, and these do not account for systematic errors introduced in the corrections procedure for long-term trends (Appendix A). RV curves for H$\beta$ and \ion{Fe}{2} profiles plotted using the derived SB2 ephemeris are shown in Appendix Figure~\ref{fig:RV_HD113120_other}. 


\placefigure{fig:spec_HD113120}

\placefigure{fig:RV_HD113120}

\begin{deluxetable*}{ccccccc}
\tablecaption{Radial Velocity Measurements of the Be stars \label{tab:RV_table}}
\tablewidth{0pt}
\tablehead{
\colhead{HD} & \colhead{Date} & \colhead{Orbital} &\colhead{$V_\mathrm{r}$} & \colhead{$\sigma_\mathrm{r}$} & \colhead{$O-C$} & \colhead{Line} \\
\colhead{Number} & \colhead{(HJD$-$2,400,000)} &\colhead{Phase$^a$} &  \colhead{(km s$^{-1}$)} & \colhead{(km s$^{-1}$)} & \colhead{(km s$^{-1}$)} & \colhead{Measured$^b$} 
}
\startdata
113120 &	58232.6079$^c$ & 	0.012 & 	\phn$-${3.9} & 	{1.0} & \nodata & H$\alpha$ \\
113120 &	58482.8629 & 	0.392 & 	$-${15.1} & 	{0.6} & 	\phs{4.1} & 	H$\alpha$ \\
113120 &	58486.8532 & 	0.414 & 	$-${16.6} & 	{0.6} & 	\phs{1.4} & 	H$\alpha$ \\
113120 &	58495.8407 & 	0.463 &  	$-${13.3} & 	{0.6} & 	\phs{1.7} & 	H$\alpha$ \\
113120 &	58497.8571 & 	0.474 & 	$-${15.6} & 	{1.0} & 	$-${1.3} & 	H$\alpha$ \\
\enddata
\tablenotetext{a}{The orbital phase of each measurement is obtained from the SB2 fits of the associated target star, except in the case of HD 157832, in which the orbital phases are from the SB1 fit. }
\tablenotetext{b}{H$\alpha$ profiles were used to measure the RVs of the target stars except for the case of HD 137387, where the \ion{He}{1} $\lambda5875$ profile was chosen for measurement. }
\tablenotetext{c}{The RV measured from this spectrum was given a zero weight and excluded from the orbital fit due to its large deviation from the derived orbital solution. }
\tablecomments{This table is available in its entirety in machine-readable form. The first five entries are shown here for guidance regarding its format and content. }
\end{deluxetable*}

\begin{deluxetable*}{lrr}
\tablecaption{Circular orbital elements of HD 113120 \label{tab:orbit_HD113120}}
\tablewidth{0pt}
\tablehead{
\colhead{Element} & \multicolumn{2}{c}{{Value}}  \\ \cline{2-3}
\colhead{} & \colhead{SB1} & \colhead{SB2} 
}
\startdata
$P$ (days)	&{183.40} $\pm$ {0.15} & {181.54} $\pm$ {0.11} \\
$T_\mathrm{sc}$ (HJD$-$2,400,000)	& {58591.96} $\pm$ {0.13} & {58593.32} $\pm$ {0.13}  \\
$K_1$ (km s$^{-1}$)		& {10.89} $\pm$ {0.06} & {10.66} $\pm$ {0.03}  \\
$K_2$ (km s$^{-1}$)		& \nodata & {74.8} $\pm$ 0.4 \\
$q$  & \nodata & {0.142} $\pm$ 0.001 \\
$\gamma_1$ (km s$^{-1}$)	&$-${12.68} $\pm$ 0.06 & $-${12.52} $\pm 0.03$   \\
$\gamma_2$ (km s$^{-1}$)	& \nodata & {15.56} $\pm$ {0.26} \\
$f(m_1)$ (\Mnor) & {0.0246} $\pm$ {0.0004} & {0.0228} $\pm$ 0.0002  \\
$f(m_2)$ (\Mnor) &\nodata & {7.88} $\pm$ {0.13} \\
$a_1\sin i$ (\Rnor)	& {39.48} $\pm$ {0.22}  & {38.23} $\pm$ {0.10}  \\
$a_2\sin i$ (\Rnor)	& \nodata  & {268.4} $\pm$ {1.5}  \\
$a\sin i$ (\Rnor)	& \nodata & {306.7} $\pm$ {1.9}  \\  
$M_1\sin^3{i}$ (\Mnor) & \nodata & {10.29} $\pm$ {0.21} \\
$M_2\sin^3{i}$ (\Mnor) & \nodata & {1.47} $\pm$ 0.04  \\
rms$_1$ (km s$^{-1}$) &  {2.6} & {2.7} \\
rms$_2$ (km s$^{-1}$) &  \nodata & {10.3} \\
\enddata
\end{deluxetable*}


\null\vspace{2 cm}
\subsection{HD 137387}\label{subsec:HD137387}
The spectra of HD 137387 are dominated by broad and strong Balmer line profiles, including the H$\alpha$ and H$\beta$ profiles (see panel (a) of Fig.~\ref{fig:vari_HD137387} in Section~\ref{subsec:VarHD137387}). The H$\alpha$ line displays profile variations transitioning from a broad absorption feature (observed on the night of HJD 2458540) to a narrower absorption superimposed on top of a wide emission profile (HJD 2459409), indicating an ongoing disk development in this Be star. Such variations prevent us from obtaining reliable RV measurements from the H$\alpha$ profiles. 

Instead, we measured RVs for the Be star from echelle orders containing H$\beta$ and \ion{He}{1} $\lambda\lambda 4713, 4921, 5015, 5875, 6678, 7065$ absorption lines by calculating the CCFs of the observed spectra with their co-added mean spectrum. The calculated CCFs are very broad, as expected from the rapid rotation of the Be star. We thus determined the Doppler shifts of the star from the wing bisector position of the CCFs as described in Section~\ref{subsec:RVHD113120}, except in the case of \ion{He}{1} $\lambda5875$, where the CCF peak velocities were obtained (using the correction constants listed in Table~\ref{tab:RV_corr} of Appendix~\ref{sec:RVcorrect}). Figure~\ref{fig:spec_HD137387} shows an example of the \ion{He}{1} $\lambda5875$ profile observed on the night of HJD 2458482 (black) and the co-added mean spectrum (green). The measured relative RV is shown as the tick mark in the figure. The RV measurements made from the \ion{He}{1} profiles generally have too much scatter to make a reliable orbital fit except for the case of \ion{He}{1} $\lambda$5875, in which the best orbital RV curve was achieved with a circular fit of the measurements (Fig.~\ref{fig:RV_HD137387}). We report the derived single-lined orbital solution of the star in column two of Table~\ref{tab:orbit_HD137387}. 

Prior FUV spectroscopy of this binary system by \citet{Wang2018,Wang2021} led to seven measurements of RVs for the sdO companion star (four from IUE and three from HST). Thus, we also determined a preliminary circular orbital solution using RVs for both stellar components, and the results are reported in column three of Table~\ref{tab:orbit_HD137387}. The RV curves with the double-lined orbital fits are shown in Figure~\ref{fig:RV_HD137387}. The corrected relative RVs and their associated errors from measurements of the \ion{He}{1} $\lambda5875$ profiles are listed in Table~\ref{tab:RV_table} together with the orbital phases and residuals from the double-lined solution, {five RV measurements made between HJD 2458526 and HJD 2458544 were assigned a zero weight due to their large deviations from the orbital fits}. RV measurements made from H$\beta$ and the other \ion{He}{1} line profiles are reported in Table~\ref{tab:RV_HD137387_HD157832_other} of Appendix~\ref{sec:RV_other}.  


\placefigure{fig:spec_HD137387}
\begin{figure*}[h]
\gridline{\fig{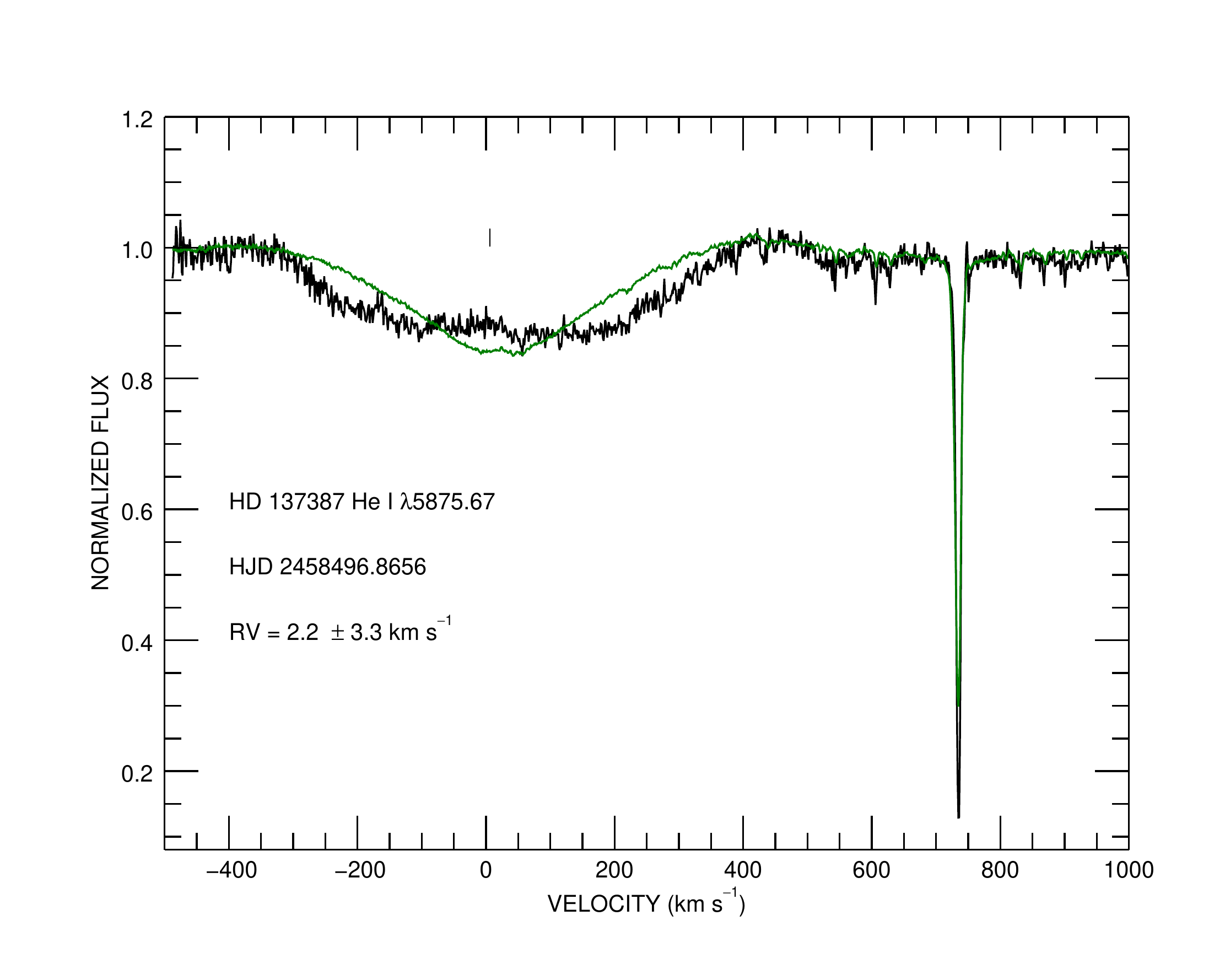}{1.0\textwidth}{} }
\caption{The \ion{He}{1} $\lambda5875$ profile observed on the night of HJD 2458496 (black) and the co-added mean spectrum (green) for HD 137387. The relative radial velocity (RV$ = 2.2 \pm 3.3$ km s$^{-1}$; shown as the tick mark) was measured from the peak position of the CCF that was calculated by cross-correlating the observed spectrum with the mean spectrum of all 46 observations. }
\label{fig:spec_HD137387}
\end{figure*} 

\pagebreak

\placefigure{fig:RVHD137387}
\begin{figure*}[h]
\gridline{
    \fig{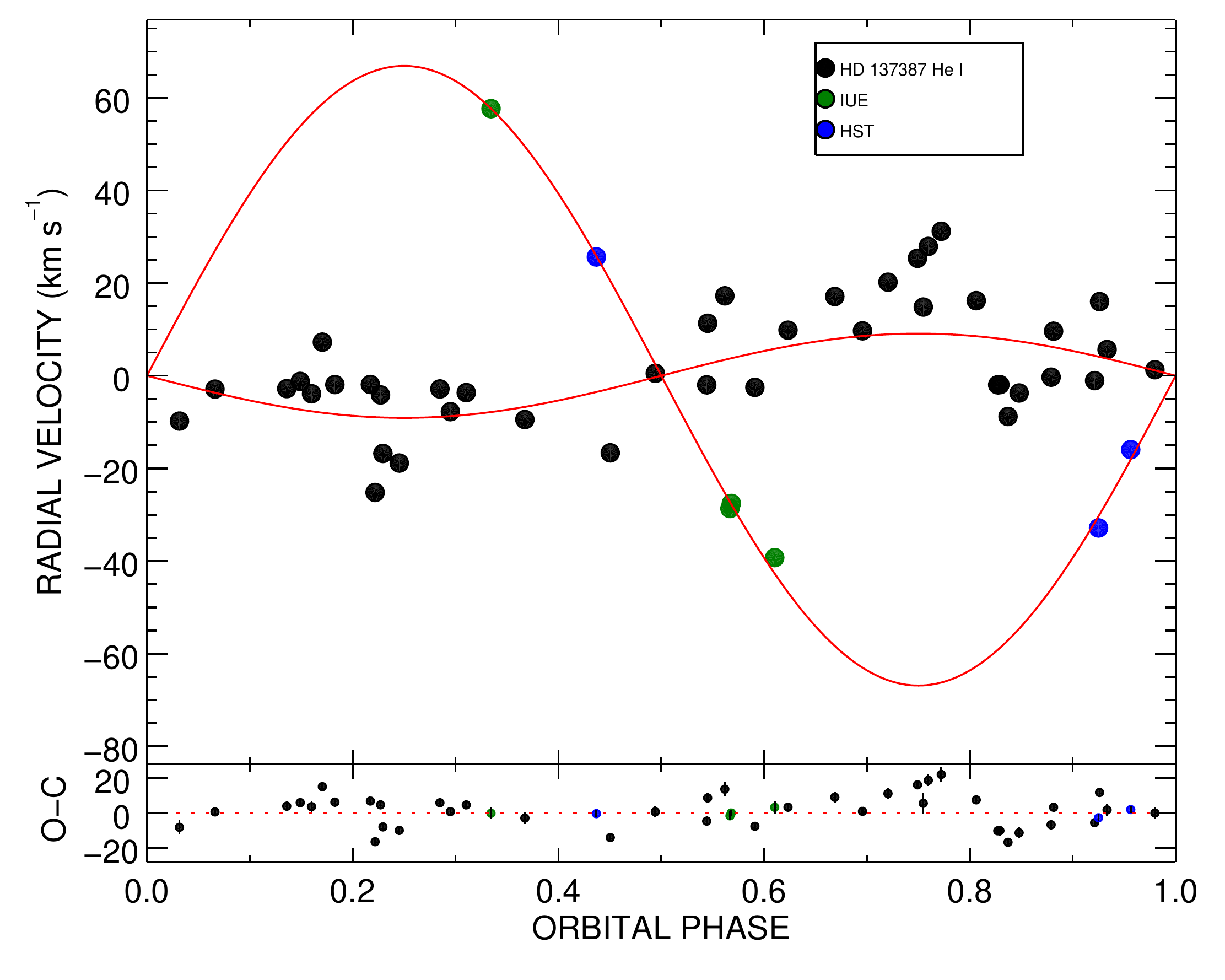}{1.0\textwidth}{}
        }
\caption{The relative radial velocity curves for the Be binary HD 137387, plotted in the same format as Figure~\ref{fig:RV_HD113120}. RVs of the Be star were measured from the \ion{He}{1} $\lambda5875$ profiles using the cross-correlation technique.  }
\label{fig:RV_HD137387}
\end{figure*}

\newpage


\begin{deluxetable*}{lrr}[ht!]
\tablecaption{Circular orbital elements of HD 137387 \label{tab:orbit_HD137387}}
\tablewidth{0pt}
\tablehead{
\colhead{Element} & \multicolumn{2}{c}{{Value}} \\\cline{2-3}
\colhead{} & \colhead{SB1} & \colhead{SB2}
}
\startdata
$P$ (days)		& {191.9} $\pm$ {0.3} & {192.1 $ \pm$ 0.1} \\
$T_\mathrm{sc}$ (HJD$-$2,400,000)	& {58402.4 $\pm$ 2.4} & {58401.9 $\pm$ 2.2}\\
$K_1$ (km s$^{-1}$)		&{9.08 $\pm$ 0.11} & {9.09 $\pm$ 0.08} \\
$K_2$ (km s$^{-1}$) & \nodata & {66.87 $\pm$ 1.84}\\
$q$ & \nodata & {0.136 $\pm$ 0.004} \\
$\gamma_1$ (km s$^{-1}$)\tablenotemark{a}	&{1.66 $\pm$ 0.12} & {1.67 $\pm$ 0.08} \\
$\gamma_2$ (km s$^{-1}$) & \nodata & {15.46 $\pm$ 0.99} \\
$f(m_1)$ (\Mnor) &{0.0149 $\pm$ 0.0006} & {0.0150 $\pm$ 0.0004} \\
$f(m_2)$ (\Mnor) &\nodata & {5.95 $\pm$ 0.49} \\
$a_1\sin i$ (\Rnor)	& {34.5 $\pm$ 0.4}  &  {34.5 $\pm$ 0.3} \\
$a_2\sin i$ (\Rnor)	& \nodata  & {253.8 $\pm$ 7.0} \\
$a\sin i$ (\Rnor)	& \nodata  & {288.3 $\pm$ 8.4} \\
$M_1\sin^3{i}$ (\Mnor) & \nodata & {7.7 $\pm$ 0.7} \\
$M_2\sin^3{i}$ (\Mnor) & \nodata & {1.0 $\pm$ 0.1} \\
rms$_1$ (km s$^{-1}$) &  {9.6} & {9.6} \\
rms$_2$ (km s$^{-1}$) &  \nodata & {1.9} \\
\enddata
\begin{center}
\tablenotetext{a}{Based upon relative RVs.}
\end{center}
\end{deluxetable*}


\newpage
\null\vspace{5 cm}

\subsection{HD 152478}\label{subsec:HD152478}
The spectra of HD 152478 display weak \ion{He}{1} absorption features and strong double-peaked emission features in the Balmer profiles of H$\alpha$ and H$\beta$. The strengths of the violet and red H$\alpha$ emission peaks varied throughout the observations (see panel (a) of Fig.~\ref{fig:vari_HD152478} in Section~\ref{subsec:VarHD152478}). Broad weak absorption profiles of \ion{He}{1} (including $\lambda\lambda 4713, 4921, 5875, 6678, 7065$) and weak metallic emission features of \ion{Fe}{2} $\lambda8451$ and \ion{Ca}{2} $\lambda8542$ appear in the spectra, but are too weak to obtain reliable Doppler shift measurements. We  measured the Doppler shifts of the Be star from the broad H$\alpha$ wings using the bisector technique. Systematic shifts appear in the measured RVs for observations obtained after HJD 2,458,800, and long-term corrections were applied to the RVs as discussed in Appendix~\ref{sec:RVcorrect}. Figure~\ref{fig:spec_HD152478} shows an example of the H$\alpha$ profile observed on HJD 2458536, and the determined bisector location is marked on the plot to indicate the RV measurement. We report the corrected RVs and the associated errors for H$\alpha$ in columns four and five of Table~\ref{tab:RV_table}. We tested both a circular and an eccentric orbital fit to the measured RVs and found that the circular solution achieves an adequate fit (given in column two of Table~\ref{tab:orbit_HD152478}). 

Five FUV observations were obtained from prior investigations (two IUE and three HST), and we used these to determine a preliminary double-lined orbital solution (column three of Table~\ref{tab:orbit_HD152478}). The derived orbital phases and calculated $O-C$ residuals obtained from the double-lined circular orbit fit are listed in columns three and six, respectively, of Table~\ref{tab:RV_table}. In Figure~\ref{fig:RV_HD152478}, we show the circular orbital fits and measured RVs of the Be component and the sdO companion star. 

We also measured RVs for H$\beta$, but used a different approach. We formed CCFs of each profile with the global average profile and then measured the bisector position for the wings of the CCFs. These relative velocities are listed in Table~\ref{tab:orbit_HD152478} and plotted versus the orbital phase from the double-lined ephemeris in panel (a) of Figure~\ref{fig:RV_HD152478_157042_157832_Hb} in Appendix~\ref{sec:RV_other}. The relative velocities for H$\beta$ have a similar appearance to the RV curve for H$\alpha$ but with greater scatter. 


\placefigure{fig:spec_HD152478}
\begin{figure*}[h]
\gridline{\fig{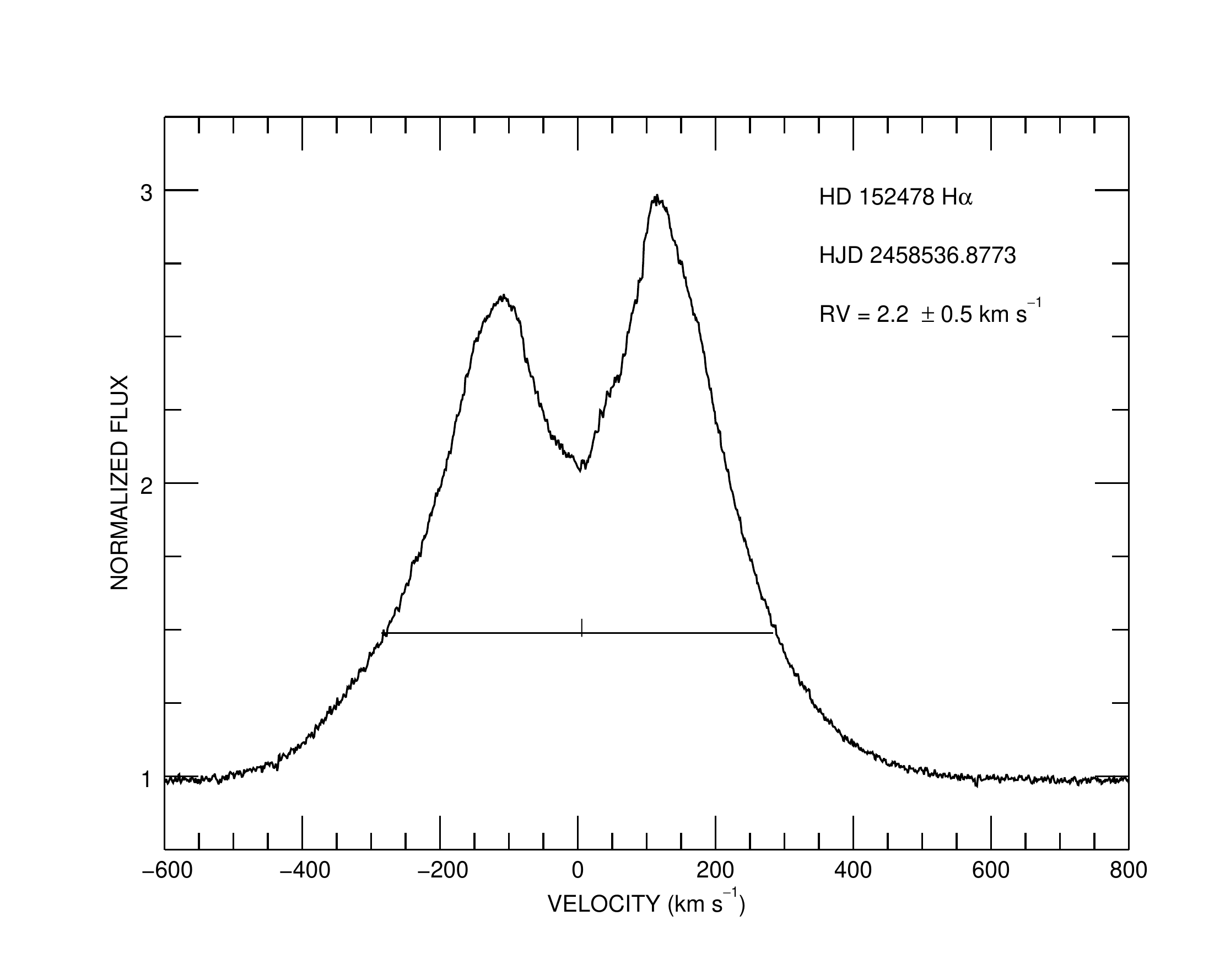}{\textwidth}{} }
\caption{The H$\alpha$ profile observed on the night of HJD 2458536 for HD 152478. The radial velocity measured from the broad wings using the bisector technique is shown as the tick mark.   }
\label{fig:spec_HD152478}
\end{figure*} 

\placefigure{fig:RV_HD152478}
\begin{figure*}[h]
\gridline{\fig{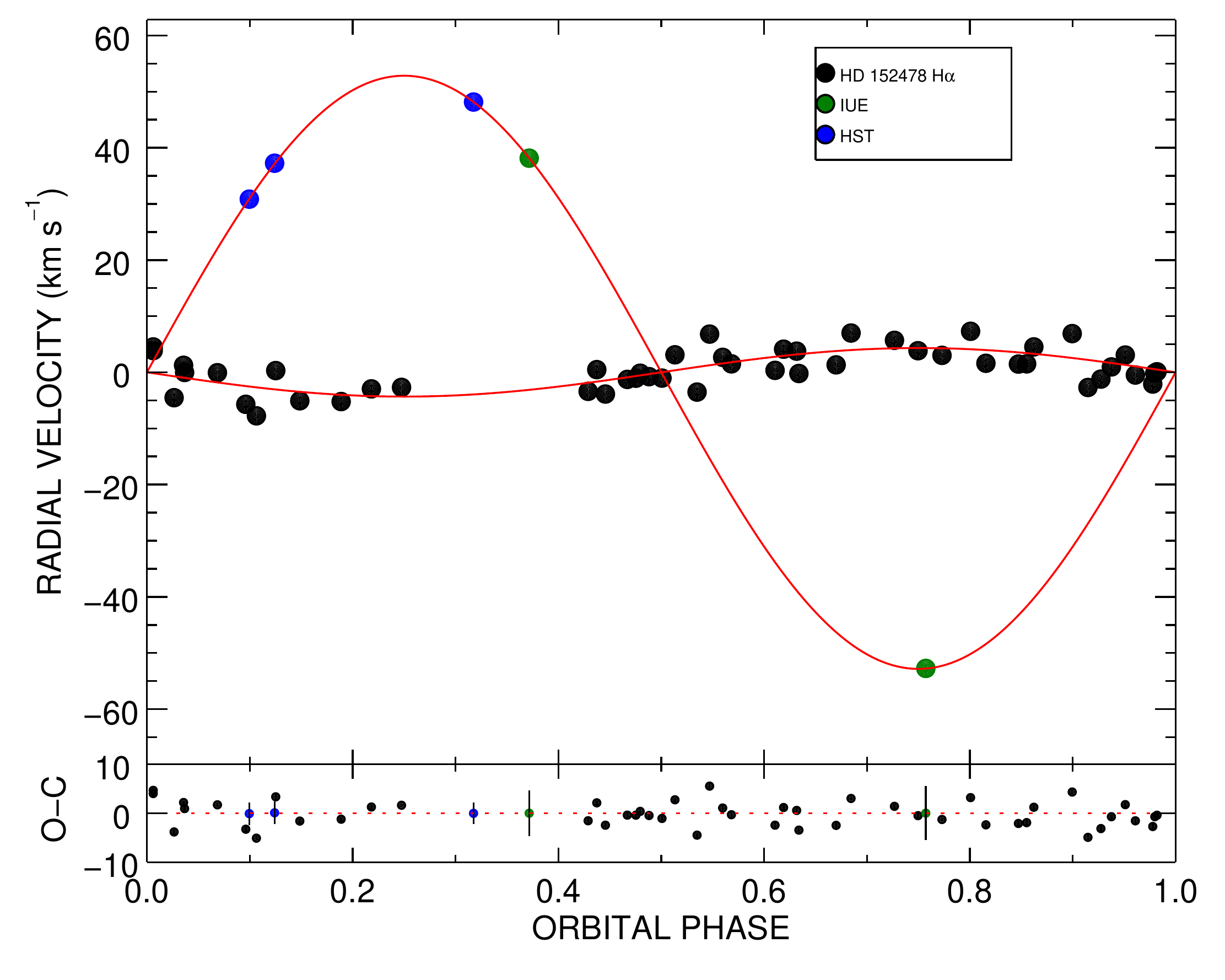}{1.0\textwidth}{} 
        }
\caption{The relative radial velocity curves for the Be binary HD 152478, plotted in the same format as Figure~\ref{fig:RV_HD113120}. RVs of the Be star were measured from the H$\alpha$ profiles using the bisector technique. The sdO RVs are collected from prior studies by \citet{Wang2018,Wang2021}. }
\label{fig:RV_HD152478}
\end{figure*}

\begin{deluxetable*}{lrr}[h]
\tablecaption{Circular orbital elements of HD 152478 \label{tab:orbit_HD152478}}
\tablewidth{0pt}
\tablehead{
\colhead{Element} & \multicolumn{2}{c}{Value} \\\cline{2-3}
\colhead{} & \colhead{SB1} & \colhead{SB2}
}
\startdata
$P$ (days)	 & {236.12 $\pm$ 0.20} & {236.50 $\pm$ 0.18}\\
$T_\mathrm{sc}$ (HJD$-$2,400,000) & {58672.25 $\pm$ 0.74} & {58672.10 $\pm$ 0.72} \\
$K_1$ (km s$^{-1}$)	 & {4.31 $\pm$ 0.06} & {4.33 $\pm$ 0.05} \\
$K_2$ (km s$^{-1}$) & \nodata & {52.83 $\pm$ 1.49} \\
$q$ & \nodata & {0.082 $\pm$ 0.003} \\
$\gamma_1$ (km s$^{-1}$) & {5.63 $\pm$ 0.07} & {5.60 $\pm$ 0.06} \\
$\gamma_2$ (km s$^{-1}$) & \nodata & {10.15 $\pm$ 1.22} \\
$f(m_1)$ (\Mnor)   &  {0.0020 $\pm$ 0.0001} &  {0.0020 $\pm$ 0.0001} \\
$f(m_2)$ (\Mnor) & \nodata &  {3.62 $\pm$ 0.30} \\
$a_1\sin{i}$ (\Rnor) &  {20.1 $\pm$ 0.3} & {20.2 $\pm$ 0.2}  \\
$a_2\sin{i}$ (\Rnor) & \nodata & {247.0 $\pm$ 6.9} \\
$a\sin{i}$ (\Rnor) & \nodata & {267.2 $\pm$ 8.2} \\
$M_1\sin^3{i}$ (\Mnor) & \nodata & {4.23 $\pm$ 0.42} \\
$M_2\sin^3{i}$ (\Mnor) & \nodata & {0.35 $\pm$ 0.05} \\
rms$_1$ (km s$^{-1}$)  &  2.6 & {2.6}\\
rms$_2$ (km s$^{-1}$)  & \nodata &  {0.1} \\
\enddata
\end{deluxetable*}


\newpage
\null\vspace{5 cm}

\subsection{HD 157042}\label{subsec:HD157042}
\begin{figure*}[h]
\gridline{\fig{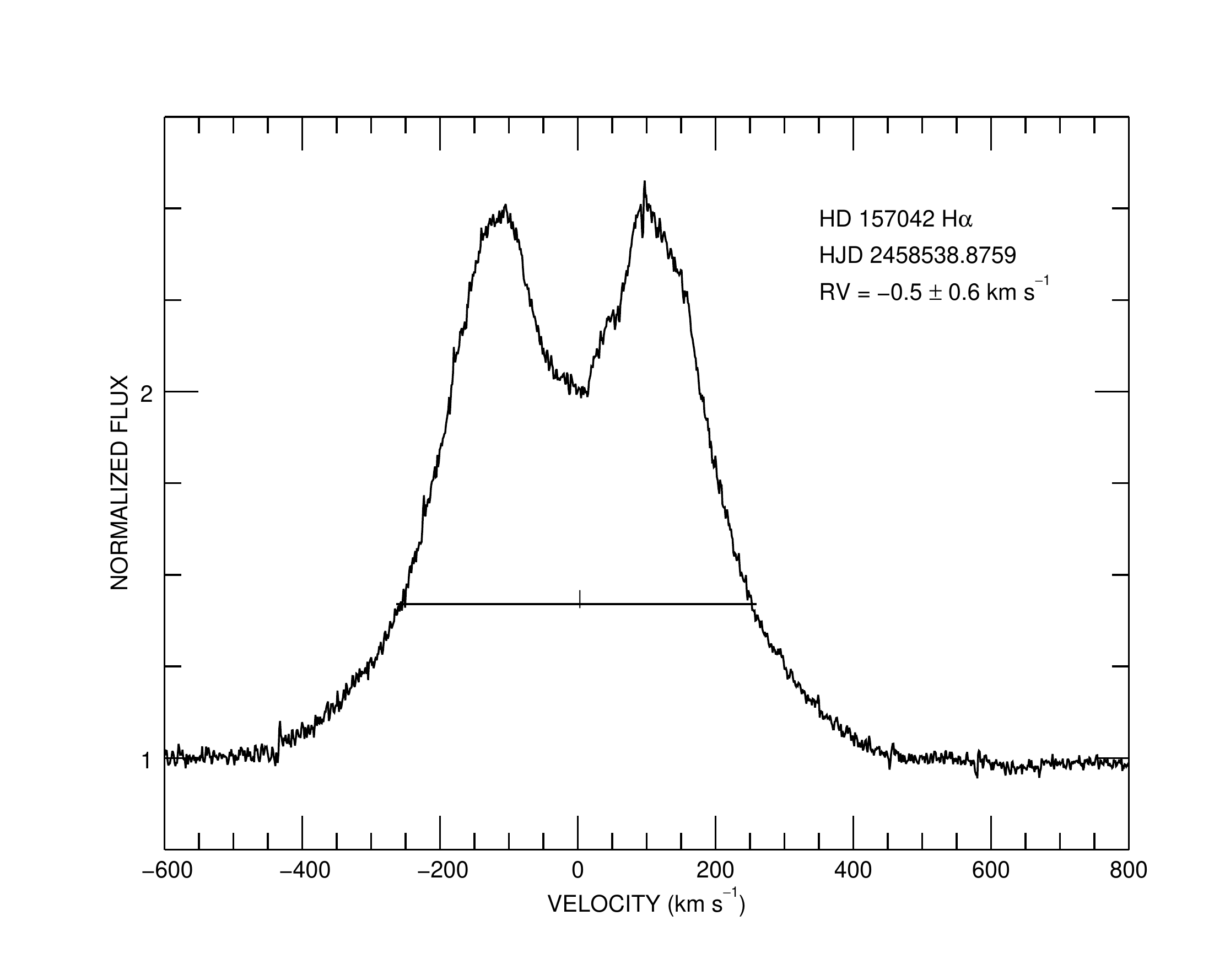}{\textwidth}{} }
\caption{The H$\alpha$ profile of HD 157042 observed on the night of HJD 2458538. The radial velocity measured from the broad wings using the bisector technique is shown as the tick mark.   }
\label{fig:spec_HD157042}
\end{figure*}

The optical spectra of HD 157042 are dominated by double-peaked emission features appearing in Balmer profiles of H$\alpha$ and H$\beta$ lines (see panel (a) of Fig.~\ref{fig:vari_HD157042} in Section~\ref{subsec:VarHD157042}).  Broad weak absorption lines of \ion{He}{1} $\lambda\lambda4713, 4921, 5015, 5875, 6678, 7065$ and metallic emission features such as \ion{Fe}{2} $\lambda\lambda8451, 8490$ and \ion{Ca}{2} $\lambda\lambda8542, 8600$ appear in the spectra. We again measured the RVs of the Be star from the broad H$\alpha$ wings using the Gaussian-sampled bisector method. Figure~\ref{fig:spec_HD157042} shows an example of the H$\alpha$ profile observed on the night of HJD 2458538 and the determined bisector location. RV measurements made from the \ion{He}{1} profiles have too much scatter to improve upon the H$\alpha$ RV curve. Thus, we restricted the orbital solution to the H$\alpha$ measurements only, and we report the long-term corrected results in Table~\ref{tab:RV_table}. 

\begin{figure*}[h]
\gridline{\fig{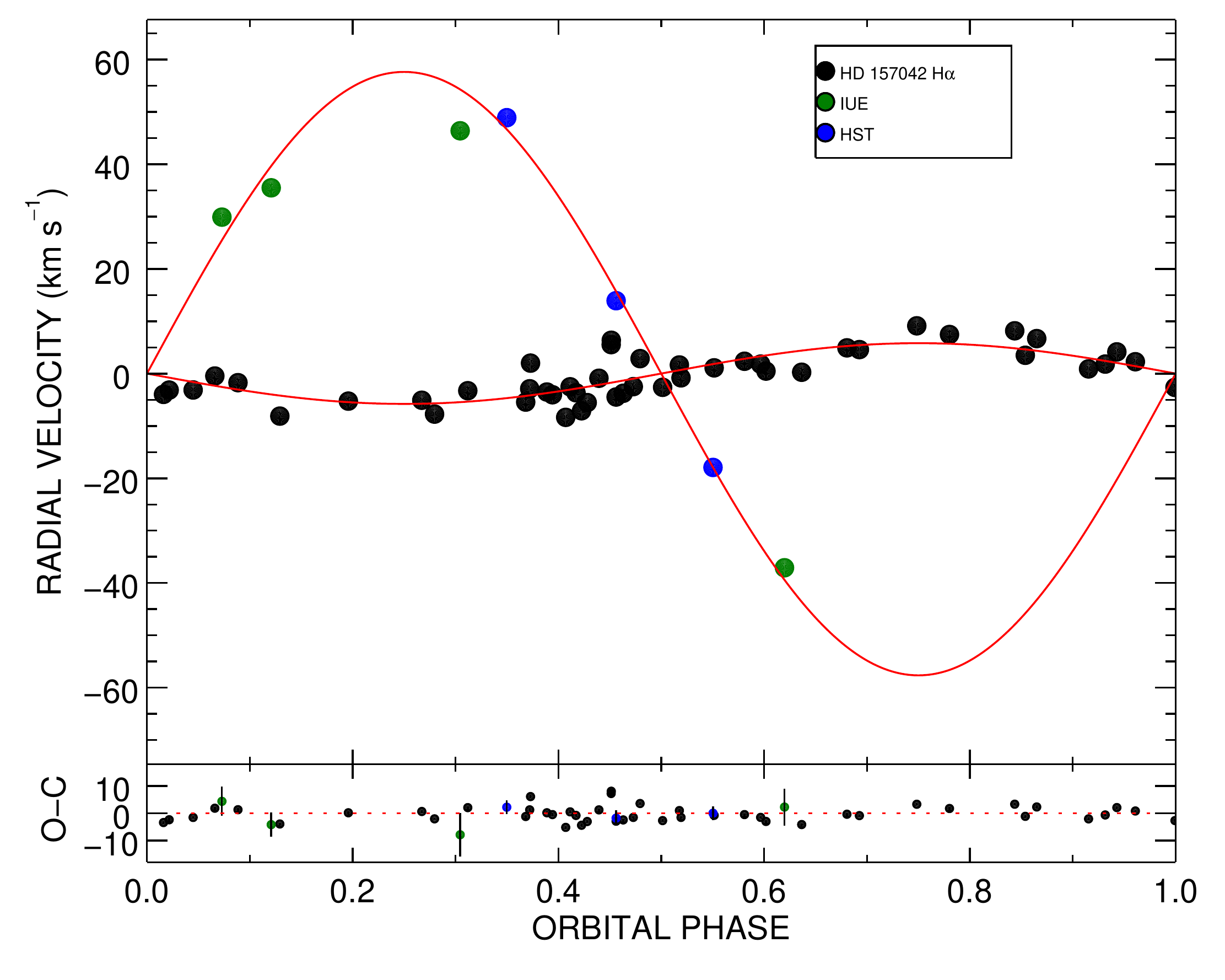}{1.0\textwidth}{}
          }
\caption{The relative radial velocity curves for HD 157042 in the same format as Figure~\ref{fig:RV_HD113120}.}
\label{fig:RV_HD157042}
\end{figure*}

We collected seven FUV measurements of the sdO RVs (four IUE and three HST) to determine a preliminary double-lined orbital solution. We tried both circular and elliptical fits to the RV measurements and found that the elliptical solution offered no significant improvement. Table~\ref{tab:orbit_HD157042} lists both the single-lined (Be) and double-lined (Be+sdO) fitting results, and the double-lined solutions are shown in Figure~\ref{fig:RV_HD157042}.

We also measured relative radial velocities for the H$\beta$ line in the same way as done for HD~152478.  These were also corrected for long-term trends.  These measurements (given in Table~\ref{tab:RV_HD152478_HD157042_Hb}) are shown as a function of the double-lined fit orbital phase in panel (b) of Figure~\ref{fig:RV_HD152478_157042_157832_Hb} in Appendix~\ref{sec:RV_other}. The H$\beta$ RV curve is similar to the H$\alpha$ case but displays more scatter. 


\placefigure{fig:spec_HD157042}

\placefigure{fig:RV_HD157042}

\begin{deluxetable*}{lrr}[h]
\tablecaption{Circular orbital elements of HD 157042 \label{tab:orbit_HD157042}}
\tablewidth{0pt}
\tablehead{
\colhead{Element} & \multicolumn{2}{c}{Value} \\\cline{2-3}
\colhead{} & \colhead{SB1} & \colhead{SB2}
}
\startdata
$P$ (days)	&   {178.64 $\pm$ 0.12} & {176.17 $\pm$ 0.04} \\
$T_\mathrm{sc}$ (HJD$-$2,400,000)	& {58650.3 $\pm$ 0.4} &  {58654.2 $\pm$ 0.5} \\
$K_1$ (km s$^{-1}$)	& {6.08 $\pm$ 0.03} & {5.80 $\pm$ 0.06} \\
$K_2$ (km s$^{-1}$) & \nodata & 57.6 $\pm$ {0.6} \\
$q$ & \nodata & {0.101} $\pm$ 0.001 \\
$\gamma_1$ (km s$^{-1}$)	& {2.24 $\pm$ 0.05} & {2.38 $\pm$ 0.06} \\
$\gamma_2$ (km s$^{-1}$) & \nodata & $-$10.2 $\pm$ {0.8}\\
$f(m_1)$ (\Mnor)   &    {0.0042} $\pm$ 0.0001 & {0.0033} $\pm$ 0.0001 \\
$f(m_2)$ (\Mnor) & \nodata & 3.50 $\pm$ {0.10} \\
$a_1\sin{i}$ (\Rnor)   &   {21.44 $\pm$ 0.12}  &  {19.61 $\pm$ 0.21}  \\
$a_2\sin{i}$ (\Rnor)   &   \nodata  & 200.7 $\pm$ {1.9} \\
$a\sin{i}$ (\Rnor) & \nodata & {220} $\pm$ 3 \\
$M_1\sin^3{i}$ (\Mnor) & \nodata & {3.88 $\pm$ 0.21} \\
$M_2\sin^3{i}$ (\Mnor) & \nodata & {0.426 $\pm$ 0.024} \\
rms$_1$ (km s$^{-1}$)  & {2.9} & {2.9}  \\
rms$_2$ (km s$^{-1}$)  & \nodata & 4.0  \\
\enddata
\end{deluxetable*}


\newpage
\null\vspace{7 cm}

\subsection{HD 157832}\label{subsec:HD157832}
The spectra of HD 157832 show strong H$\alpha$ and H$\beta$ emission-line profiles. We observed weak double-peaked emission features in \ion{He}{1} $\lambda\lambda4921, 5015$, while \ion{He}{1} $\lambda5875$ appears as a broad weak absorption line. Metallic emission profiles occur in \ion{Fe}{2} $\lambda\lambda4583$, 5020, 5169, 5197, 5234, 5276, 5284, 5316, 5362, 6148, 6383, 6385, 6456, 7516, 8451, \ion{N}{2} $\lambda$4630, \ion{N}{2} or \ion{Fe}{2} $\lambda5535$, \ion{Si}{2} $\lambda\lambda$6239, 6347, 6371, \ion{Al}{2} $\lambda$6243,  \ion{O}{1} $\lambda\lambda7771$, 7774, 7775, 8446, and \ion{Ca}{2} $\lambda\lambda$8542, 8600 (\ion{Ca}{2} $\lambda\lambda8498$, 8662 were positioned too close to the echelle order boundaries). Doppler shifts of H$\alpha$ and H$\beta$ were measured from the broad emission wings, and the RVs were corrected for the systematic shifts reported in Appendix~\ref{sec:RVcorrect}. Figure~\ref{fig:spec_HD157832} shows an example of the H$\alpha$ profile observed on the night of HJD 2458541 with the marked position of the measured bisector wing velocity. We report the corrected Doppler shifts and their associated errors from the H$\alpha$ profiles in Table~\ref{tab:RV_table}. 

Additional relative RV measurements were made using the CCF approach to measure the CCF peak RVs for weak emission line features observed in \ion{He}{1},  \ion{Fe}{2} $\lambda\lambda5197$, 5234, 5276, 5284, and \ion{O}{1} $\lambda8446$. For the \ion{Fe}{2} $\lambda6456$ and \ion{Ca}{2} $\lambda\lambda8542$, 8600 profiles, the Doppler shifts were determined from the broad CCF wings. Radial velocity curves of these lines were plotted from the ephemeris reported in Table~\ref{tab:orbit_HD157832} and are shown in Figure~\ref{fig:RV_HD157832_other} of Appendix~\ref{sec:RV_other}. The RV measurements are similar to those of H$\alpha$ but display more scatter. These measurements are included in Table~\ref{tab:RV_HD137387_HD157832_other} of Appendix~\ref{sec:RV_other}. 

\citet{Wang2018} reported the detection of a weak signal of the sdO companion star based upon two FUV spectra from IUE. However, later HST FUV spectroscopy did not confirm this detection \citep{Wang2021}. Thus, we restricted the fit to the Be star RV measurements to obtain a single-lined orbital solution of this binary system from a circular orbital fit of the corrected velocities, and the elements are given in Table~\ref{tab:orbit_HD157832}. The epoch $T_{SC}$ refers to the time of Be star superior conjunction. 
Seven RV measurements observed between the nights of HJD 2458635 and HJD 2458667 were assigned zero weight for the orbital fit. These spectra displayed large deviations from the fit and showed a steep drop in intensity in the blue wings of H$\alpha$ (see Fig.~\ref{fig:spec_omit_HD157832} of Appendix~\ref{sec:RV_other}). Figure~\ref{fig:RV_HD157832} shows the orbital velocities and the derived fit. Additional RV plots with more scatter are given for H$\beta$ in panel (c) of Figure~\ref{fig:RV_HD152478_157042_157832_Hb} and for \ion{He}{1} $\lambda5015$ and \ion{Fe}{2} $\lambda5020$, \ion{Fe}{2} $\lambda\lambda5197$, 5234, 5276, \ion{O}{1} $\lambda8446$, and \ion{Ca}{2} $\lambda\lambda 8542$, 8600 in Figure~\ref{fig:RV_HD157832_other} of Appendix~\ref{sec:RV_other}.    

We can use the orbital mass function and an estimate of the Be star mass to find a mass range for the companion.  \citet{Lopes2011} determined $T_{\rm eff} = 25000\pm1000$ K from a spectral energy distribution fit, and Be stars with similar temperature in the list compiled by \cite{Zorec2016} have a mass of about 11 \Mnor. Table~\ref{tab:sdO_HD157832} shows the estimated masses of the companion for several values of orbital inclination. 

\placefigure{fig:spec_HD157832}
\begin{figure*}[h]
\gridline{\fig{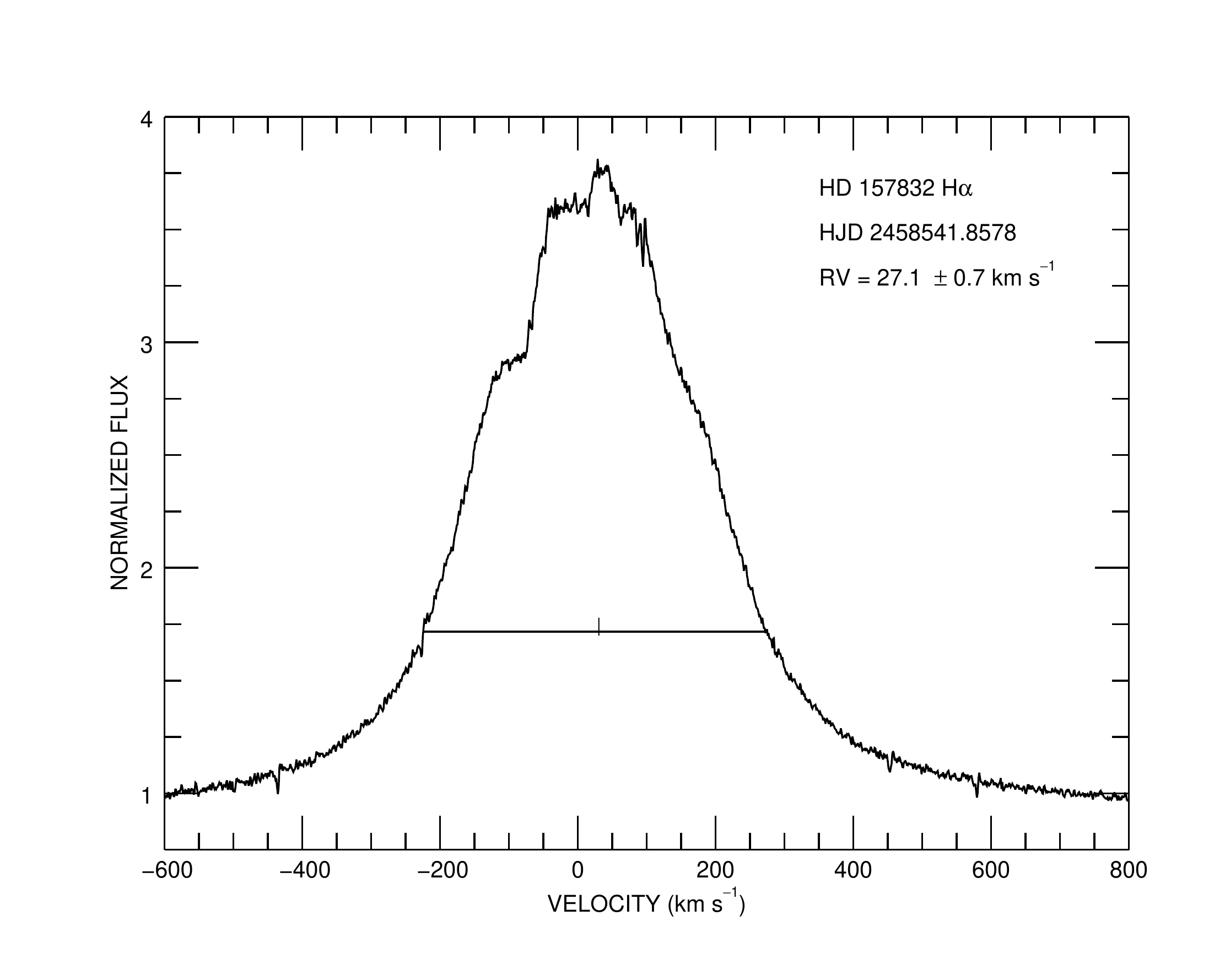}{\textwidth}{} }
\caption{The H$\alpha$ profile of HD 157832 observed on the night of HJD 2458541. The radial velocity measured from the broad wings using the bisector technique is shown as the tick mark. }
\label{fig:spec_HD157832}
\end{figure*} 

\placefigure{fig:RV_HD157832}
\begin{figure*}[h]
\gridline{\fig{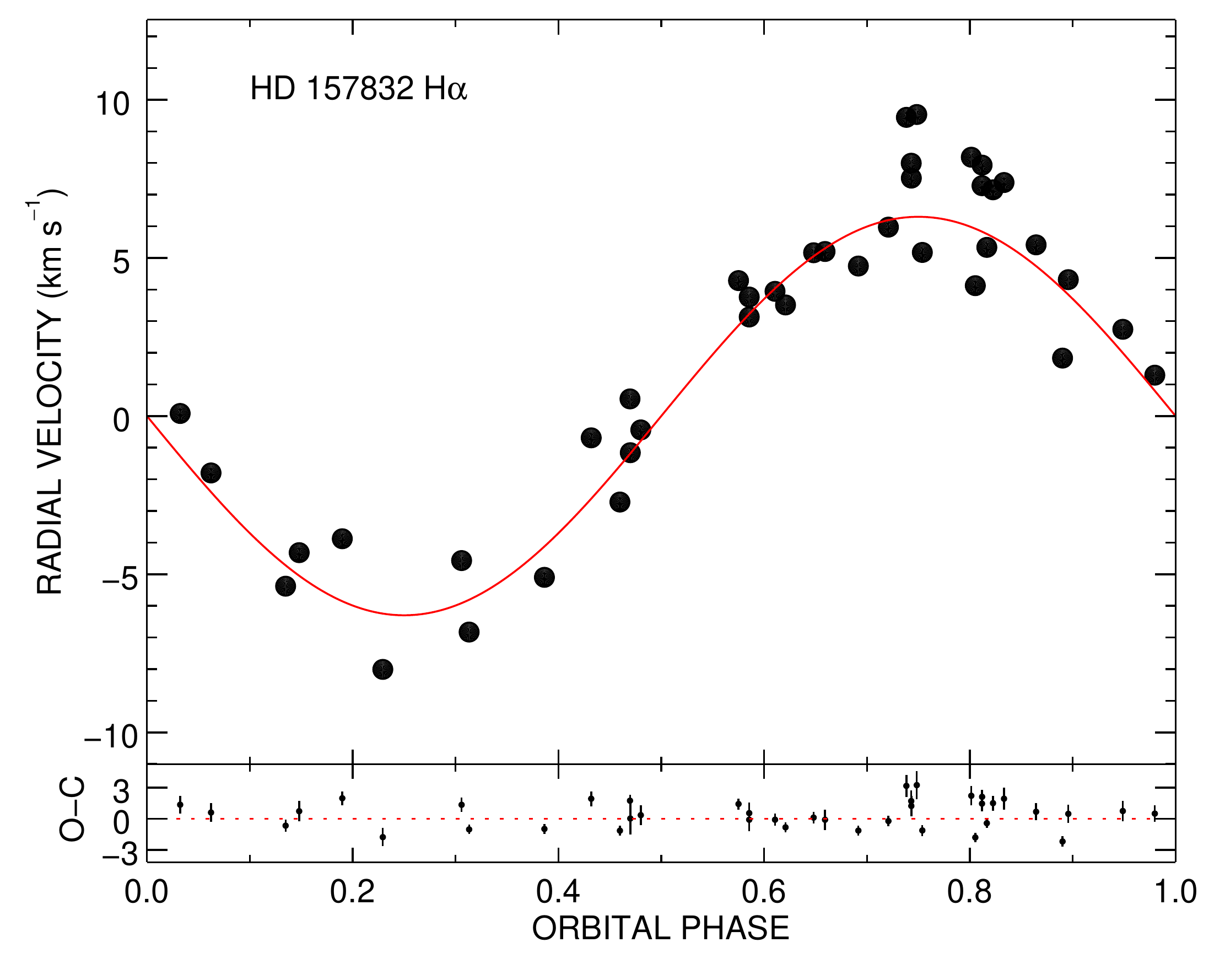}{1.0\textwidth}{}
        }
\caption{The relative radial velocity curve (red) for the Be star HD 157832 from the H$\alpha$ measurements (black) shown in the same format as Figure~\ref{fig:RV_HD113120}. }
\label{fig:RV_HD157832}
\end{figure*}

\pagebreak

\begin{deluxetable*}{lr}[h!]
\tablecaption{Circular orbital elements of HD 157832 \label{tab:orbit_HD157832}}
\tablewidth{0pt}
\tablehead{
\colhead{Element} & \colhead{Value}
}
\startdata
$P$ (days)		& {95.23 $\pm$ 0.07} \\
$T_\mathrm{sc}$ (HJD$-$2,400,000)	&{58566.3 $\pm$ 0.4}\\
$K_1$ (km s$^{-1}$)		&{6.25 $\pm$ 0.20}\\
$\gamma$ (km s$^{-1}$)	&{18.85 $\pm$ 0.10}\\
$f(m)$ (\Mnor) &{0.0024 $\pm$ 0.0002}\\
$a_1\sin i$ (\Rnor)	& {11.76 $\pm$ 0.37} \\
rms (km s$^{-1}$) &  {1.0} \\
\enddata
\end{deluxetable*}

\begin{deluxetable*}{lrr}[h]
\tablecaption{Companion Star Mass for HD 157832 \label{tab:sdO_HD157832}}
\tablewidth{0pt}
\tablehead{
\colhead{Inclination} & \colhead{$q$} & \colhead{$M_2$} \\
\colhead{(deg)} & \colhead{($M_2/M_1$)} & \colhead{(\Mnor)}
}
\startdata
60     & {0.073}    & {0.80} \\ 
65     & 0.070    & 0.77 \\
70     & 0.067    & 0.74 \\
75     & {0.065}    & 0.72 \\
80     & 0.064    & {0.70} \\
\enddata
\end{deluxetable*}

\newpage
\null\vspace{8 cm}

\section{Spectral Variability of the B\lowercase{e} stars} \label{sec:Variability}
The observed Balmer profiles and metallic lines display spectral variations in the target stars. Here we document the line variability appearing in the spectra and measure their equivalent width values ($W_\lambda$) by adopting the convention of positive $W_\lambda$ values for stellar absorption profiles and negative values for emission lines.  

\subsection{HD 113120} \label{subsec:VarHD113120}
\begin{figure*}[h!]
\gridline{
    \fig{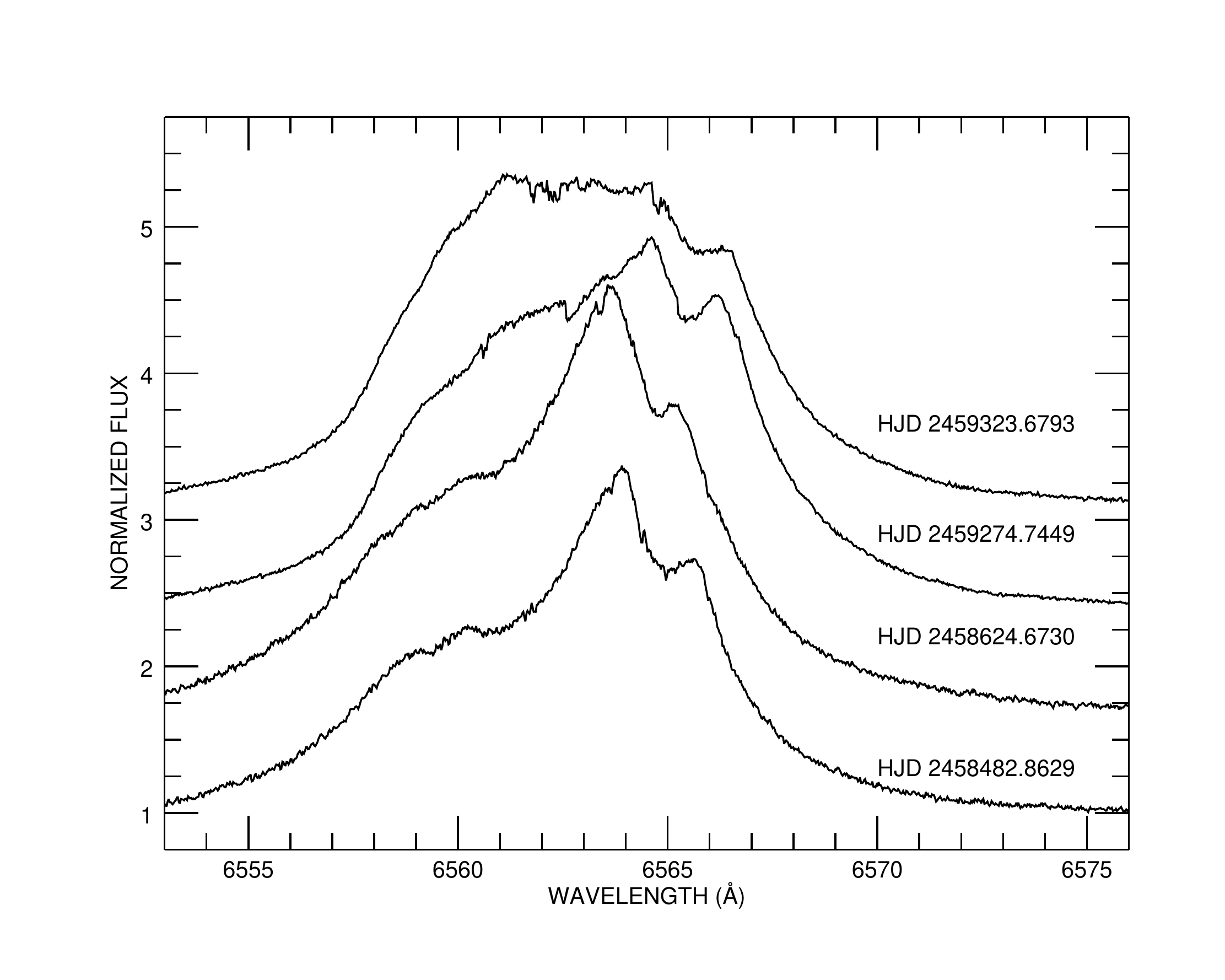}{0.5\textwidth}{ (a) H$\alpha$}
    \fig{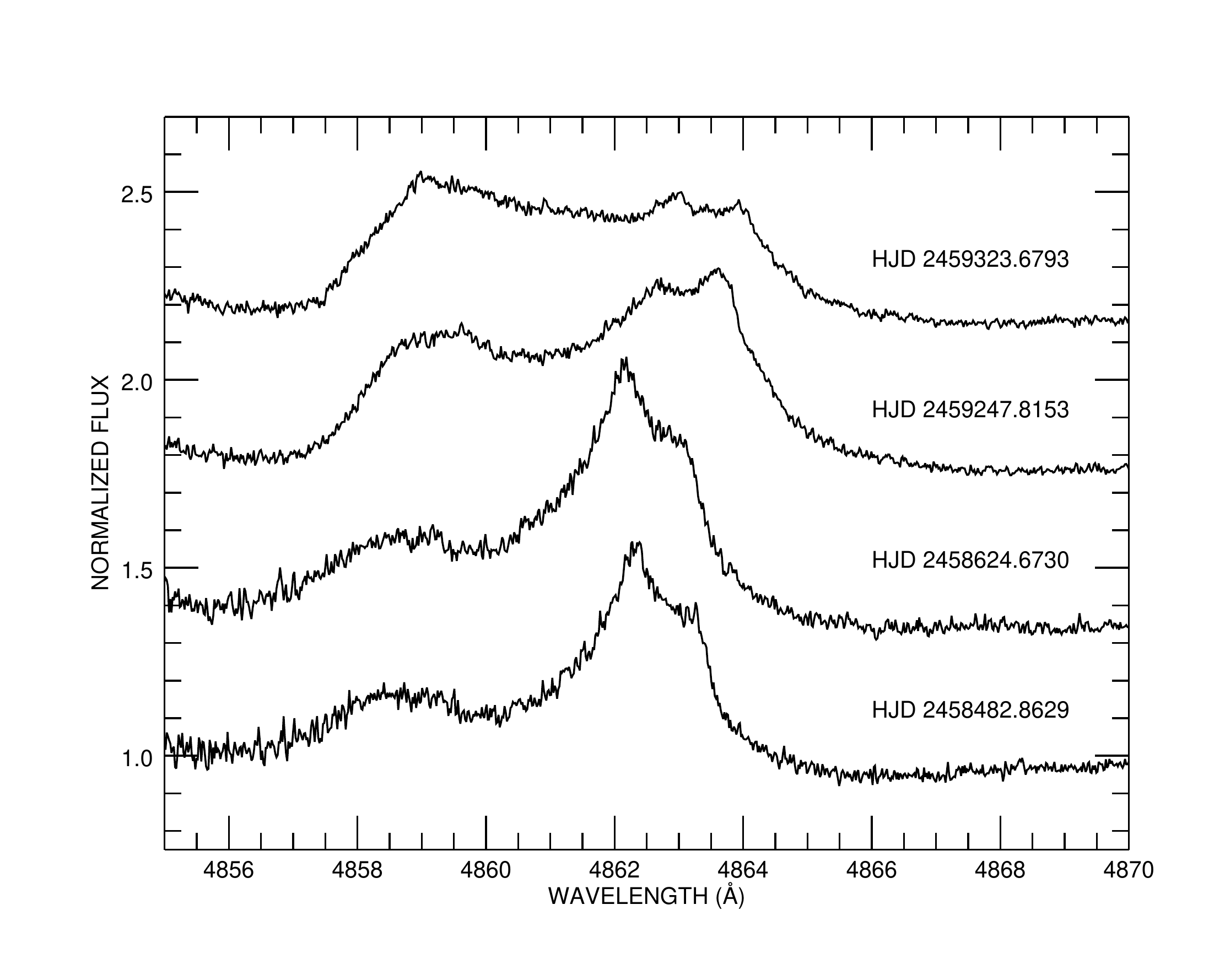}{0.5\textwidth}{ (b) H$\beta$}
        }
\caption{Emission profile variations of HD 113120 for H$\alpha$ (left) and H$\beta$ (right). This sub-sample is arranged with vertical offsets that increase with time.}
\label{fig:vari_HD113120}
\end{figure*}

The emission profiles of H$\alpha$ and H$\beta$ displayed striking variations in shape through the observing campaign. Figure~\ref{fig:vari_HD113120} shows a subset of the profiles to illustrate the changes on several timescales.  At the beginning of the observing sequence (HJD 2458482), the profiles are marked by a strong central peak plus sub-peaks in the red and far-blue.  After 142 days, the central peak increased in strength, but was almost absent some 650 days later, while the red and blue peaks continued to grow. In the final spectrum (HJD 2459323), the central region appears as a plateau.  The basic variations seen in H$\alpha$ are also present in H$\beta$, but the latter is more influenced by the changes in the sub-peaks. 
These sub-peaks also appear as a narrow emission component in the \ion{He}{1} $\lambda 6678$ line.

We measured the $W_\lambda$ values of H$\alpha$, H$\beta$, and \ion{He}{1} $\lambda\lambda 6678, 7065$ profiles, and these are collected in Table~\ref{tab:EW_all} of Appendix~\ref{sec:EW_other}. Time plots of the measured $W_\lambda$ values are given in Figure~\ref{fig:EW_HD113120}.  All four lines show a rapid increase in emission over the first hundred days of observation.  Slower increases followed for H$\alpha$ and H$\beta$, while the \ion{He}{1} emission generally declined.  These changes are consistent with an episode of increased mass ejection into the base of the disk followed by an outward expansion of a density enhancement. 

\placefigure{fig:spec_vari_HD113120}

\newpage
\null\vspace{0 cm}

\subsection{HD 137387} \label{subsec:VarHD137387}
\begin{figure*}[h!]
\gridline{
    \fig{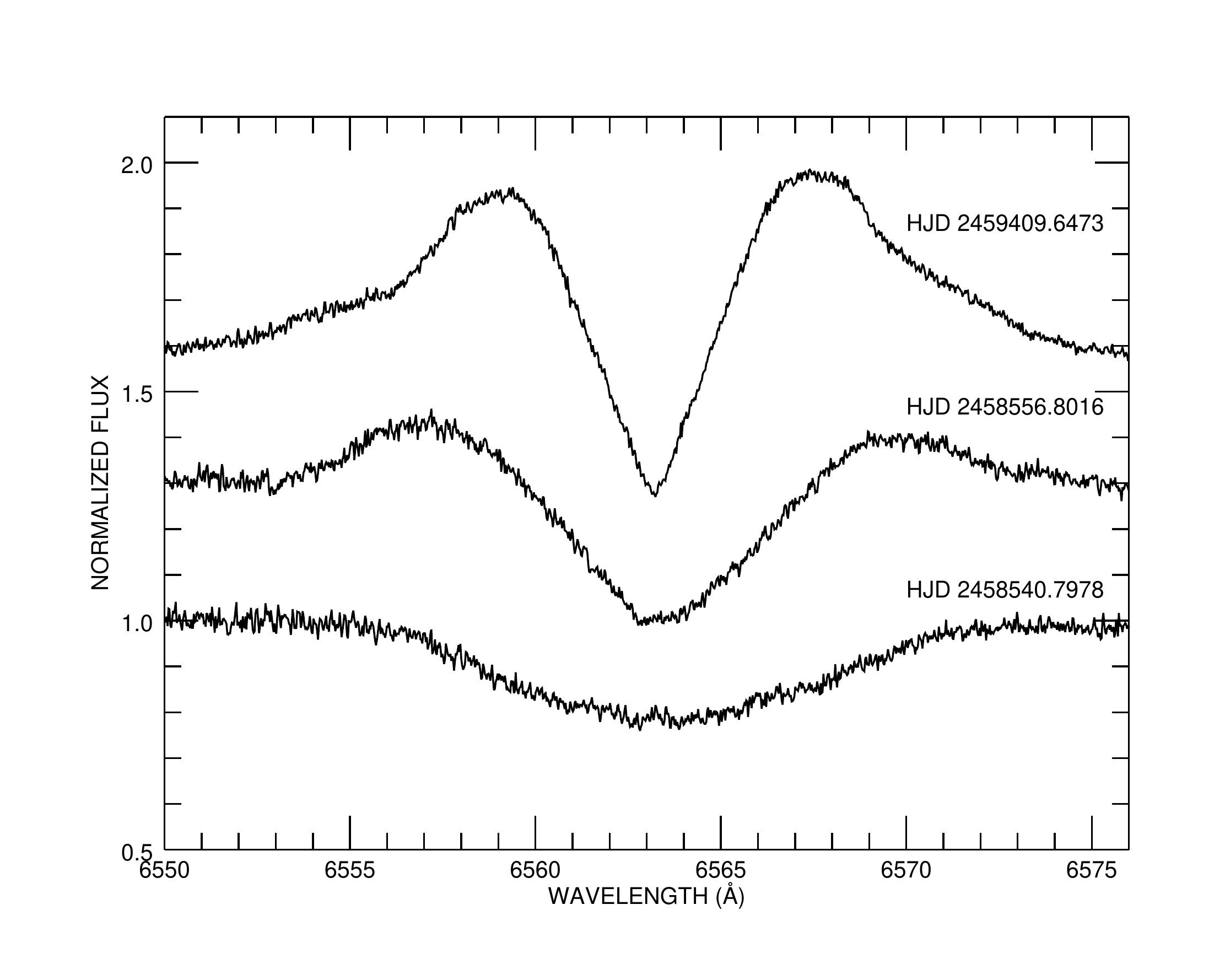}{0.5\textwidth}{(a) H$\alpha$}
        \fig{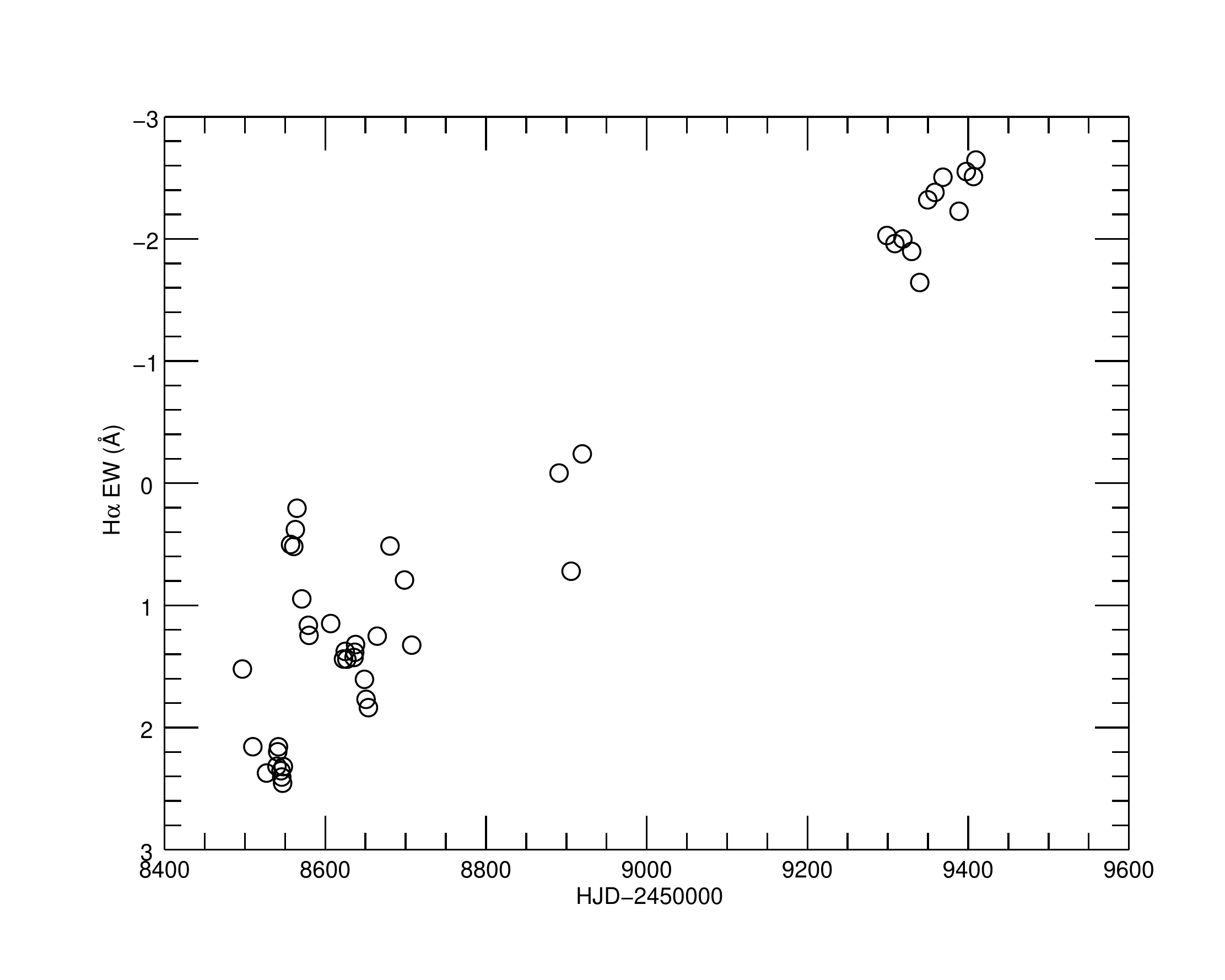}{0.5\textwidth}{(b) $W_{\rm H\alpha}$}
        }
\caption{H$\alpha$ variations of HD 137387. Panel (a) (left) shows the growth of new emission on short and long timescales with offsets of +0.3 and +0.6 in normalized flux. Panel (b) (right) shows the time evolution of equivalent width $W_\lambda$ (summed over the wavelength range from 6552 to 6578 \AA\ ).}
\label{fig:vari_HD137387}
\end{figure*}

The spectra of HD 137387 showed broad absorption in the H$\beta$, \ion{He}{1}, and metallic lines.  However, in the case of H$\alpha$, the profile changed from broad absorption (see spectrum observed on HJD 2458540 in Fig.~\ref{fig:vari_HD137387}) to a narrower shell absorption profile  superimposed upon broad emission (see HJD 2459409). The associated change in H$\alpha$ equivalent width is shown in panel (b) of Figure~\ref{fig:vari_HD137387}. The $W_\lambda$ measurements of H$\alpha$, H$\beta$, \ion{He}{1} $\lambda\lambda4921, 5015, 5875, 6678, 7065$, and \ion{Ca}{2} $\lambda8542$ are given in Table~\ref{tab:EW_all} of Appendix~\ref{sec:EW_other}. The temporal variation of these $W_\lambda$ measurements are shown in Figure~\ref{fig:EW_HD137387}.    

\placefigure{fig:spec_HD137387}

\pagebreak

\subsection{HD 152478} \label{subsec:VarHD152478}
\begin{figure*}[h!]
\gridline{
        \fig{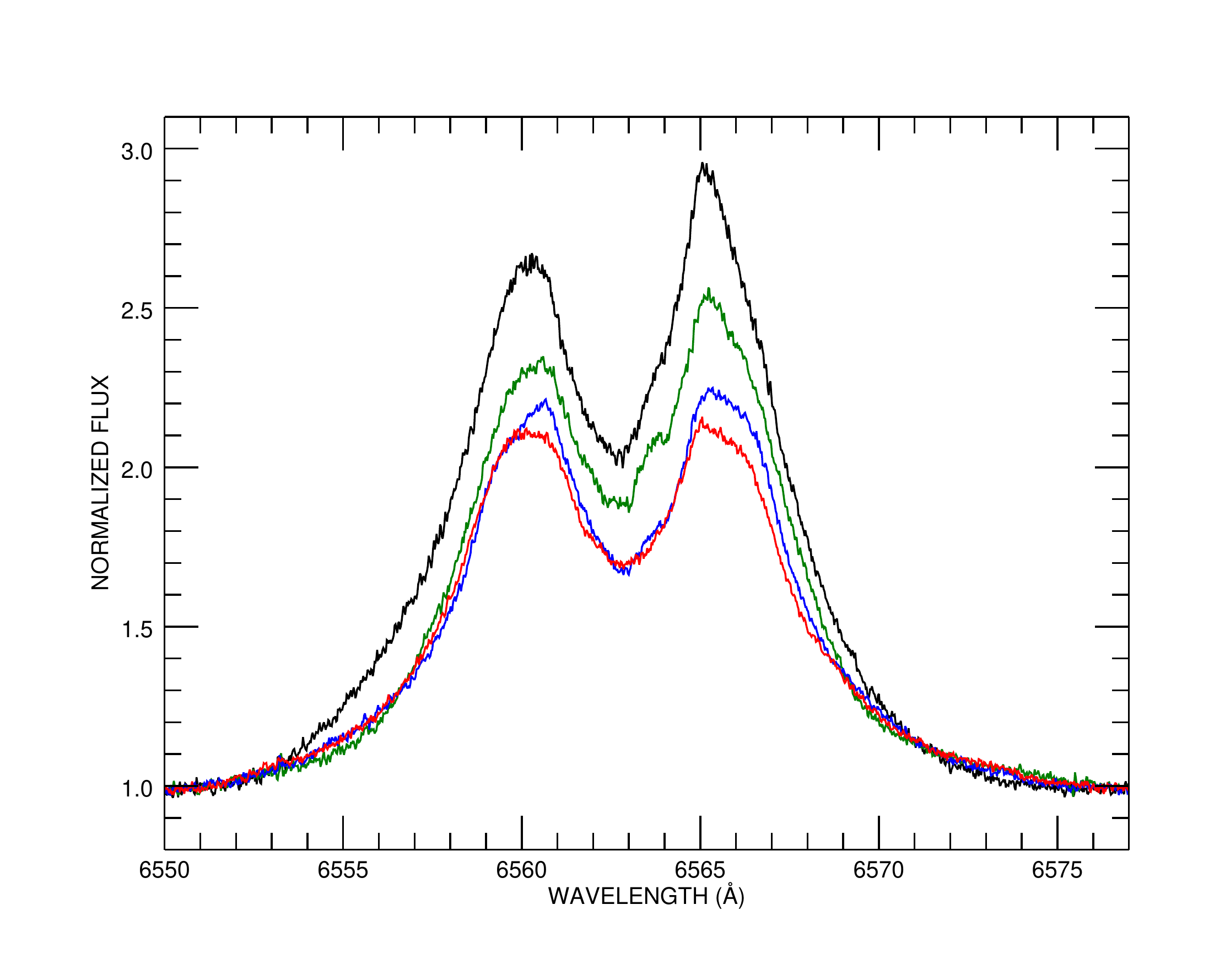}{0.5\textwidth}{(a) H$\alpha$}
        \fig{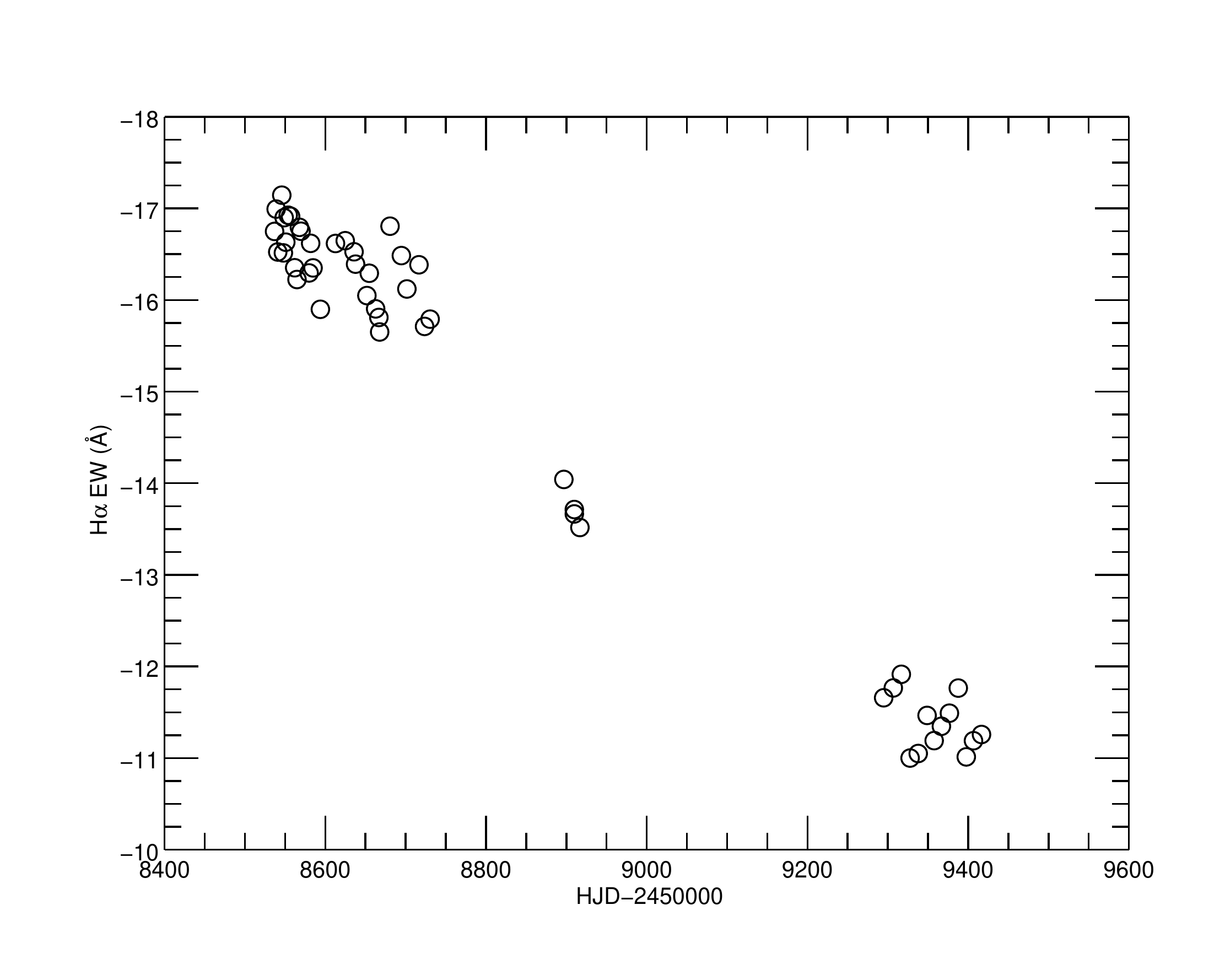}{0.5\textwidth}{(b) $W_{\rm H\alpha}$}
        }
\caption{(a) (left) The H$\alpha$ variations of HD 152478. The red peak dominated at the beginning of the observing campaign (black, HJD 2458536). The profile gradually decreased in strength (green, HJD 2458916, and blue, HJD 2459294). The red and violet emission peaks reached a comparable and lower strength towards the end of the observations (red, HJD 2459416). (b) (right) Temporal evolution of the H$\alpha$ equivalent width (integrated over 6552 to 6577 \AA ). }
\label{fig:vari_HD152478}
\end{figure*}

The double-peaked emission features of H$\alpha$ for HD 152478 displayed a change in $V/R$. At the beginning of the observing campaign, the profile had a  red peak that was stronger than the violet peak (HJD 2458536, black in panel (a) of Fig.~\ref{fig:vari_HD152478}). The asymmetry in the peak strength diminished over 380 days (HJD 2458916, green), and the two emission peaks attained comparable strength by HJD 2459294 (blue). The overall emission strength decreased to a minimum at the end of the campaign (HJD 2459416, red). The time evolution of the H$\alpha$ equivalent width is shown in panel (b) of Figure~\ref{fig:vari_HD152478}. The $W_\lambda$ measurements are collected in Table~\ref{tab:EW_all} of Appendix~\ref{sec:EW_other}, together with measurements made of H$\beta$ and \ion{He}{1} $\lambda\lambda 4921, 6678, 7065$.

\placefigure{fig:spec_HD152478}

\pagebreak

\subsection{HD 157042} \label{subsec:VarHD157042}
\begin{figure*}[h!]
\gridline{
    \fig{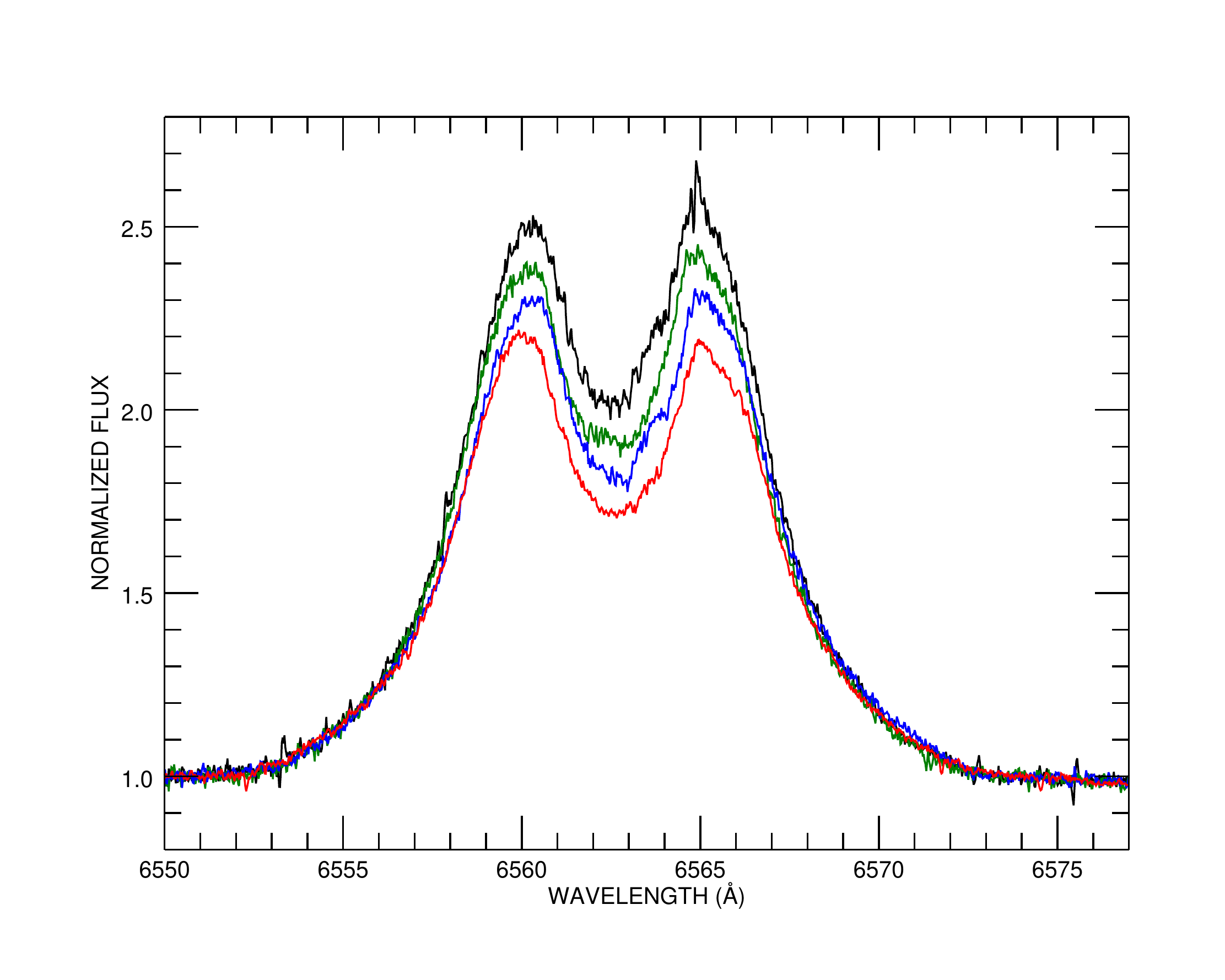}{0.5\textwidth}{(a) H$\alpha$}
        \fig{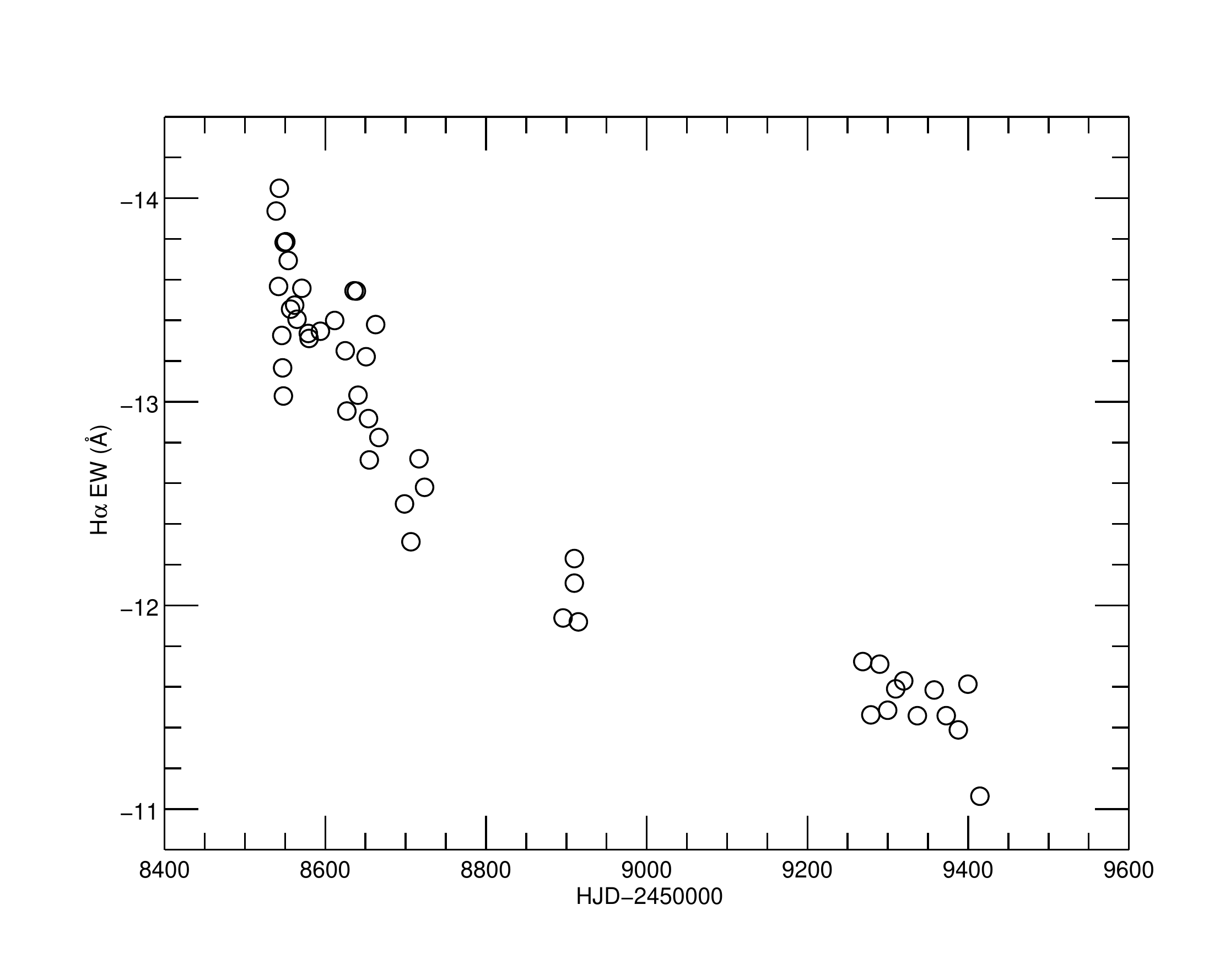}{0.5\textwidth}{(b) $W_{\rm H\alpha}$}
        }
\caption{(a) (left) H$\alpha$ profiles of HD 157042 displayed a gradual weakening throughout the observing campaign. Observations made on the nights of HJD 2458538, 2458716, 2458909, and 2459414 are plotted in black, green, blue, and red, respectively. (b) (right) The temporal variations of H$\alpha$ equivalent width (integrated over the range 6551 to 6574 \AA ). }
\label{fig:vari_HD157042}
\end{figure*}

The $V$ and $R$ peaks of the H$\alpha$ emission profiles of HD 157042 displayed comparable relative strength over the time span of the observations. However, the whole profile showed a gradual decrease in strength over time (see panel (a) of Fig.~\ref{fig:vari_HD157042}). No significant spectral variations appeared in H$\beta$, \ion{He}{1}, and metallic line profiles. We report the measured $W_\lambda$ values of H$\alpha$, H$\beta$, and \ion{He}{1} $\lambda\lambda4921$, 6678, 7065 in Table~\ref{tab:EW_all} of Appendix~\ref{sec:EW_other}.

\placefigure{fig:spec_HD157042}

\null\vspace{0 cm}

\subsection{HD 157832} \label{subsec:VarHD157832}
\begin{figure*}[h]
\plottwo{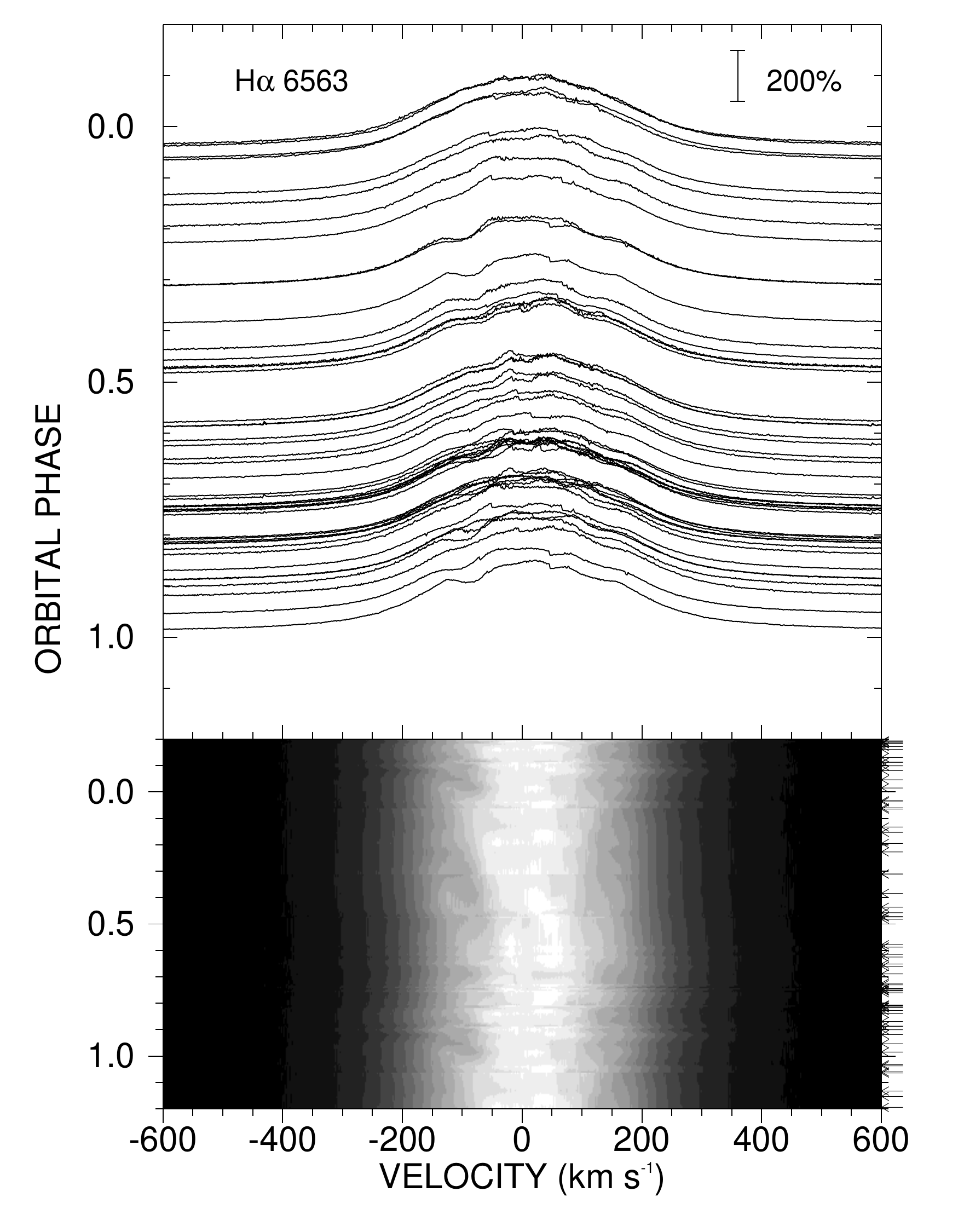}{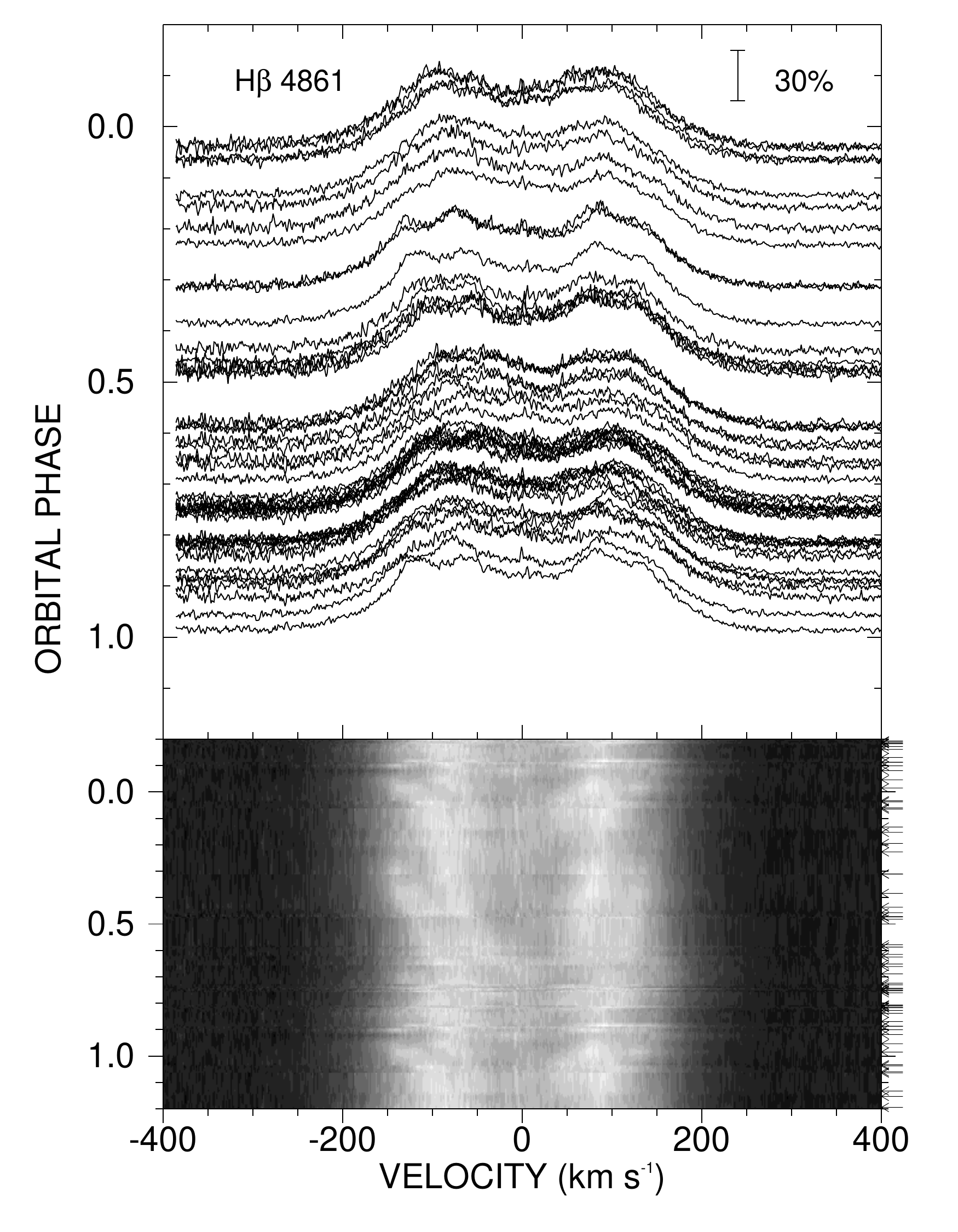}
\caption{Left: H$\alpha$ profiles of HD 157832 plotted as a function of the orbital phase and Doppler shift. The continuum of each profile is aligned with orbital phase in the top panel, and the flux scale relative to the continuum is indicated by the vertical bar in the upper right. A gray-scale version of the profiles appears in the bottom panel in which the brightness scales with emission flux above the continuum. The arrows indicating the orbital phases of the observations are marked along the right ordinate. Right: The H$\beta$ profiles plotted in the same format as the left panel, except for different flux scaling.}
\label{fig:gray_HD157832}
\end{figure*}

The H$\alpha$ emission profiles of HD 157832 are broad and strong, and they are marked by narrower sub-features that appear to vary with orbital phase. The H$\alpha$ and H$\beta$ profiles are plotted as functions of Doppler shift and orbital phase in Figure~\ref{fig:gray_HD157832}. The upper panels show the profiles offset according to orbital phase while the lower panels portray the emission flux as a grayscale image. These show that profiles exhibit migrating sub-features.  For example, a weak depression appears in the blue wing of H$\alpha$ near $V_{r}=-75$ km~s$^{-1}$ at $\phi=0.3$, and this sub-feature moves slightly redward to persist for a part of the orbit. The sub-features are more striking in H$\beta$, and excess emission appears in sub-features at orbital phases $\phi=0.0$ and $0.5$ that progress from both line wings towards the line center (right panel of Fig.~\ref{fig:gray_HD157832}). We give the $W_\lambda$ measurements for H$\alpha$ and H$\beta$ in Table~\ref{tab:EW_all} of Appendix~\ref{sec:EW_other}.  The appearance of coherent patterns of emission in spectra from many different orbital cycles offers independent support for the period derived from the radial velocities (Section \ref{subsec:HD157832}).

Such phase-locked spectral variability appears in the 
Balmer profiles of other Be binaries.  For example, 
the variations of the H$\beta$ profiles of HD~157832  
are very similar to those observed in HD~55606 by 
\citet{Chojnowski2018} (see the bottom panels of their Fig.~7). 
\citet{Chojnowski2018} argue that the variations are 
related to a two-armed spiral density pattern in the Be star 
disk that results from the tidal pull of the companion 
(see the models of \citealt{Panoglou2018}).  The portion 
of a spiral arm that is aligned with the line of sight will
concentrate emission flux at a Doppler shift associated 
with the Keplerian velocity at the distance of the arm 
from the Be star. We first observe the excess flux from 
the inner part of the arm at large speed, and then we sample 
the line-of-sight portions of the arm at larger radii and 
lower speed as the orbit progresses.  This is observed as
an emission sub-feature that is seen first in the line wing and 
then progresses towards line center.  In a two-armed spiral, 
this occurs twice each orbit with the motions of one arm 
mirrored in velocity space by the other arm.

\placefigure{fig:spec_HD157832}


\null\vspace{8 cm}

\section{Secondary Lines in the Visual Spectrum} \label{sec:4686}
The spectral signature of the hot sdO companions of the targets were discovered in the ultraviolet, where a host of narrow absorption lines are detected through cross-correlation analysis \citep{Wang2021}. We made a visual inspection for similar narrow-lined features in the visible spectra recorded by CHIRON, but none were found. This is not surprising given the expected faintness of the sdO spectrum in the visible region (see below). Thus, direct measurements of the radial velocities of the hot companions are still limited to those from ultraviolet spectroscopy from IUE and HST that we used in the previous section for preliminary double-lined orbital solutions.

Nevertheless, we can predict the spectral appearance of the sdO stars in the visible band using the stellar parameters and flux ratio determined from the analysis of the ultraviolet spectrum. Here we focus on the spectral region in the immediate vicinity of the \ion{He}{2} $\lambda 4686$ transition. This feature is prominent in all hot, O-type spectra while absent in the cooler B-type spectra of the Be star components.  Indeed the first discovery of a Be+sdO binary was made thanks to the presence of \ion{He}{2} $\lambda 4686$ in the spectrum of $\phi$~Per \citep{Poeckert1981}.

\begin{figure*}[h!]
\gridline{\fig{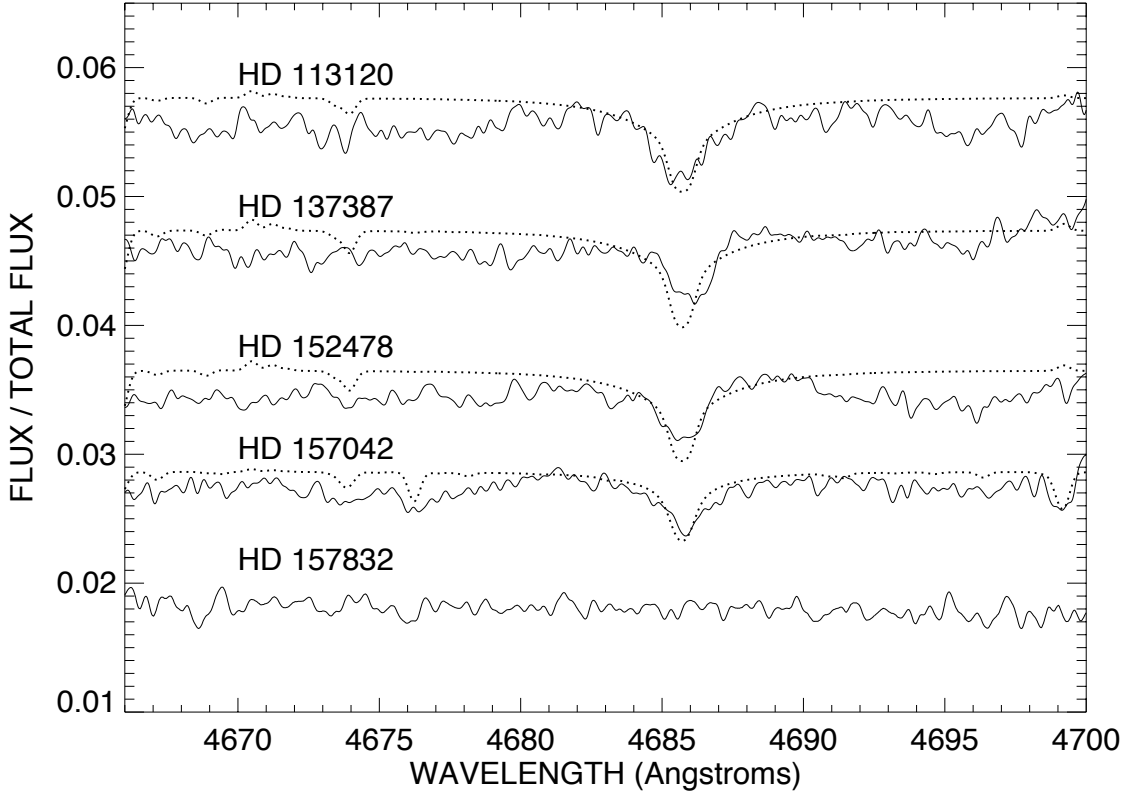}{1.0\textwidth}{}}
\caption{The spectra of the sdO components from Doppler tomography reconstructions of the CHIRON spectroscopy (solid lines) that were made using the preliminary, double-lined, orbital velocity solutions. There is no double-lined orbit for HD~157832, so we assumed reflex motion of the companion for a mass ratio of $M_2/M_1= 0.1$ and then plotted the reconstructed spectrum for a supposed flux contribution of $f_2/(f_1 + f_2) = 0.018$. TLUSTY model spectra predicted from analysis of the HST ultraviolet spectra \citep{Wang2021} are overplotted as dotted lines. The reconstructed and model spectra are shown with vertical offsets of 0.04, 0.03, 0.02, and 0.01 for HD~113120, HD~137387, HD~152478, and HD~157042, respectively, for ease of comparison. The reconstructed spectra reveal the presence of the expected \ion{He}{2} $\lambda 4686$ line in all but the case of HD~157832. }
\label{fig:4686}
\end{figure*}

\placefigure{fig:4686}

Our goal in this section is to compare the predicted visible spectrum of the sdO component with the CHIRON spectroscopy.  We used the TLUSTY, solar-metallicity, atmosphere models from \citet{Lanz2003} and \citet{Lanz2007} to create model spectra for the hotter sdO and cooler Be components, respectively. The first step was to estimate the monochromatic flux ratio of the stars $f_2/f_1$ in the vicinity of the \ion{He}{2} $\lambda 4686$ line. We adopted the stellar parameters (effective temperature, gravity) of the Be stars from \citet{Wang2021} to create model spectral energy distributions (SEDs).  The SEDs for the sdO stars were set according to the temperatures derived by \citet{Wang2021} with the gravity set to $\log g = 4.75$, the largest value available in the TLUSTY grid but probably lower than the actual gravities. We formed the ratio of the sdO to Be fluxes by normalizing the ratio to the derived value in the 1450 -- 1460 \AA\ region \citep{Wang2021}. We then evaluated the wavelength-dependent flux ratio $f_2/f_1$ in the continuum adjacent to \ion{He}{2} $\lambda 4686$. The expected sdO flux contributions are low: $f_2/f_1 =$ 0.018, 0.018, 0.017, and 0.019 for HD~113120, HD~137387, HD~152478, and HD~157042, respectively (there was no detection and hence no flux ratio estimate for HD~157832). Finally, the detailed spectrum of the sdO star was derived from the TLUSTY models for the same wavelength grid as that of the CHIRON echelle order recording \ion{He}{2} $\lambda 4686$, and this model spectrum was normalized to its expected net flux contribution, $f_2/(f_1 + f_2)$.

We used a Doppler tomography algorithm \citep{Bagnuolo1994} with the preliminary orbital velocity solutions to reconstruct the individual spectral components of the Be and sdO stars from the entire set of CHIRON spectra for each target. The reconstructed spectra have a much higher S/N ratio than that of the individual spectra.  The distribution of flux between components is set in advance, and we arbitrarily assigned the sdO fluxes to the predicted values extrapolated from the ultraviolet flux ratios.  The final results are plotted in Figure~\ref{fig:4686} in units of the combined flux $f_2/(f_1 + f_2)$. The solid lines depict the reconstructed sdO spectra that were smoothed to a resolving power of $R=8800$ to improve the spectral S/N ratio. These are compared to the predicted model spectra (also smoothed) shown as dotted lines. We see that the \ion{He}{2} $\lambda 4686$ absorption line is present and appears similar to the model predictions in all the sdO spectra except for the case of HD~157832 where no \ion{He}{2} line is detected. 
The reconstructed spectrum of the secondary of HD~157832
shown in Figure~\ref{fig:4686} was obtained using Doppler shifts 
based upon a trial mass ratio of $M_2/M_1=0.1$, but 
we arrived at similar non-detections of \ion{He}{2} $\lambda 4686$
using a range of mass ratios (0.05 to 0.13) for the reconstruction.
This is consistent with the lack of detection of the sdO lines in the ultraviolet spectrum of HD~157832 from HST \citep{Wang2021}.  This suggests that the companion of HD~157832 is either too cool to create a \ion{He}{2} line or is much fainter than found in the other cases.  The velocity registration of the observed and model profiles is generally good except in the case of HD~137387, which suggests that the systemic velocity derived from the preliminary double-lined orbit may need adjustment for the \ion{He}{2} $\lambda 4686$ line.

The presence of the \ion{He}{2} $\lambda 4686$ line in the reconstructed spectra of the faint companions confirms that the spectral signal of the sdO stars is found in the visible spectrum. However, their flux contribution in the visible is small, and it would require very high S/N spectroscopy to detect the \ion{He}{2} $\lambda 4686$ line in an individual spectrum.  The success of the tomographic reconstructions in finding such faint features attests to the reliability of our preliminary orbital solutions that were the basis of the tomographic reconstructions.


\section{Discussion} \label{sec:dis}
Since the first detection of the stripped helium star in the $\phi$ Per binary system, multi-wavelength search efforts using optical and FUV spectroscopy and near-IR ground-based interferometry have led to the detection and characterization of the orbital and physical properties of eight confirmed Be+sdO binaries. 
Our work has provided preliminary double-lined orbits for four additional systems, so it is worthwhile to review the orbital and physical properties of the current sample.  We list in Table~\ref{tab:Be+sdO} our results for these double-lined systems plus HD~157832.  We give estimates of the Be star mass in Table~\ref{tab:Be+sdO} from prior work by  \citet{Zorec2016} and \citet{Tetzlaff2011}.  The values $M_{\rm Be}$ shown are the means of the estimates from these two works with uncertainties that span the ranges from both. Unfortunately, there are no prior estimates for the mass of HD~157832, so we simply list an estimate from Be stars of similar effective temperature \citep{Lopes2011} in the compilation by \citet{Zorec2016}. Then the masses of the companions $M_{\rm sdO}$ follow from the mass ratios quoted in the orbital solutions given in Section \ref{sec:RVs} (or from Table \ref{tab:sdO_HD157832} in the case of HD~157832).

\begin{deluxetable*}{lllllll}[h!]
\tabletypesize{\scriptsize}
\tablecaption{Be+sdO Binary Systems with Orbital Periods\label{tab:Be+sdO} }
\tablewidth{0pt}
\tablehead{
\colhead{HD} & \colhead{Star} & \colhead{Orbital} & \colhead{$K_1$} & \colhead{$M_\mathrm{Be}$} & \colhead{$M_\mathrm{sdO}$} & \colhead{Reference} \\
\colhead{Number} & \colhead{Name} & \colhead{Period (d)} & \colhead{(km s$^{-1}$)} & \colhead{(\Mnor)} & \colhead{(\Mnor)} & \colhead{} 
}
\startdata
\multicolumn{7}{c}{This work} \\
\hline
113120 & LS Mus       & {181.54 $\pm$ 0.11} & {10.66 $\pm$ 0.03} & $10.1\pm 2.2$  & {1.43 $\pm$ 0.31}   & \nodata \\
137387 & $\kappa$ Aps & {192.1 $\pm$ 0.1}  & {9.09 $\pm$ 0.08} & $11.8\pm 1.0$  & {1.60 $\pm$ 0.14}   & \nodata \\
152478 & V846 Ara     & {236.50 $\pm$ 0.18} & {4.33 $\pm$ 0.05}  & $6.5\pm 1.3$   & {0.53 $\pm$ 0.11} & \nodata \\
157042 & $\iota$ Ara  & 176.17 $\pm$ {0.04} & {5.80 $\pm$ 0.06}  & $10.5\pm 2.9$  & {1.06 $\pm$ 0.29} & \nodata \\
157832 & V750 Ara     & {95.23 $\pm$ 0.07}  & {6.25 $\pm$ 0.20}  & 11$^a$         & $0.7^b$  & \nodata \\
\hline
\multicolumn{7}{c}{FUV confirmed Be+sdO binary systems} \\
\hline 
\phn10516   & $\phi$ Per & $126.6982 \pm 0.0035$ & $10.2 \pm 1.0$ & $9.6 \pm 0.3$ & $1.2 \pm 0.2$ & \citet{Mourard2015} \\
\phn41335   & HR 2142 & $80.913 \pm 0.018$ & $7.1 \pm 0.5$ & 9 & 0.7 & \citet{Peters2016} \\
\phn55606   & BD$-$01 1603 & $93.76 \pm 0.02$ & $10.74 \pm 1.17$ & {6.0$-$6.6} & {0.83$-$0.90} & \citet{Chojnowski2018}  \\
\phn58978   & FY CMa & $37.253 \pm 0.007$ & $17.4 \pm 0.9$ & {10$-$13} & {1.1$-$1.5} & \citet{Peters2008} \\
109387  & $\kappa$ Dra & 61.5496 $\pm$ 0.0058 & 6.90 $\pm$ 0.15 & 3.65 $\pm$ 0.48 & 0.426 $\pm$ 0.043 & \citet{Klement2022b} \\
194335  & V2119 Cyg & $63.146 \pm 0.003$ & \nodata & $8.65 \pm 0.35$ & $1.62 \pm 0.28$ & \citet{Klement2022a} \\
200120 & 59 Cyg & $28.1871 \pm 0.0011$ & $11.7 \pm 0.9$ & {6.3$-$9.4} & {0.62$-$0.91} & \citet{Peters2013}  \\
200310 & 60 Cyg & $147.68 \pm 0.03$ & $11.6 \pm 1.2$ & $7.3 \pm 1.1$ & $1.2 \pm 0.2$ & \citet{Klement2022a} \\
\hline
\multicolumn{7}{c}{Be+sdO/sdB binary candidates identified from optical spectroscopy} \\
\hline
\phn37202 & $\zeta$ Tau & $132.987 \pm 0.050$ & $7.43 \pm 0.46$ & 11 & {0.87$-$1.02} & \citet{Ruzdjak2009} \\
\phn63462 & $o$ Pup & $28.903 \pm 0.004$ & $10.3 \pm 9.6$ & {11$-$15} & $0.7 \pm 1.0$ & \citet{Koubsky2012} \\
\phn68980 & MX Pup & $5.1526 \pm 0.0011$ & 0.90 & 15 & {0.6$-$6.6} & \citet{Carrier2002} \\
148184 & $\chi$ Oph & $138.8 \pm 1.3$ & $15.0 \pm 1.8$ & 10 & {1.7$-$2.0} & \citet{Abt1978} \\
161306 & MWC 271 & $99.90 \pm 0.50$ & $4.90 \pm 1.53$ & 15 & 0.9 & \citet{Koubsky2014} \\
167128 & HR 6819 & $40.334 \pm 0.005$ & $3.9 \pm 0.7$ & 6 & {0.4$-$0.8} & \citet{Gies2020}\\
183537 & 7 Vul & $69.4212 \pm 0.0034$ & $8.86 \pm 0.62$ & {5.47$-$8.98} & {0.56$-$0.91} & \citet{Harmanec2020} \\
184279 & V1294 Aql & $192.91 \pm 0.18$ & $6.26 \pm 0.61$ & 16.9 & {1.171$-$1.361} & \citet{Harmanec2022} \\
\nodata & LB$-$1 & $78.7999 \pm 0.0097$ & $11.2 \pm 1.0$ & $7 \pm 2$ & $1.5 \pm 0.4$ & \citet{Shenar2020} \\
\hline
\multicolumn{7}{c}{$\gamma$\,Cas-analog Be binaries$^b$} \\
\hline
\phn\phn5394 & $\gamma$\,Cas & $203.52 \pm 0.08$ & $4.297 \pm 0.090$ & 13 & 0.98 & \citet{Nemravova2012} \\
\phn12882 & V782 Cas & $122.0 \pm 1.5$ & $5.2 \pm 0.9$ & 9 & {0.6$-$0.7} & \citet{Naze2022} \\
\phn45995 & BD$+11$ 1204 & $103.1 \pm 1.0$ & $6.7 \pm 0.4$ & 10 & $1.0 \pm 0.1$ & \citet{Naze2022} \\
183362 & V558 Lyr & $83.3 \pm 1.8$ & $8.2 \pm 1.1$ & 8 & {0.7$-$0.8} & \citet{Naze2022} \\
212571 & $\pi$ Aqr & $84.07 \pm 0.02$ & $16.7 \pm 0.2$ & $15 \pm 3$ & $2.4 \pm 0.5$ & \citet{Bjorkman2002} \\
220058 & V810 Cas & $75.8 \pm 0.7$ & $6.4 \pm 0.7$ & 12.5 & {0.7$-$0.8} & \citet{Naze2022} \\
\nodata & SAO 49725 & $26.11 \pm 0.08$ & $2.8 \pm 0.5$ & 13 & {0.2$-$0.5} & \citet{Naze2022} \\
\nodata & V2156 Cyg & $126.6 \pm 2.0$ & $5.5 \pm 0.7$ & 11 & {0.7$-$0.8} & \citet{Naze2022} \\
\enddata
\tablenotetext{a}{Based upon mass estimates for Be stars of similar effective temperature.}
\tablenotetext{b}{Companion may not be sdO type.}
\end{deluxetable*}

 Table~\ref{tab:Be+sdO} also collects results on the orbital period, Be star semi-amplitude $K_1$, and stellar masses of the other confirmed Be+sdO binaries, candidate systems, and the recent detections of Be binaries among the $\gamma$~Cas analog stars with hard X-ray emission \citep{Naze2022}.
 The periods and masses of the systems discussed here are similar to those of the other confirmed binaries.  There are now three systems, HD 113120, HD 137387, and HD 194335, that have sdO masses above the Chandrasekhar limit, and these are potentially the progenitors of hydrogen-poor supernovae that will leave a neutron star remnant in a Be X-ray binary (BeXRB). 
 
 The periods of the five systems discussed here tend to be longer than those from prior investigations.  For example, HD~152478 has the longest orbital period of {237} days in the known population.  We suspect that this is a selection effect.  Determinations of the earlier binary solutions were more favorable in shorter period systems where the Be star semiamplitudes $K_1$ are larger.  On the other hand, our sample was selected from detections of the sdO FUV spectrum in a larger survey of the ultraviolet spectra \citep{Wang2018}, so the longer periods in the current sample are probably more representative of Be+sdO binaries as a whole.
 
 We now have a sample of 12 systems (8 prior plus 4 new) with known mass ratios, and the mean ratio is $q=M_2/M_1 =$ {0.124} $\pm$ {0.059} or $Q=M_1/M_2 =$ {8.1} $\pm$ {3.8}, where the error is the standard deviation of the sample.  The scatter in $q$ between systems is larger than the individual uncertainties, so there are significant differences in mass ratio among the systems.  However, there is no apparent dependence of mass ratio on the Be star mass.  Thus, if the mass ratio after stripping of the envelope of the donor is independent of mass, then we may expect that sdO remnants with mass greater than the Chandrasekhar limit will generally be found in binaries with Be star masses greater than {8.1} $\times 1.4 =$ {11.3} $M_\odot$.  This is consistent with the finding that the descendants of the massive Be+sdO systems, the BeXRBs, have Be star masses that are never less than $8\ M_\odot$ \citep{Reig2011}.

 It is useful to compare estimates of the orbital inclination with those for the Be star and its disk, because on evolutionary grounds we expect that mass transfer transformed orbital angular momentum into spin angular momentum. 
 Future interferometric observations should yield the orbital (and possibly disk) inclination \citep{Klement2022b,Klement2022a}. 
 However, we can estimate the orbital inclination from the $M_1 \sin^{3} i$ products given in Section 3 and the estimated masses of the Be star given in Table~\ref{tab:Be+sdO}.
These orbital inclination estimates are given in column two of Table~\ref{tab:inclination} for the four systems with double-lined orbital solutions. The orbital inclination values generally agree within errors with the stellar rotational inclinations derived by \citet{Zorec2016} (column three). Thus, the inclination estimates for these four systems agree with prediction that the orbital and Be star spin inclinations are the same.
 
\begin{deluxetable*}{lcc}[h!]
\tablecaption{Orbital and Rotational Inclinations \label{tab:inclination}}
\tablewidth{0pt}
\tablehead{
\colhead{HD} & \colhead{$i$(orbital)} & \colhead{$i$(rotational)} \\
\colhead{Number} & \colhead{($^\circ$)} & \colhead{($^\circ$)}
}
\startdata
113120     & 74 $\pm$ {4}    & $63 \pm 15$ \\ 
137387     & {60 $\pm$ 4}     & $50 \pm 12$ \\ 
152478     & {60 $\pm$ 7}    & $60 \pm 15$ \\ 
157042     & {46 $\pm$ 6}     & $59 \pm 14$ \\ 
\enddata
\end{deluxetable*}

\citet{Shao2014} performed theoretical calculations to simulate the population of Be binary systems in the local Galactic environment using a binary population synthesis (BPS) code. The authors investigated the population distribution considering both stable mass transfer and a common envelope channel during the early stage of the binary interaction. They concluded that Be binaries with  helium star companions are likely formed through stable mass transfer, and they avoid a spiral-in associated with the common envelope phase. A tailored BPS calculation was conducted by \citet{Shao2021} to predict the population distribution of Galactic Be+He binaries and to compare the results with observational data on confirmed Be+sdO binaries determined from FUV spectroscopy and the recently proposed Be+He binaries of LB$-1$ \citep{Shenar2020} and HR~6819 \citep{Gies2020}. The simulations map the parameter space (including [$M_\mathrm{He}$, $M_\mathrm{Be}$] and [$P_\mathrm{orb}$, $M_\mathrm{Be}$]) for the Be binary systems that have evolved through either Case A or Case B scenarios\footnote{Case A mass transfer refers to an interacting binary system in which the donor is a main-sequence star in the core hydrogen burning phase when it fills its Roche-lobe; in Case B, mass transfer occurs when the donor is in or evolving to the red giant phase with hydrogen shell burning following core exhaustion; Case C refers to the case when the donor is in an advanced phase after helium core exhaustion.} and include several cases of mass transfer efficiency. They conclude that Case B mass transfer is likely responsible for forming the most of the population of Be+He binaries.  Furthermore, models that assume almost conservative mass transfer (i.e., no systemic mass loss) yield mass ratios after stripping that agree with observations (see their Fig.~1, lower right panel).   

It is difficult at present to compare the model and observed distributions of mass and orbital period, because the sample is small and subject to selection effects.  \citet{Shao2021} find that most of the post-mass transfer binaries have a mass distribution that attains a maximum around $\sim 3-5$ \Mnor\ for the Be star and $\sim 0.3-0.6$ \Mnor\ for the He star. These are lower than found among the known systems, with the exception of the case of $\kappa$~Dra that has lower masses and cooler temperatures than the rest \citep{Klement2022a}.  The model orbital period distribution peaks over the range {25$-$100} days, and this is significantly lower than $95-${237} day range  of the new sample. 

The current models may need further development to capture all the important processes. For example, Be stars may have experienced rotation-dependent mass accretion during their early accretion history \citep{Stancliffe2009}. In addition, recent theoretical studies suggest that giant donor stars may enable stable mass transfer and avoid a common envelope stage. \citet{Ge2020} utilized an adiabatic mass-loss model to simulate the evolutionary state of binary stars, in which the donor stars reach the red giant branch (RGB) and asymptotic giant branch (AGB). They conclude that these binaries may significantly expand the parameter space of systems that undergo stable mass transfer than previously suggested by the BPS calculations. A similar result is obtained by \citet{Temmink2022} from 1D MESA simulations of the mass transfer stability for a wide range of interacting binaries. These findings open up the possibility of the formation of the Be+sdO systems through interactions with a more evolved giant donor star. Future improvements in the simulations, such as careful treatments of stellar winds, tidal interactions, and stellar rotation in the mass transfer and accretion histories, may be needed to reproduce the observed sample.

\section{Conclusions}\label{sec:conc}
Growing evidence suggests that the rapid rotation of Be stars is likely a consequence of past mass and angular momentum accretion from a companion star. If so, many Be stars will have hot, helium star companions that are the stripped remnants of the donor stars. Searches for the signature of the stripped helium companion stars are best accomplished in the FUV region of the spectra, and a number of prior studies have utilized IUE and HST FUV spectroscopy to detect the spectral signatures and properties of the stripped helium stars. However, two of the most important evolutionary pillars, the orbital period and the stellar mass of the helium star, are still missing in many cases. In this work, we have carried out a three-year observing campaign of high S/N and high resolving power spectroscopy to determine these missing quantities for five Be+sdO binary systems visible in the southern sky. The Be star spectra have lines that are much broader than the orbital Doppler shifts and that are often blended with emission components from the disks, so Doppler shift measurements are difficult and orbital solutions require long sequences of observations. Based upon a careful spectral line inspection, we succeeded in measuring the Doppler shifts of the Be stars, and combining these with velocities for the sdO components from prior FUV spectroscopy, we determined the orbital solutions of these binary systems.  

The binaries in our sample have long orbital periods ($P$ from {95 to 237} days) with small Be star semi-amplitudes ($K_{1}<11$ km~s$^{-1}$).  Using the derived semi-amplitude ratios and independent estimates of Be star mass, we determined preliminary mass estimates of the subdwarf companion stars. The sdO stars have estimated masses in the range from $0.5$ to $1.7\ \Mnor$, which is consistent with the masses reported in other Be+sdO binaries from prior investigations.  The stripped helium stars in HD 113120 and HD 137387 have estimated masses greater than the Chandrasekhar mass limit of 1.4 $M_\odot$, indicating that the sdO stars in these systems may be progenitors of Type Ib and Ic supernovae. We also documented the line variations appearing in the spectra on orbital and non-orbital timescales. One of the stars, HD 157832, displays phased-locked sub-features that appear in the H$\alpha$ and H$\beta$ emission profiles. Such features may result from a two-armed spiral density wave pattern in the circumstellar disk that is due to the tidal pull of the companion star. 

We also made a search for spectral features of the faint sdO stars by making a Doppler tomography reconstruction of the component spectra using the new orbital solutions.  The reconstructed spectra of the secondary show the presence of a \ion{He}{2} $\lambda 4686$ absorption line in four of the targets with a strength consistent with model predictions for the sdO stars from the analysis of their FUV spectra.  This confirms the presence of a hot stripped companion in these cases.  However, the \ion{He}{2} line is absent from the reconstructed secondary spectrum in the case of HD~157832, indicating that any hot companion must be very faint.  This agrees with the lack of a hot spectral signature in the FUV spectra of this star from HST \citep{Wang2021}, and it casts doubt on the marginal detection of the sdO signal in the IUE spectrum \citep{Wang2018}.

Theoretical simulations by \citet{Shao2021} and others predict that the population of the Be+He binaries should be abundant but often hidden from detection in the local Galactic environment. These binaries likely experienced stable mass transfer during their earlier Case B interaction, and models that assume conservative mass transfer provide the best match to the observed ratios of the He star to Be star mass. Recent adiabatic mass-loss models suggest a new approach to simulate the formation of these binary systems, in which the former donor stars (now observed as the stripped sdO stars) may have attained a more evolved stage as an RGB or AGB star. 

Our multi-year spectroscopic investigation has revealed the orbital solutions for five key systems, and this information will complement future FUV observations to constrain the helium stars' masses, atmospheric properties, and compositions. This is vital step in painting a complete evolutionary portrait of the lives of these binary systems. 

\begin{acknowledgments}
 This work is based on observations obtained at Cerro Tololo Inter-American Observatory, NSF’s NOIRLab (NOIRLab Prop.\ ID 2019A-5152 and ID 2020A-5169 PI: L.\ Wang), which is managed by the Association of Universities for Research in Astronomy (AURA) under a cooperative agreement with the National Science Foundation. We thank the members of the SMARTS consortium, Todd Henry, Hodari James, Wei-Chun Jao, and Leonardo Paredes, for their effort in operating the 1.5-m telescope and CTIO staff members Rodrigo Hinojosa and Roberto Aviles for carrying out the observations. {We acknowledge an anonymous referee for his or her constructive comments to improve the manuscript.} This work was supported by the National Natural Science Foundation of China under programs Numbers 12103085, 12090040, and 12090043.
 The work was also supported by the U.S.\ National Science Foundation under grant No. AST-1908026.
 This research is based on observations made with the NASA/ESA Hubble Space Telescope obtained from the Space Telescope Science Institute, which is operated by the Association of Universities for Research in Astronomy, Inc., under NASA contract NAS 5–26555. These observations are associated with program HST-GO-15659.
 Institutional support was provided by the GSU College of Arts and Science. This work has made use of the SIMBAD database, operated at CDS, Strasbourg, France.  
\end{acknowledgments}

%

\vspace{5mm}
\facilities{CTIO:1.5m}





\appendix
\section{RV corrections}\label{sec:RVcorrect}
\begin{figure*}[ht!]
\plotone{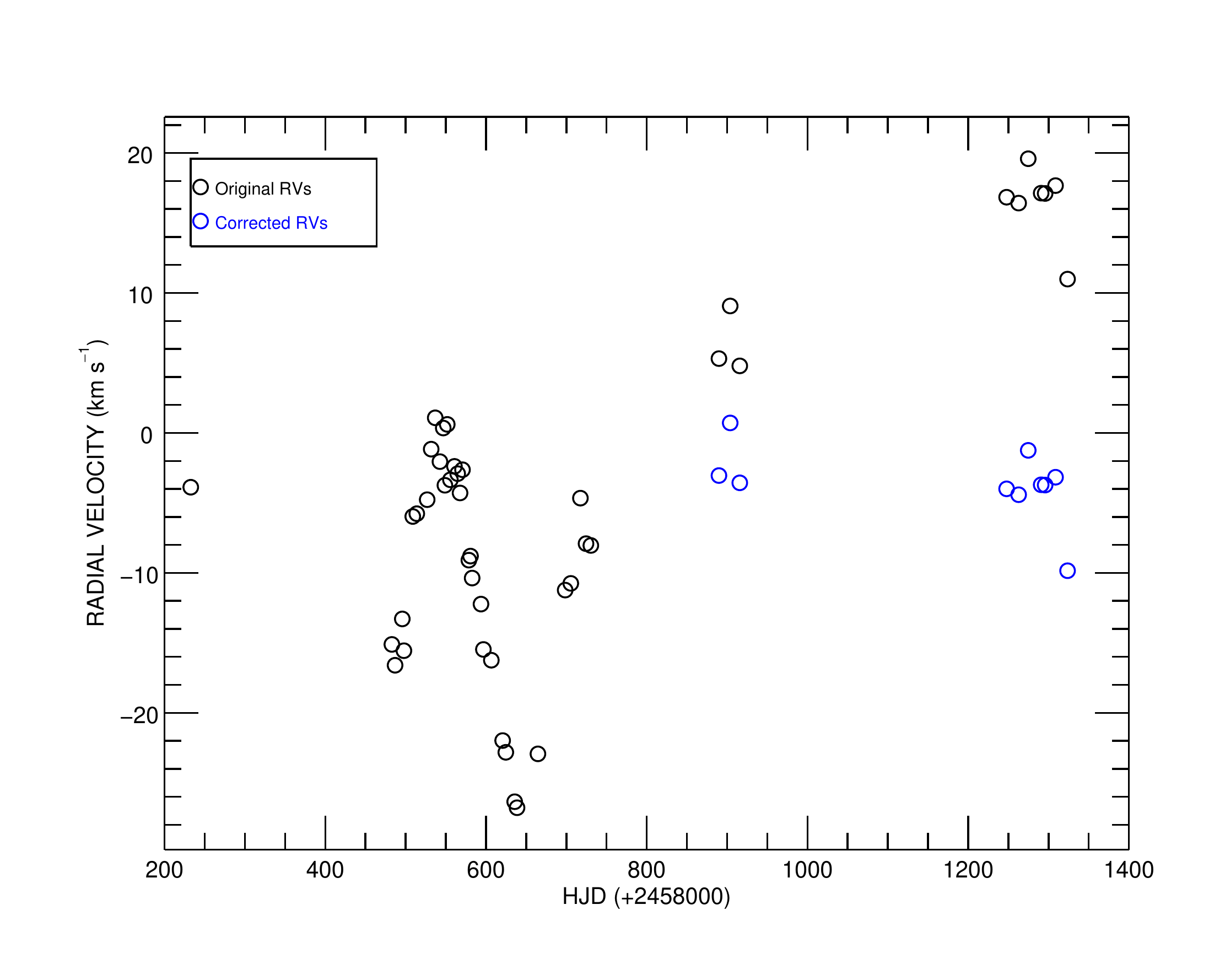}      
\caption{The measured RVs from the H$\alpha$ profiles of HD 113120. Long-term variations resulted in shifts appearing in the measured RVs obtained after HJD 2458800 (black). Corrected RVs are shown in blue.  }
\label{fig:RV_HD113120_ncorr}
\end{figure*}

Long-term variations in the spectra resulted in measured radial velocities that appear to show systemic shifts in all five target stars on a timescale of about a year. Thus, we removed such an effect from the measured RVs before obtaining the orbital solutions. Taking the star HD 113120 as an example, Figure~\ref{fig:RV_HD113120_ncorr} shows a time plot of RVs measured from H$\alpha$ following the procedures described in Section~\ref{subsec:RVHD113120}. Measurements made from the spectra on the nights after HJD 2458800 displayed an increasing trend in the RVs (black). We corrected such shifts by dividing the dataset into three subsets based on the dates, i.e., spectra observed before HJD 2458800 (Set 1), between HJD 2458800 and 2459000 (Set 2), and after HJD 2459000 (Set 3), and then we made a preliminary circular orbital fit to the measured RVs in Set 1 (with the most observations) to obtain the systemic velocity $\gamma_1$. We then applied the circular orbital fits to RVs in Set 2 (Set 3) by fixing the $P$ and $K_{1}$ parameters obtained from the circular fit from Set 1 to update the systemic velocities $\gamma_2$ ($\gamma_3$). The difference between the systemic velocities $\Delta\gamma=\gamma_1-\gamma_2$ ($\Delta\gamma=\gamma_1-\gamma_3$) indicates the deviation of measured RVs from those in Set 1. We applied this difference to the RVs in Set 2 (Set 3) to correct for any systematic shifts. The corrected RVs of HD 113120 are shown in blue in Figure~\ref{fig:RV_HD113120_ncorr}. Table~\ref{tab:RV_corr} lists the number of spectra in each of the subsets, $\gamma$ velocity obtained from the orbital fitting of each subset for each target star, and the correction factors $\Delta\gamma$ applied to the associated RVs in each subset.

\placefigure{fig:RV_HD113120_ncorr}
\restartappendixnumbering

\newpage

\begin{deluxetable*}{lccc}
\tablecaption{RV Corrections \label{tab:RV_corr}}
\tablewidth{0pt}
\tablehead{
\colhead{Data} & \colhead{No.} & \colhead{$\gamma$} &\colhead{$\Delta\gamma$} \\
\colhead{Subset} & \colhead{Observations} &\colhead{(km s$^{-1}$)} &  \colhead{(km s$^{-1}$)} 
} 
\startdata
\multicolumn{4}{c}{HD 113120}\\
\hline
1 & 34      &     $-${13.49}  & \nodata      \\ 
2 & \phn3   & \phn{$-$5.14}  &     \phn$-${8.36} \\
3 & \phn7   & \phs\phn{7.34}  &     $-${20.84} \\
\hline
\multicolumn{4}{c}{HD 137387$^a$} \\
\hline
1 & {26}      &  {\phs2.65}  & \nodata      \\ 
2 & \phn3   &  {$-$3.85}  &  {6.50} \\
3 & 12      &  {$-$6.45}  & {9.10} \\
\hline
\multicolumn{4}{c}{HD 152478}\\
\hline
1 & {32}      & {\phn5.68}  & \nodata      \\ 
2 & \phn4   & {13.37}  & \phn$-${7.69} \\
3 & 13      & {16.40}  & $-${10.72} \\
\hline
\multicolumn{4}{c}{HD 157042}\\
\hline
1 & 31      & {\phs1.56}  & \nodata      \\ 
2 & \phn4   & {$-$3.70}  & {\phs5.26} \\
3 & 12      & {\phs6.74}  & $-${5.18} \\
\hline
\multicolumn{4}{c}{HD 157832}\\
\hline
1 & 30      &   {18.02}  & \nodata      \\ 
2 & \phn7   &   {23.32}  & $-${5.29}  \\
3 & 10      &   {20.88}  & $-${2.86}  \\
\enddata
\tablenotetext{a}{The relative systemic velocity for each data subset was obtained from the orbital solution of RVs measured by cross-correlating the observed spectra with the co-added mean spectrum of the \ion{He}{1} $\lambda5875$ profile. }
\end{deluxetable*}

\newpage
\null\vspace{1 cm}

\section{Other RV measurements}\label{sec:RV_other}
We made a number of ancillary radial velocity measurements beyond the basic sets that were made of the H$\alpha$ profiles (or \ion{He}{1} $\lambda5875$ profiles for HD~137387) as described in Section~\ref{sec:RVs}. No corrections of long-term variations were applied to these measurements. Here we list the RVs shifts measured from other lines for HD~113120 (Table \ref{tab:RV_HD113120_other}), HD~137387 (Table \ref{tab:RV_HD137387_HD157832_other}), HD~152478 (Table \ref{tab:RV_HD152478_HD157042_Hb}), HD~157042 (Table \ref{tab:RV_HD152478_HD157042_Hb}), and HD~157832 (Table \ref{tab:RV_HD137387_HD157832_other}).  These are plotted against orbital phase (from Section~\ref{sec:RVs}) in Figure B2 (HD~113120), Figure B3 (HD~152478, HD~157042, HD~157832), and Figure B4 (HD~157832). Figure B1 shows an episode of blue wing variability in the H$\alpha$ line of HD~157832 that influenced the measured RVs.

\restartappendixnumbering
\begin{deluxetable*}{lcccccccc}
\tablecaption{Relative RVs for HD 113120 \label{tab:RV_HD113120_other}}
\tablewidth{0pt}
\tablehead{
\colhead{Date} &\colhead{$V_{\rm{H\beta}}^a$} & \colhead{$\sigma$} & \colhead{$V_{\rm{FeII}}^b$} & \colhead{$\sigma$} &
\colhead{$V_{\rm{FeII}}^c$} & \colhead{$\sigma$} & \colhead{$V_{\rm{FeII}}^d$} & \colhead{$\sigma$} \\
\colhead{(HJD$-$2,400,000)}  &  \colhead{(km s$^{-1}$)} & \colhead{(km s$^{-1}$)} & \colhead{(km s$^{-1}$)} & \colhead{(km s$^{-1}$)} & 
\colhead{(km s$^{-1}$)} & \colhead{(km s$^{-1}$)} & \colhead{(km s$^{-1}$)} & \colhead{(km s$^{-1}$)}
}
\startdata
58482.8629 & 	$-${5.4} & {0.7} &	$-${6.2} & 0.7 &	$-${0.7} & 0.3 &	\phn$-${7.5} & {0.5} \\
58486.8532 &	$-${4.0} & {0.7} &	$-${1.3} & 0.4 &	$-$2.0 & {0.4} &	\phn$-${9.1} & {0.4} \\
58495.8407 &	$-${2.1} & {0.7} &	$-${2.3} & 0.4 &	$-${1.3} & 0.4 &	\phn\phs{6.8} &	0.4 \\
58497.8571 &	$-${0.5} & {0.6} &	\phs{0.6}& 	0.4 &	\phs{0.1} &	0.4 &	\phn$-${3.4} & {0.3} \\
58508.8668 & \phs{0.5} & {0.6} &	\phs1.7 &	0.4 &	\phs{1.8} &	0.4 &	\phs{10.5} &	{0.3} \\
\enddata
\tablenotetext{a}{RVs were measured for the H$\beta$ profiles by calculating the wing bisectors of CCFs of the observed spectra with the co-added mean spectrum. }
\tablenotetext{b}{\ion{Fe}{2} $\lambda\lambda$5276, 5316 RVs were measured together for these profiles that were recorded in the same echelle order. }
\tablenotetext{c}{\ion{Fe}{2} $\lambda\lambda$ 5316, 5362 RVs were measured together for these profiles in the same echelle order. }
\tablenotetext{d}{\ion{Fe}{2} $\lambda$8451.}
\tablecomments{This table is available in its entirety in machine-readable form. The first five entries are shown here for guidance regarding its format and content. }
\end{deluxetable*}

\begin{deluxetable*}{lccc}
\tabletypesize{\scriptsize}
\tablecaption{RVs for HD 152478 and HD 157042 from H$\beta$ \label{tab:RV_HD152478_HD157042_Hb}}
\tablewidth{0pt}
\tablehead{
\colhead{Star} & \colhead{Date} &\colhead{$V_{\rm{H\beta}}^a$} & \colhead{$\sigma$} \\
\colhead{Name} & \colhead{(HJD$-$2,400,000)}  &  \colhead{(km s$^{-1}$)} & \colhead{(km s$^{-1}$)} 
}
\startdata
152478 &	58536.8773 & 	$-${4.2} & {2.9} \\
152478 &	58538.8683 &	$-${0.7} &  {0.8} \\ 
152478 &	58540.8518 & 	$-${2.1} &  {0.5}  \\
152478 &	58545.8775 & 	$-${0.9} &  {0.6}  \\
152478 &	58547.8603 & 	$-${0.1} &  {0.5}  \\
\enddata
\tablenotetext{a}{The radial velocities were measured from two adjacent echelle spectral orders that recorded H$\beta$, and the mean values of these measurements and their associated uncertainties are given here. }
\tablecomments{This table is available in its entirety in machine-readable form. The first five entries are shown here for guidance regarding its format and content. }
\end{deluxetable*}

\newpage

\begin{deluxetable*}{lccccccccccccc}
\rotate
\tabletypesize{\fontsize{5}{13} \selectfont}
\tablecaption{RVs for HD 137387 and HD 157832 \label{tab:RV_HD137387_HD157832_other}}
\tablewidth{0pt}
\tablehead{
\colhead{} & \colhead{} &\colhead{$V_{\rm{H\beta}}$} & \colhead{$V_{\rm He\ I}^a$} & \colhead{$V_{\rm He\ I}^b$} & \colhead{$V_{\rm He\ I}^c$} &
\colhead{$V_{\rm Fe\ II}^d$} & \colhead{$V_{\rm Fe\ II}^e$} & \colhead{$V_{\rm Fe\ II}^f$} & \colhead{$V_{\rm Fe\ II}^g$} & \colhead{$V_{\rm He\ I}^h$} & \colhead{$V_{\rm He\ I}^i$} &
\colhead{$V_{\rm O\ I}^j$} & \colhead{$V_{\rm Ca\ II}^k$}\\
\colhead{Star} & \colhead{Date} &\colhead{$\lambda4861$} & \colhead{$\lambda4713$} & \colhead{$\lambda4921$} & \colhead{$\lambda5015$} &
\colhead{$\lambda5197$} & \colhead{$\lambda5234$} & \colhead{$\lambda5276$} & \colhead{$\lambda6456$} & \colhead{$\lambda6678$} & \colhead{$\lambda7065$} &
\colhead{$\lambda8446$} & \colhead{$\lambda8542$}\\
\colhead{Name} & \colhead{(HJD$-$2,400,000)}  &  \colhead{(km s$^{-1}$)} & \colhead{(km s$^{-1}$)} & \colhead{(km s$^{-1}$)} & \colhead{(km s$^{-1}$)} & 
\colhead{(km s$^{-1}$)} & \colhead{(km s$^{-1}$)} & \colhead{(km s$^{-1}$)} & \colhead{(km s$^{-1}$)} & \colhead{(km s$^{-1}$)} & \colhead{(km s$^{-1}$)} & \colhead{(km s$^{-1}$)} & \colhead{(km s$^{-1}$)}
}
\startdata
137387 &	58496.8656 & \phs{11.0} $\pm$ {0.6} & $-$15.3 $\pm$ 0.8 & \phs{22.8} $\pm$ {0.3} &  $-${32.5} $\pm$ 1.1 & \nodata &		\nodata &	\nodata &	\nodata & \phs{21.7} $\pm$ 0.5 & 	\phn{34.2} $\pm$ {0.4} & \nodata & \nodata \\	
137387 &	58509.8566 & \phs\phn{3.9} $\pm$ {1.1} & $-$16.8 $\pm$ 0.6 & \phs\phn{4.2} $\pm$ 0.4 & 	$-${16.5} $\pm$ 0.7 & \nodata &	\nodata & \nodata &	\nodata & \phs{13.5} $\pm$ {1.4} & 	\phn{44.8} $\pm$ {0.6} &	\nodata	 &	\nodata \\	
137387 &	58526.8796 &  $-${34.6} $\pm$ 1.1 & $-$81.6 $\pm$ 0.8 & $-${17.1} $\pm$ {0.7} & $-${68.9} $\pm$ 0.8 & 	\nodata & \nodata &	\nodata & \nodata &	$-${22.4}$\pm$ {1.1} & \phn{37.8} $\pm$ {0.3} & \nodata & \nodata \\	
137387 &	58539.8429 & $-${13.0} $\pm$ 0.6 & $-$30.3 $\pm$ 0.5 & \phs\phn{1.3} $\pm$ {0.4} & 	$-${20.5} $\pm$ {1.1} & 	\nodata & \nodata &	\nodata	 &	\nodata & \phs{21.8} $\pm$ {1.1} & \phn{74.3} $\pm$ {0.5} & 	\nodata & \nodata	\\
137387 &	58540.7978 &  \phn$-${6.1} $\pm$ {0.7} & $-$18.8 $\pm$ 0.6 & \phs{24.6} $\pm$ {0.5} & $-${33.3} $\pm$ {1.0} & 	\nodata &	\nodata & \nodata &	\nodata & \phs{14.4} $\pm$ 0.5 & 	{103.8} $\pm$ {0.6} &  \nodata  & \nodata \\	
\enddata
\tablecomments{HD 137387: RVs were measured from the broad CCF wings by cross-correlating the observed spectra with the co-added mean spectrum. RVs were measured over the wavelength range of {4852$-$4866} for H$\beta$; {4704$-$4724} \AA\ for {He} \rom{1} $\lambda$4713$^a$; {4910$-$4933} \AA\ for {He} \rom{1} $\lambda$4921$^b$; {5007$-$5024} \AA\ for {He} \rom{1} $\lambda$5015$^c$; {6662$-$6693} \AA\ for {He} \rom{1} $\lambda$6678$^h$; {7058$-$7077} \AA\ for {He} \rom{1} $\lambda$7065$^i$. \\
HD 157832: RVs were measured using CCFs for the He \rom{1}  and metallic line profiles. CCF peak velocities were determined for He \rom{1} $\lambda4921^b$ (over the wavelength range of {4915$-$4930} \AA), $\lambda5015^c$ (measured together with $\lambda5020$ in the same echelle spectral order over the wavelength range of {5012$-$5026} \AA), Fe \rom{2} including $\lambda5197^d$ (over the wavelength range of {5191$-$5205} \AA), $\lambda5234^e$ (over the range of {5227$-$5243} \AA), $\lambda5276^f$ ({5268$-$5293} \AA), and O \rom{1} $\lambda8446^j$ (over the range of {8409$-$8489} \AA). CCF wing velocities were measured for Fe \rom{2} $\lambda6456^g$ (over the wavelength range of {6404$-$6464} \AA) and Ca \rom{2} $\lambda8542^k$ (over the range of {8534$-$8623} \AA). \\
This table is available in its entirety in machine-readable form. The first five entries are shown here for guidance regarding its format and content. }
\end{deluxetable*}

\pagebreak

\begin{figure*}
\gridline{
    \fig{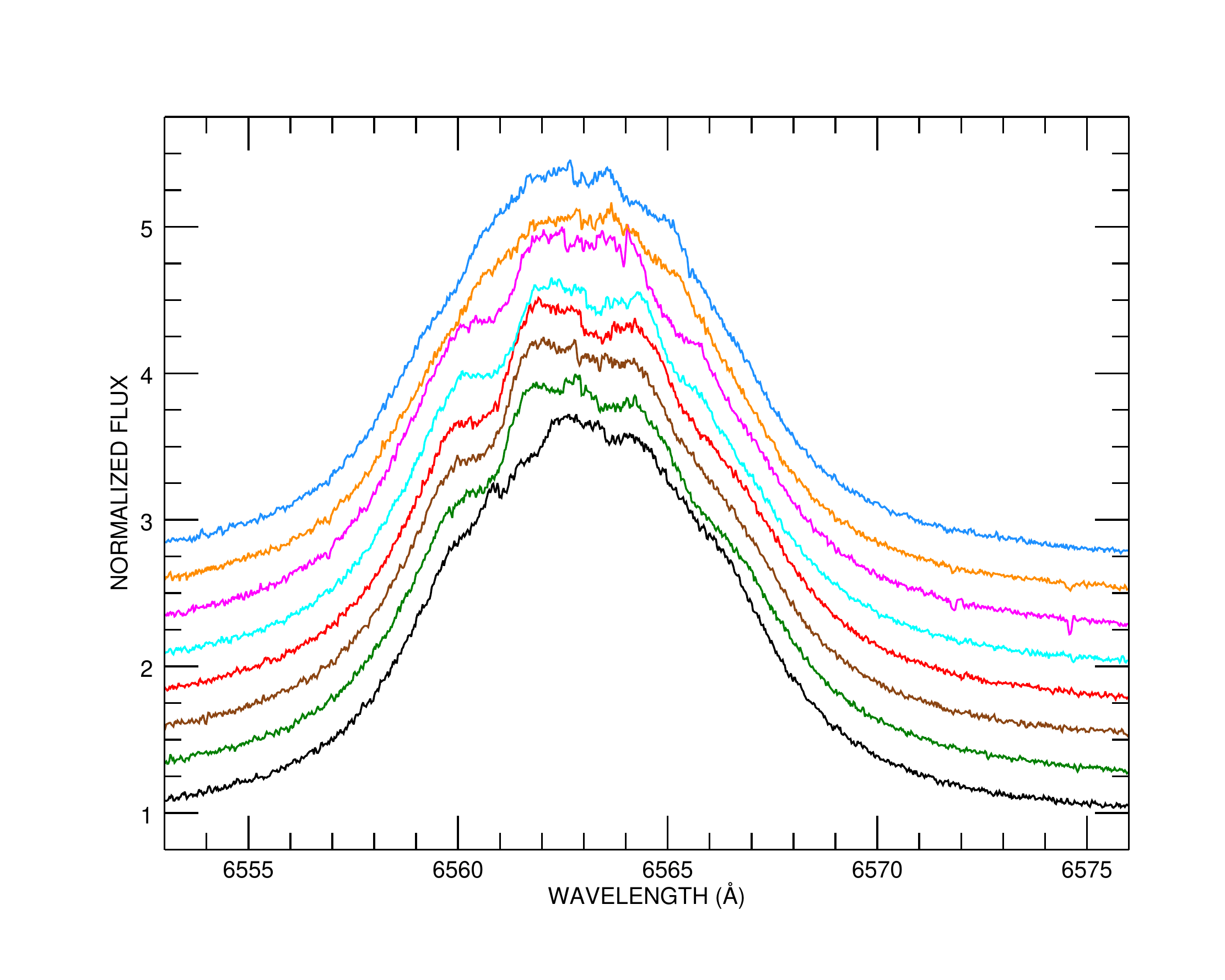}{1.0\textwidth}{(a)}
        }
\caption{H$\alpha$ profiles that were omitted in the orbital fit for HD 157832. The RVs measured from the seven spectra displayed a significant deviation from the derived orbital fit, and these spectra showed a steep drop in intensity in the blue wing of the peak profile. These include observations made on the nights of HJD 2458635 (green), HJD 2458637 (saddle brown), HJD 2458638 (red), HJD 2458650 (cyan), HJD 2458653 (magenta), HJD 2458664 (dark orange), and HJD 2458667 (dodger blue). The spectrum observed on the night of HJD 2458625 with a measured RV in accord with the orbital solution is plotted in black for comparison. }
\label{fig:spec_omit_HD157832}
\end{figure*}

\placefigure{fig:spec_omit_HD157832}

\null\vspace{5 cm}

\placefigure{fig:RV_HD113120}
\begin{figure*}
\gridline{
    \fig{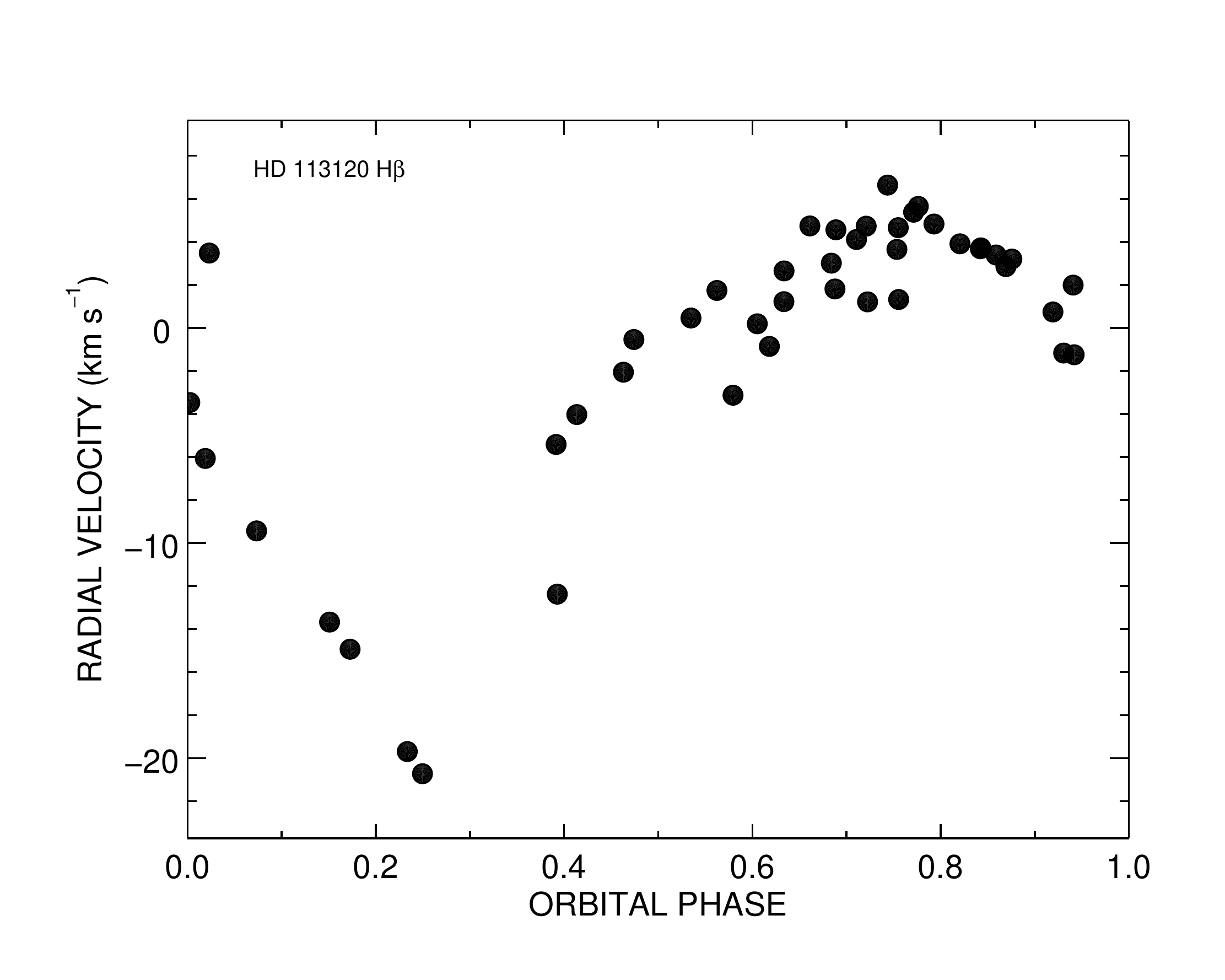}{0.5\textwidth}{(a)}
    \fig{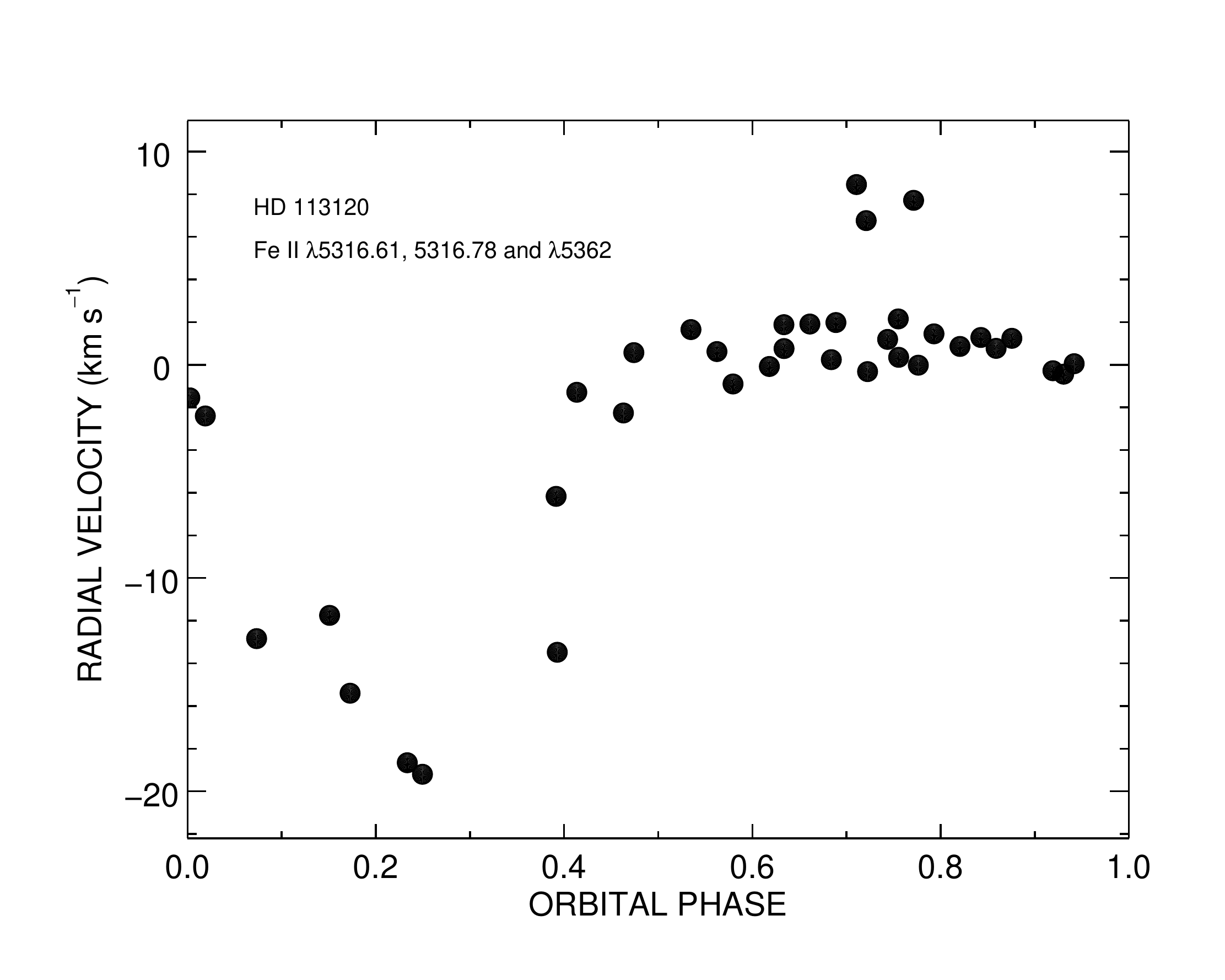}{0.5\textwidth}{(b)}
        }
\gridline{
       \fig{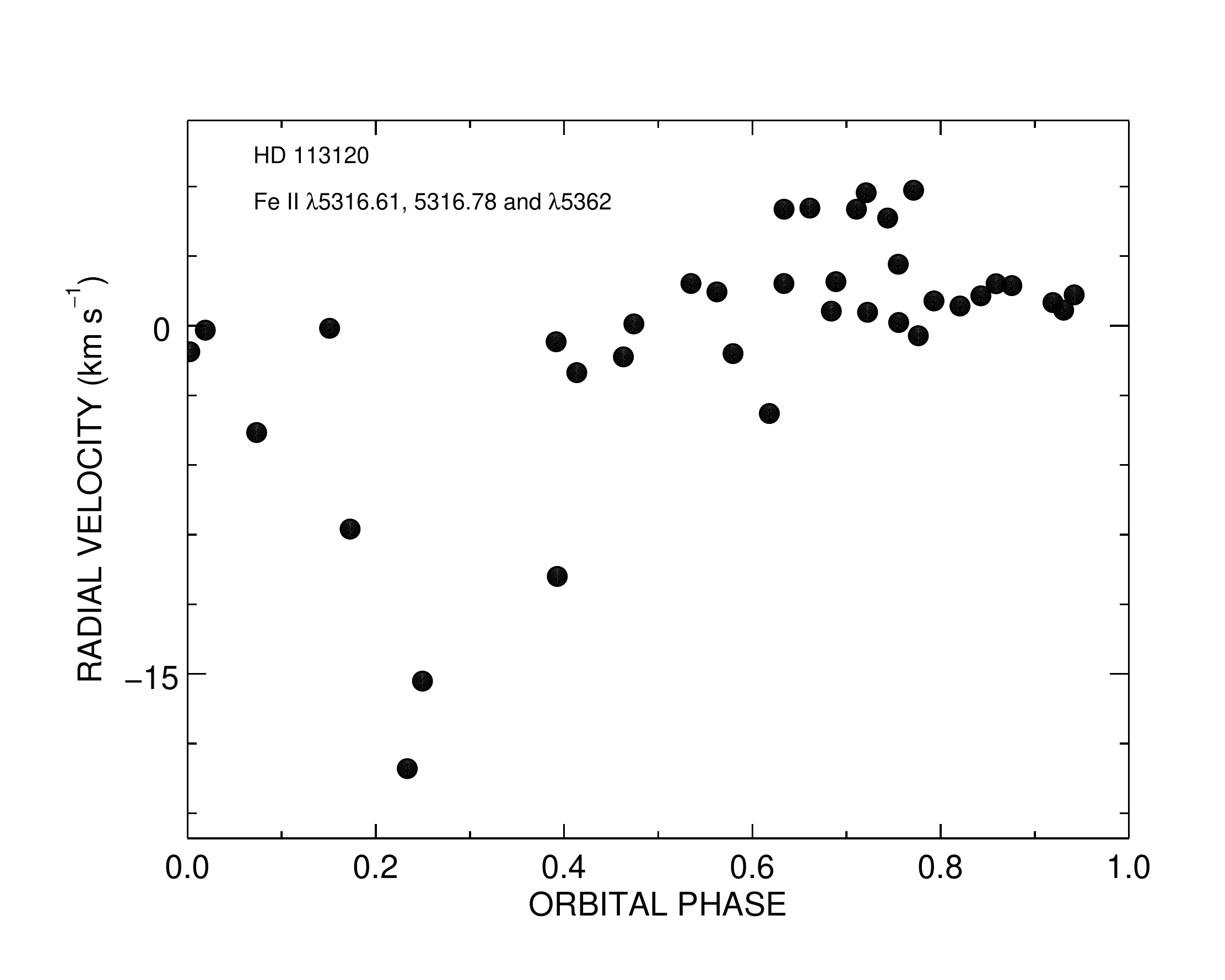}{0.5\textwidth}{(c)}
       \fig{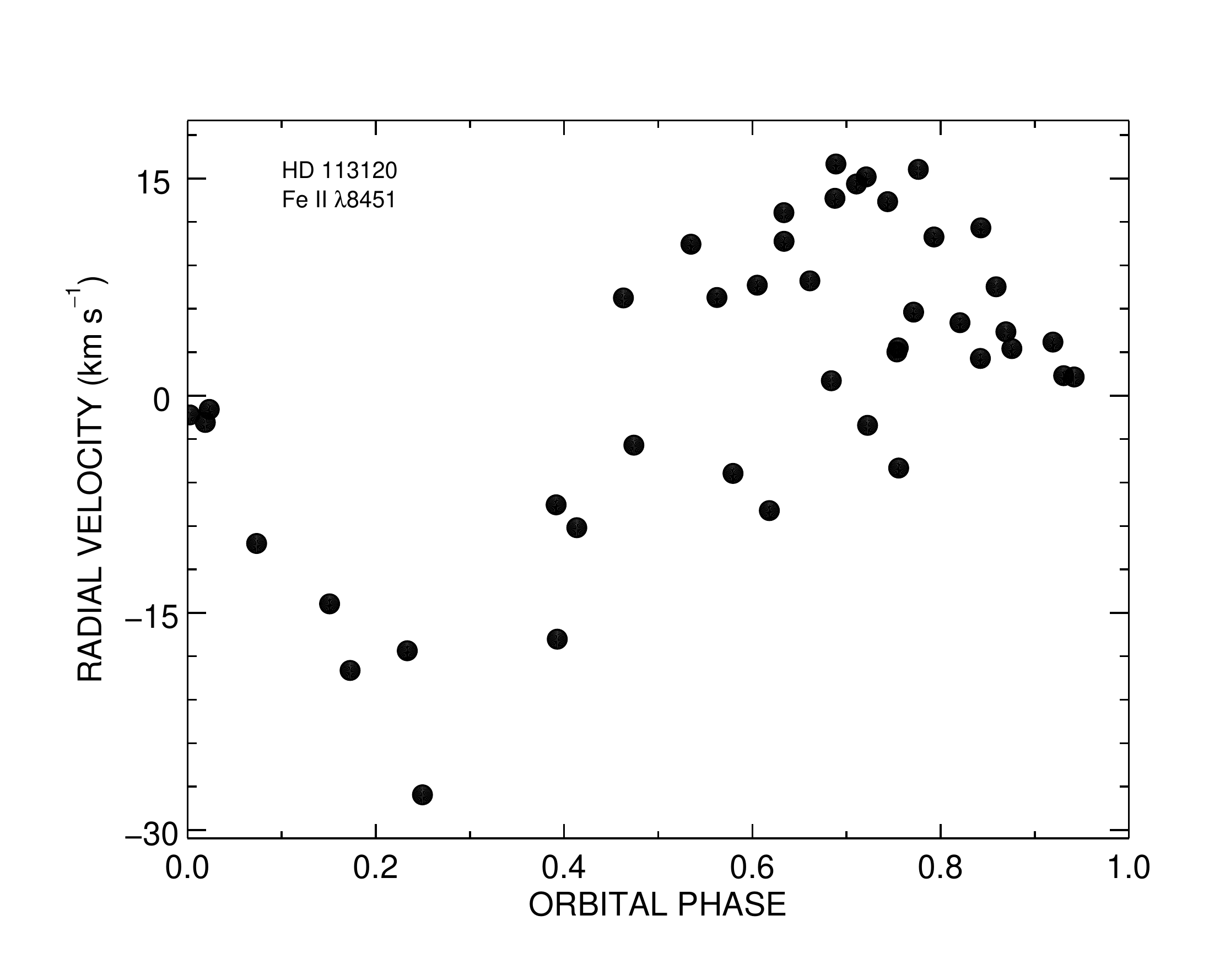}{0.5\textwidth}{(d)}
        }
\caption{RV curves of HD 113120 for H$\beta$ (panel a), \ion{Fe}{2} $\lambda$5276, 5316 (panel b, all line features were measured together in the same echelle spectral order using the CCF approach), \ion{Fe}{2} $\lambda\lambda$5316, 5362 (panel c, all line features were measured together in the same echelle spectral order), and \ion{Fe}{2} $\lambda$8451 (panel d), plotted for phases from ephemeris reported in Table~\ref{tab:orbit_HD113120}. }
\label{fig:RV_HD113120_other}
\end{figure*}

\null\vspace{5 cm}

\begin{figure*}
\gridline{
        \fig{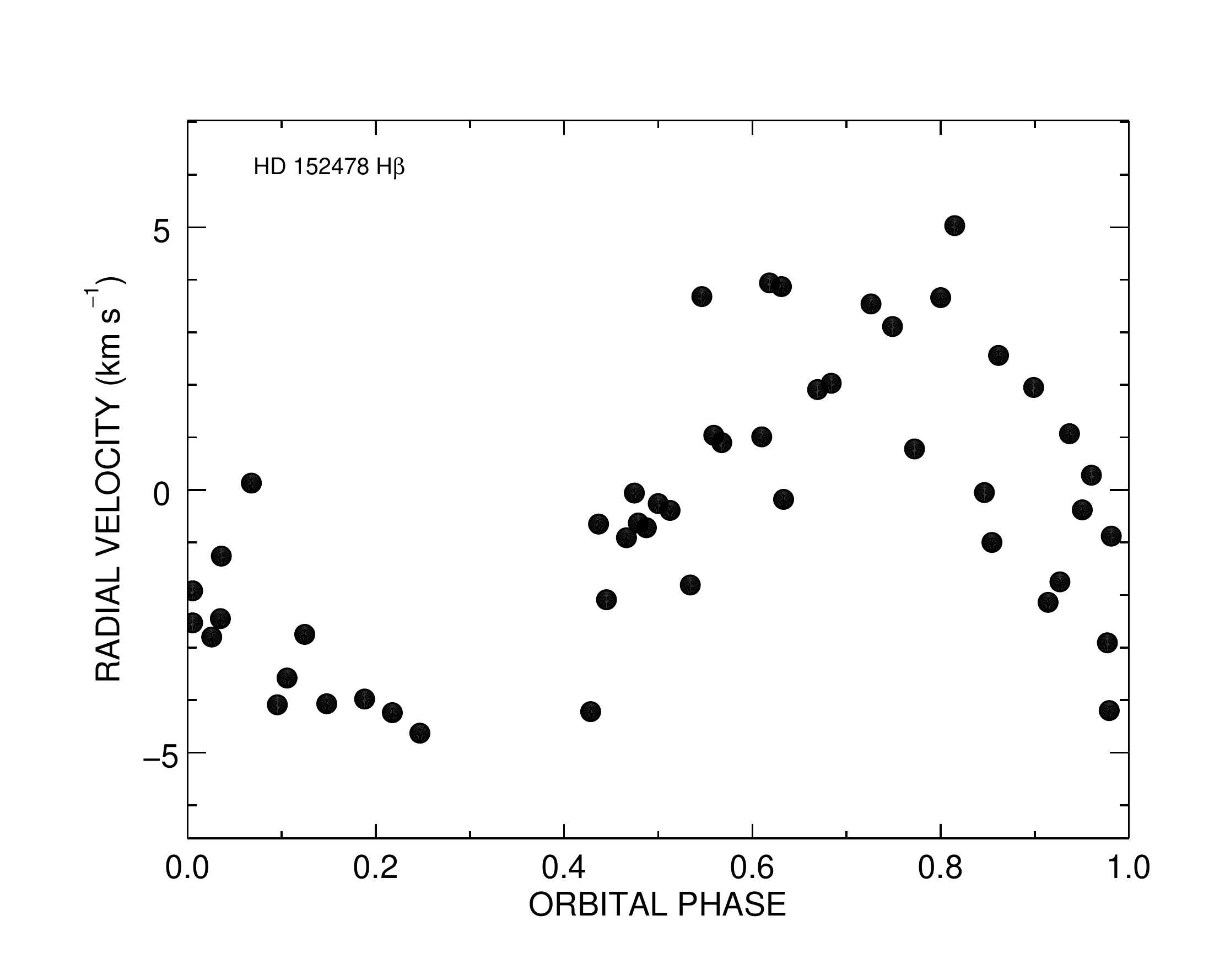}{0.5\textwidth}{(a) HD 152478}
        \fig{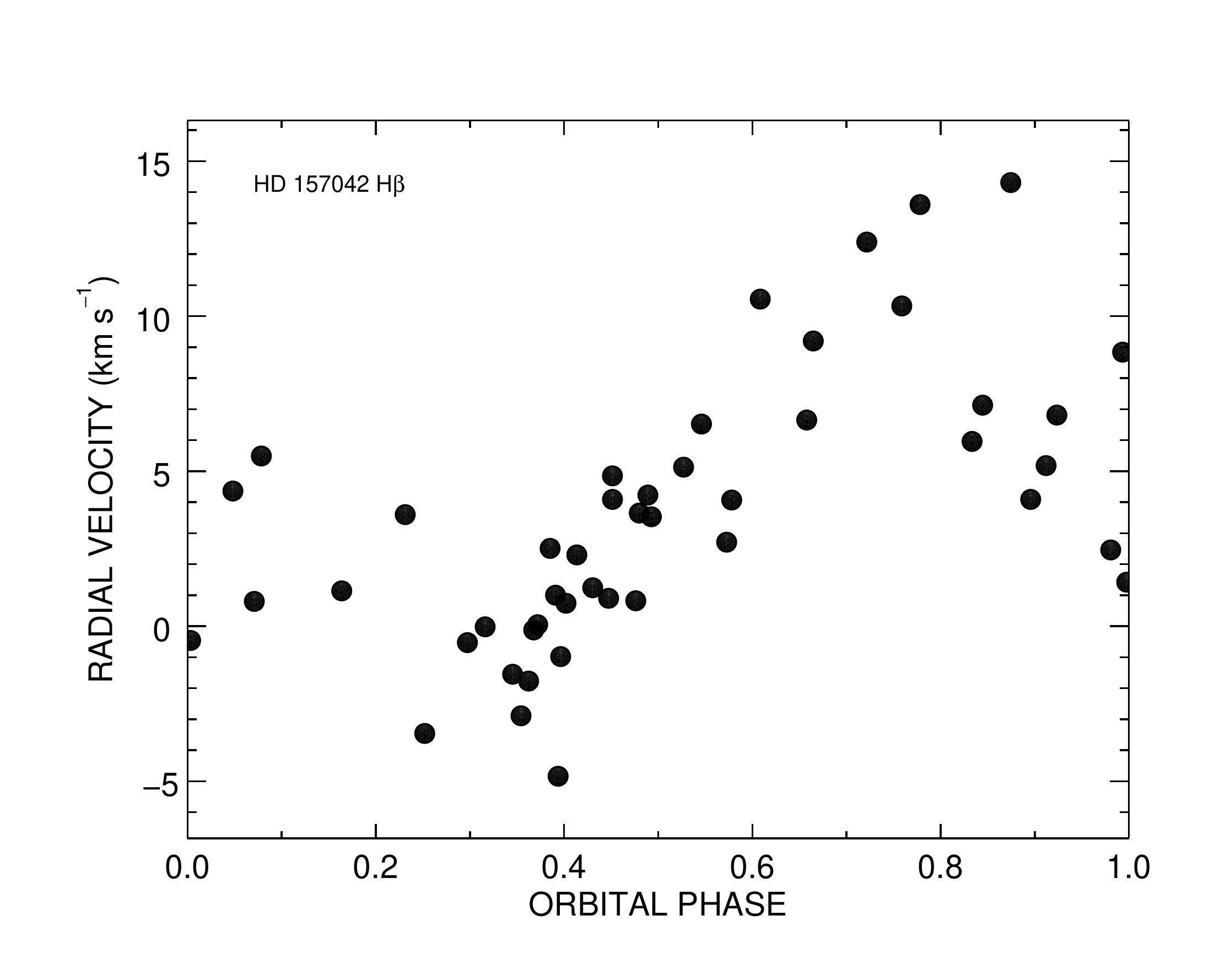}{0.5\textwidth}{(b) HD 157042}
        }
\gridline{
            \fig{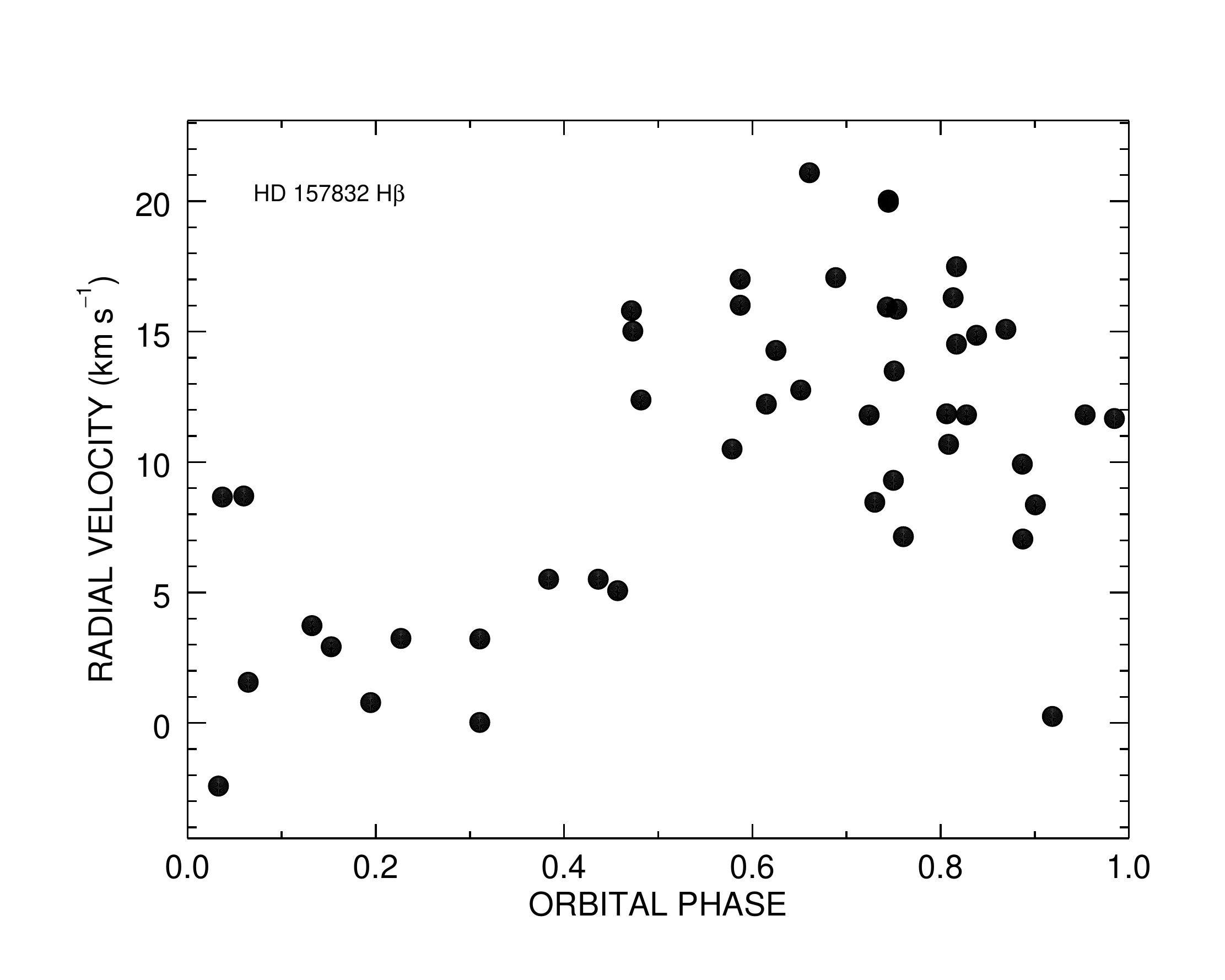}{0.5\textwidth}{(c) HD 157832}
            }
\caption{Radial velocity curves from H$\beta$ for the Be binary systems HD 152478 (panel a), HD 157042 (panel b), and HD157832 (panel c). Relative RVs of the Be stars HD 152478 and HD 157042 were determined from the {peak} locations of the CCFs constructed by cross-correlating the observed spectra with the co-added mean H$\beta$ profile. Alternatively, the absolute H$\beta$ velocities of HD 157832 were measured directly from the wings using the bisector technique from \citet{Shafter1986}. The phases are obtained from ephemeris reported in Table~\ref{tab:orbit_HD152478} for HD 152478 (Table~\ref{tab:orbit_HD157042} for HD 157042, and Table~\ref{tab:orbit_HD157832} for HD 157832).  }
\label{fig:RV_HD152478_157042_157832_Hb}
\end{figure*}

\placefigure{fig:RV_HD152478_157042_157832_Hb}

\null\vspace{5 cm}

\placefigure{fig:RV_HD157832}
\begin{figure*}
\gridline{\fig{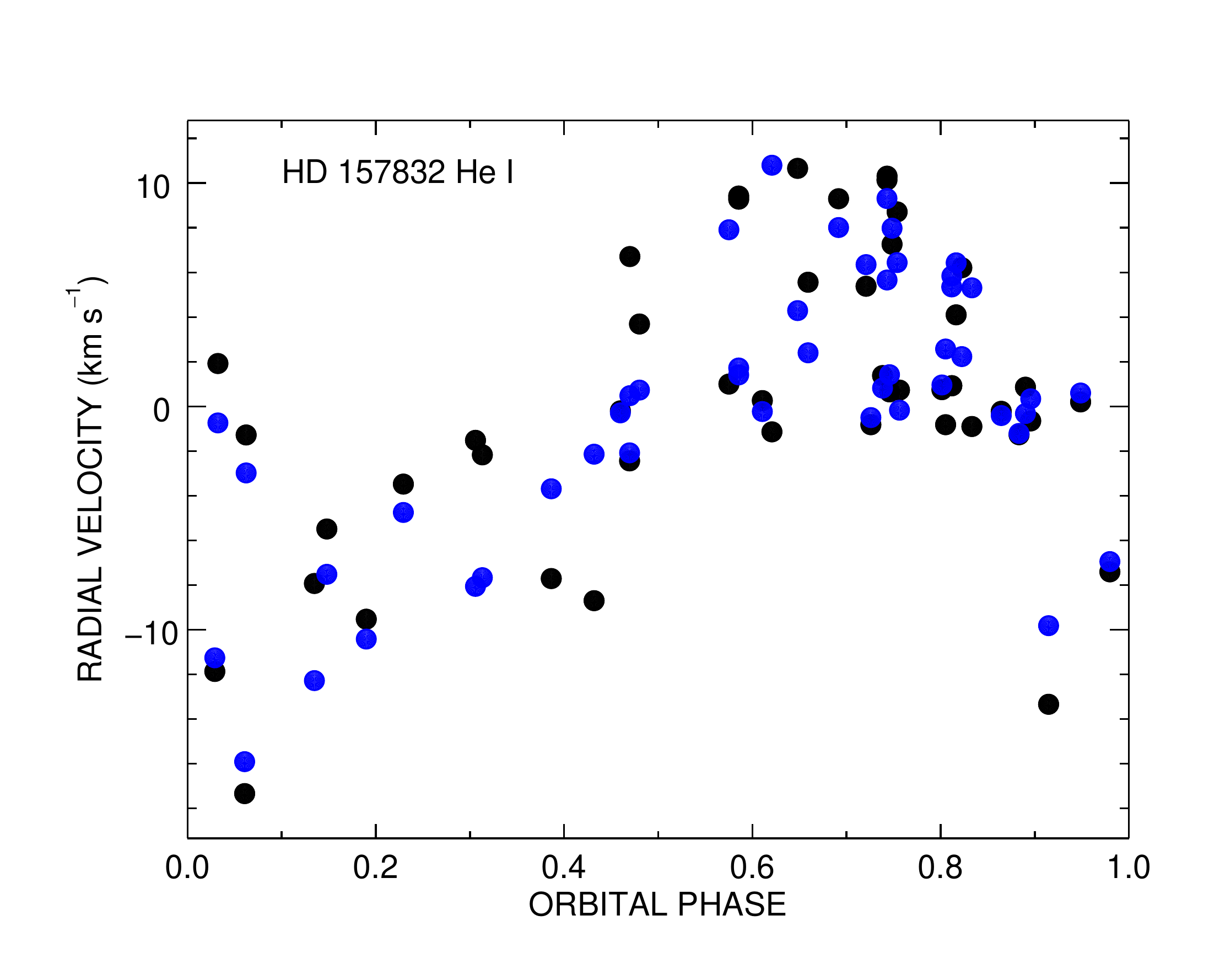}{0.5\textwidth}{(a) \ion{He}{1}}
    \fig{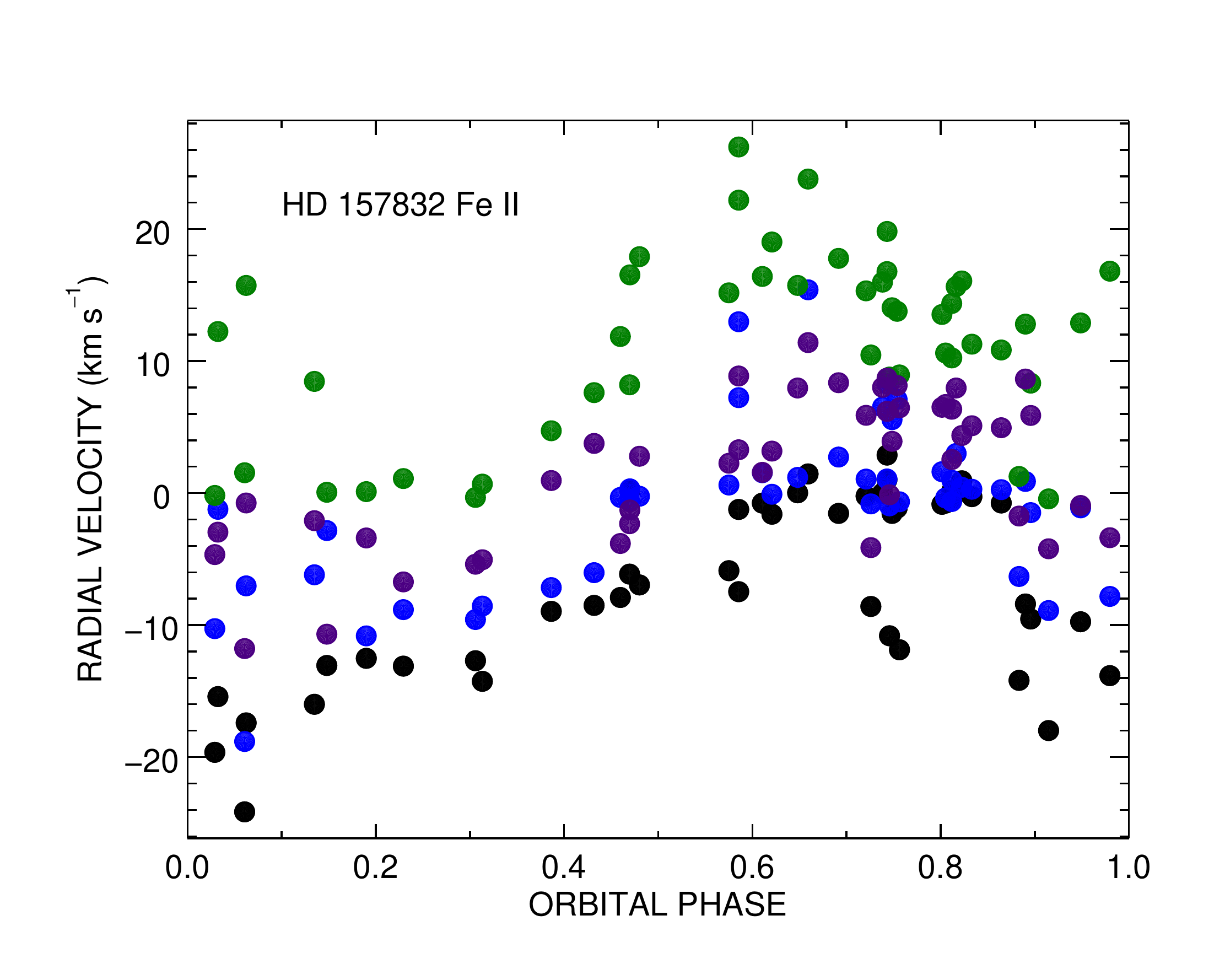}{0.5\textwidth}{(b) \ion{Fe}{2} }
    }
\gridline{\fig{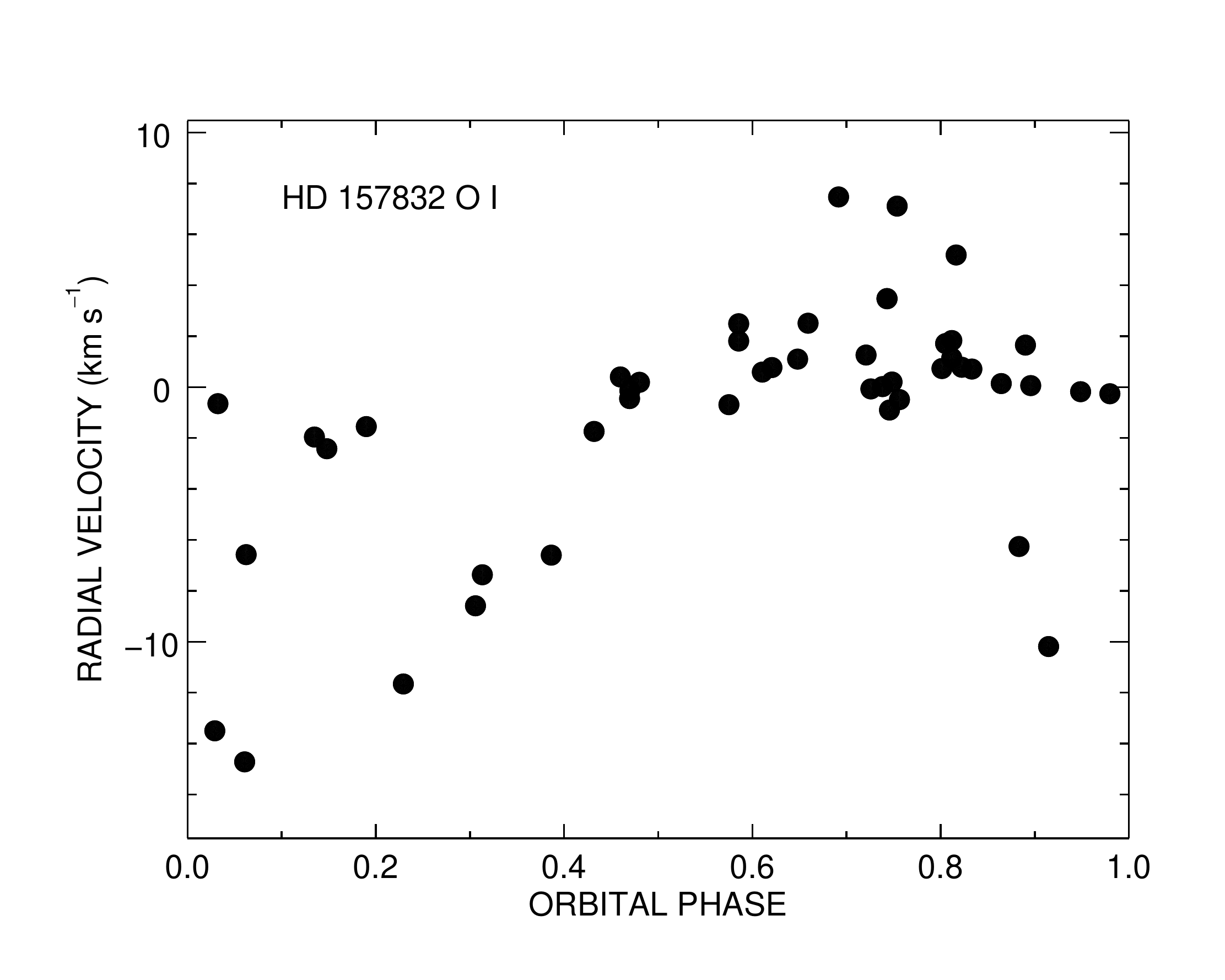}{0.5\textwidth}{(c) \ion{O}{1}}
    \fig{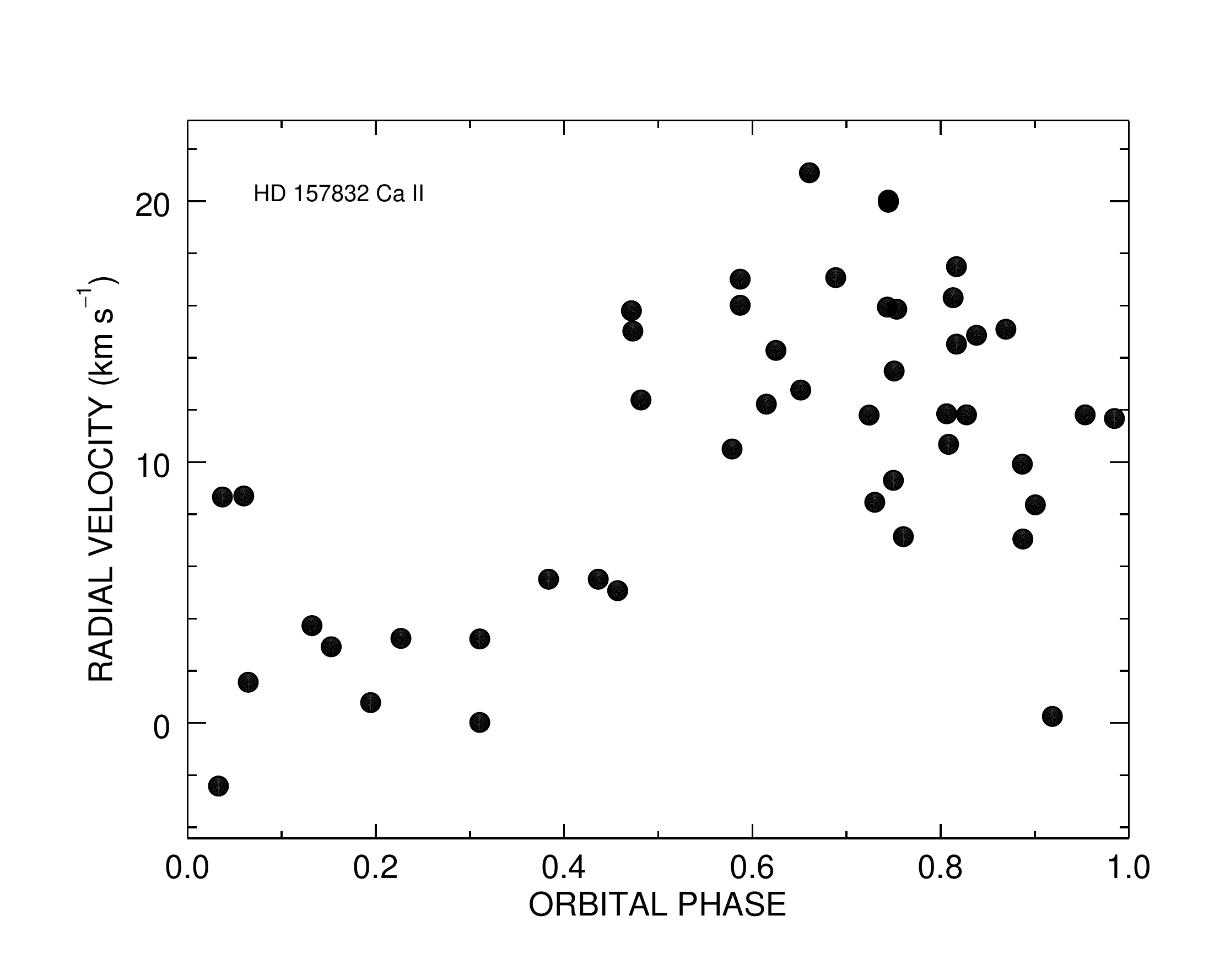}{0.5\textwidth}{(d) \ion{Ca}{2}}
        }
\caption{Radial velocity plots of the Be star HD 157832 measured from \ion{He}{1} and metallic lines, plotted from the ephemeris reported in Table~\ref{tab:orbit_HD157832}. Relative RVs of these profiles were determined from the peak location of the calculated CCFs of observed spectra with the co-added mean spectrum of the associated profile, except in the case of \ion{Fe}{2} $\lambda6456$ and \ion{Ca}{2} $\lambda\lambda8542,8600$ profiles, in which the RVs were measured from the CCF wings using the bisector technique. Panel (a): RVs for \ion{He}{1} $\lambda4921$ (black). Velocities were measured for both \ion{He}{1} $\lambda5015$ and \ion{Fe}{2} $\lambda5020$ profiles together in the same echelle spectral order (blue). Panel (b): RVs for \ion{Fe}{2} $\lambda5197$ (black), $\lambda5234$ (blue), $\lambda\lambda5276, 5284$ (green), and $\lambda6456$ (indigo). Panel (c): RVs for \ion{O}{1} $\lambda8446$. Panel (d): RVs for \ion{Ca}{2} $\lambda\lambda8542, 8600$.  }
\label{fig:RV_HD157832_other}
\end{figure*}

\section{Equivalent width measurements}\label{sec:EW_other}
Equivalent width ($W_\lambda$) values were measured over H$\alpha$, H$\beta$, and \ion{He}{1} profiles, the selected spectral ranges for each measured profile are given in Table~\ref{tab:EW_range}, and the $W_{\rm \lambda}$ measurements are collected in Table~\ref{tab:EW_all}. Subsets of these measurements are plotted in Figures C1 and C2. 

\restartappendixnumbering
\begin{deluxetable*}{lcccccccc}
\tabletypesize{\tiny}
\tablecaption{Wavelength Ranges for Equivalent Width Measurements \label{tab:EW_range} }
\tablewidth{0pt}
\tablehead{
\colhead{Star}  & \colhead{$W_{\rm H\alpha}$} & \colhead{$W_{\rm H\beta}$} &
\colhead{$W_{\rm He\ \rom{1}} (\lambda4921)$} & 
\colhead{$W_{\rm He\ \rom{1}} (\lambda5015)$} & 
\colhead{$W_{\rm He\ \rom{1}} (\lambda5875)$} & 
\colhead{$W_{\rm He\ \rom{1}} (\lambda6678)$} & \colhead{$W_{\rm He\ \rom{1}} (\lambda7065)$} &
\colhead{$W_{\rm Ca\ \rom{2}} (\lambda8542)$} \\
\colhead{Name}  & \colhead{(\AA)} & \colhead{(\AA)} & \colhead{(\AA)} & \colhead{(\AA)} & \colhead{(\AA)} & \colhead{(\AA)} & \colhead{(\AA)} & \colhead{(\AA)} 
}
\startdata
113120   & {6551$-$6577}  & {4857$-$4866} & \nodata &	\nodata	&	\nodata	&	{6675$-$6685} & {7054$-$7069} & \nodata	\\
137387  & {6552$-$6578}  &	{4855$-$4871} & {4914$-$4931} & {5007$-$5020}  & {5868$-$5885} & {6667$-$6690} & {7053$-$7075} & {8535$-$8561} \\
152478  & {6551$-$6577}  &  {4867$-$4871}	 & {4913$-$4931} &	\nodata	& \nodata	&	{6666$-$6689} & {7055$-$7076} & \nodata	\\
157042  & {6551$-$6574}  & {4866$-$4872}	 & {4914$-$4929} &	\nodata	&	\nodata	&	{6667$-$6688} & {7055$-$7075} &	\nodata	\\
157832  & {6550$-$6577} &	{4856$-$4866} & \nodata &	\nodata	&	\nodata	& \nodata & \nodata & \nodata \\
\enddata
\end{deluxetable*}

\null\vspace{0 cm}

\begin{deluxetable*}{lccccccccc}
\rotate
\tabletypesize{\tiny}
\tablecaption{Equivalent Width Measurements \label{tab:EW_all} }
\tablewidth{0pt}
\tablehead{
\colhead{Star} & \colhead{Date} & \colhead{$W_{\rm H\alpha}$} & \colhead{$W_{\rm H\beta}$} & 
\colhead{$W_{\rm He\ \rom{1}} (\lambda4921)$} & 
\colhead{$W_{\rm He\ \rom{1}} (\lambda5015)$} & 
\colhead{$W_{\rm He\ \rom{1}} (\lambda5875)$} & 
\colhead{$W_{\rm He\ \rom{1}} (\lambda6678)$} & \colhead{$W_{\rm He\ \rom{1}} (\lambda7065)$} &
\colhead{$W_{\rm Ca\ \rom{2}} (\lambda8542)$} \\
\colhead{Name} & \colhead{(HJD$-$2400000)} & \colhead{(\AA)} & \colhead{(\AA)} & \colhead{(\AA)} & \colhead{(\AA)} & \colhead{(\AA)} & \colhead{(\AA)} & \colhead{(\AA)} & \colhead{(\AA)} 
}
\startdata
113120  &	58482.8629 & $-${16.584} $\pm$ {0.025}  & 	$-$1.345 $\pm$ 0.028 & \nodata &	\nodata	&	\nodata	&	$-${0.081} $\pm$ {0.011} & $-${0.254} $\pm$ {0.015} & \nodata	\\
113120 &	58486.8532 & $-${16.651} $\pm$ {0.026}  &	$-$1.156 $\pm$ 0.025 & \nodata & \nodata  & \nodata &	$-${0.145} $\pm$ {0.010} & 	$-${0.109} $\pm$ 0.016 & \nodata \\
113120 &	58495.8407 & $-${15.899} $\pm$ {0.025}  & 	$-$1.152 $\pm$ 0.024 & \nodata &	\nodata	& \nodata	&	$-${0.036} $\pm$ {0.010} & 	$-${0.136} $\pm$ {0.018} & \nodata	\\
113120 &	58497.8571 & $-${16.327} $\pm$ {0.040}  & 	$-$1.115 $\pm$ 0.042 & \nodata &	\nodata	&	\nodata	&	\phs{0.045} $\pm$ {0.023} & \phs{0.152} $\pm$ {0.036} &	\nodata	\\
113120 &	58508.8668 & $-${15.690} $\pm$ {0.025}  &	$-$1.266 $\pm$ 0.028 & \nodata &	\nodata	&	\nodata	& $-${0.061} $\pm$ {0.011} & $-${0.029} $\pm$ {0.020} & \nodata \\
\enddata
\tablecomments{This table is available in its entirety in machine-readable form. The first five entries are shown here for guidance regarding its format and content.}
\end{deluxetable*}

\newpage

\begin{figure*}
\gridline{
        \fig{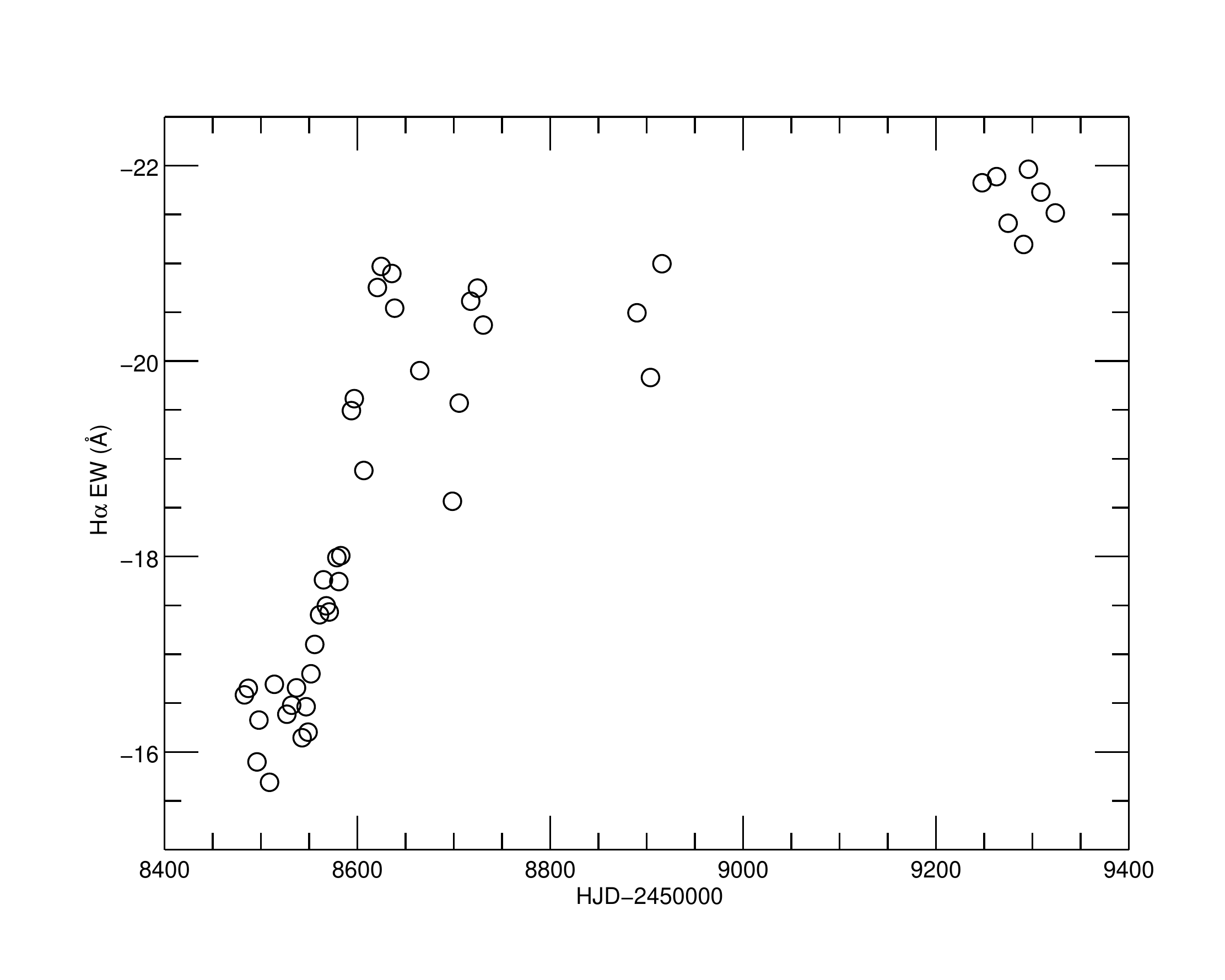}{0.5\textwidth}{(a) H$\alpha$ profiles} 
        \fig{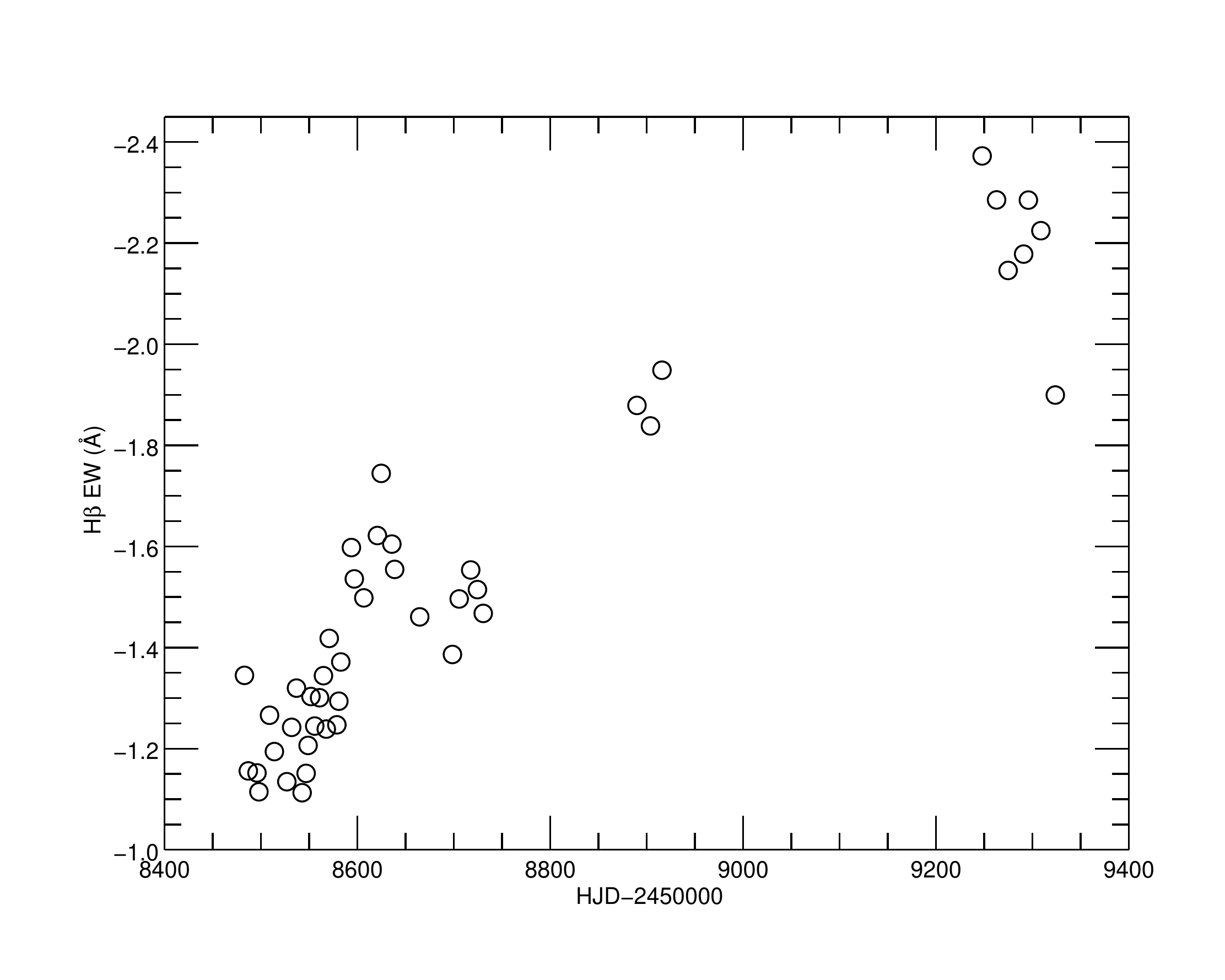}{0.5\textwidth}{(b) H$\beta$ profiles}
          }
\gridline{
        \fig{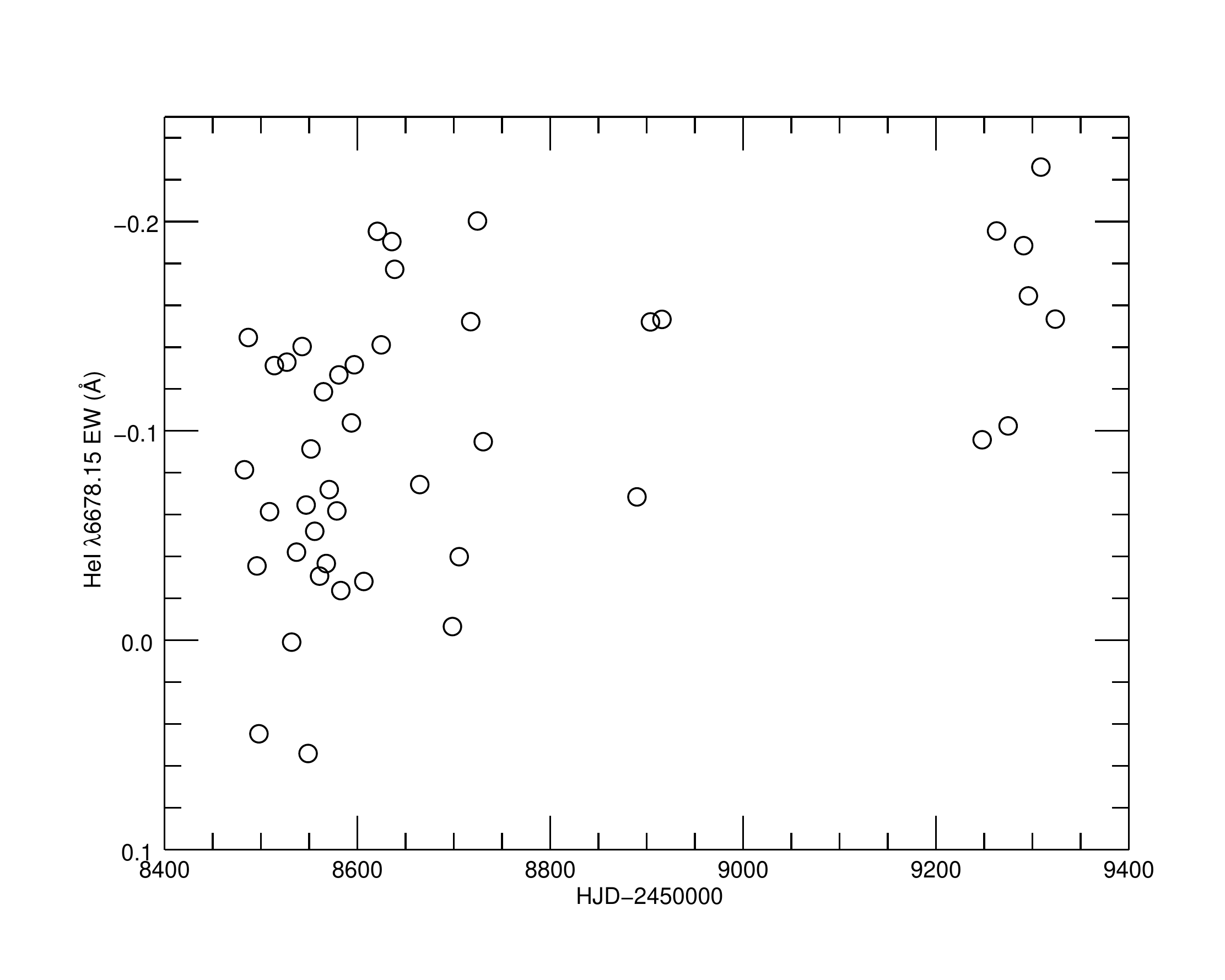}{0.5\textwidth}{(c) \ion{He}{1} $\lambda6678$ profiles}
        \fig{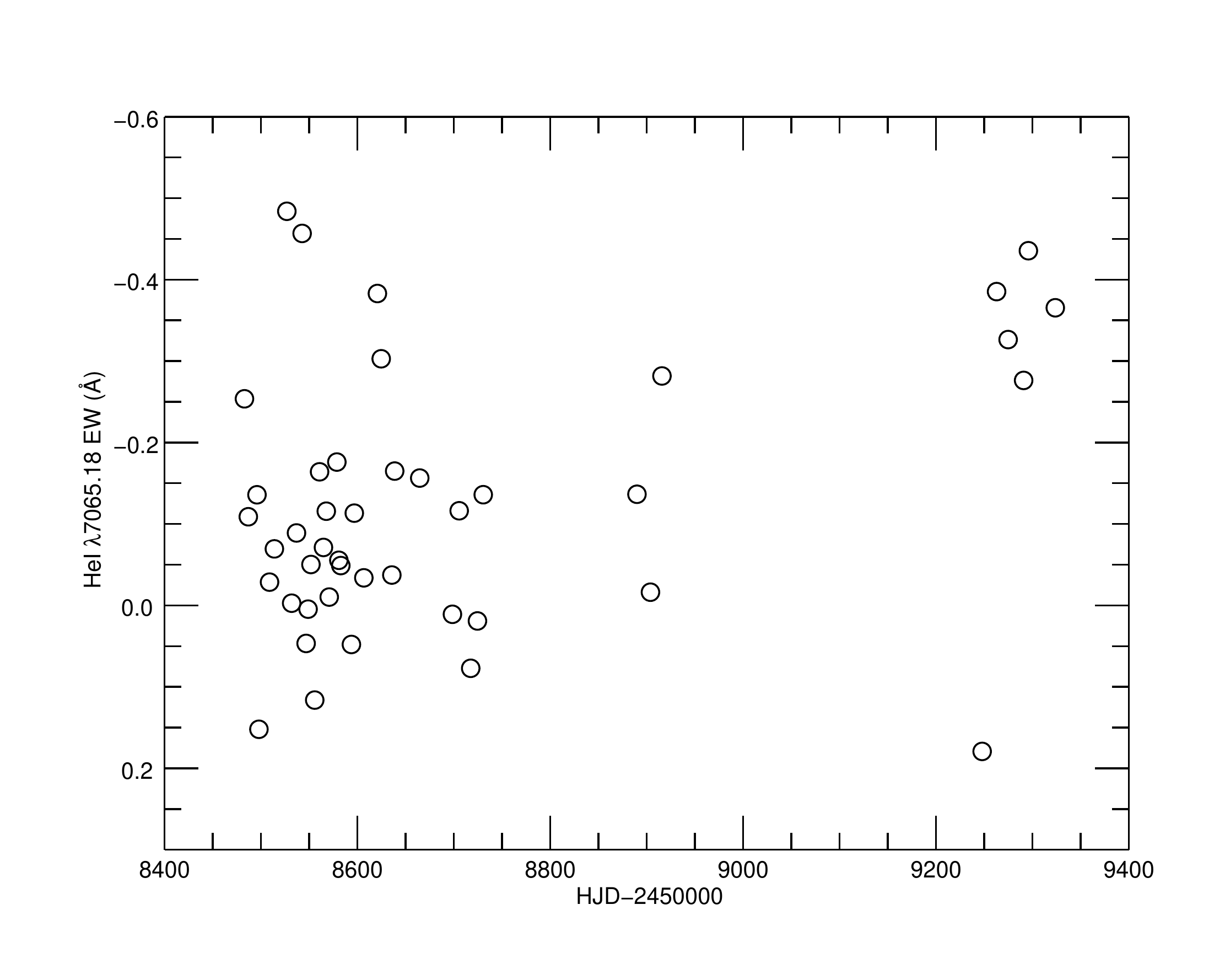}{0.5\textwidth}{(d) \ion{He}{1} $\lambda7065$  profiles}
        }       
\caption{The time plot of $W_\lambda$ measured for H$\alpha$ (panel a), H$\beta$ (panel b), \ion{He}{1} $\lambda6678$ (panel c), and \ion{He}{1} $\lambda7065$ (panel d) for HD 113120. }
\label{fig:EW_HD113120}
\end{figure*}

\placefigure{fig:EW_HD113120}

\null\vspace{5 cm}

\begin{figure*}
\gridline{
        \fig{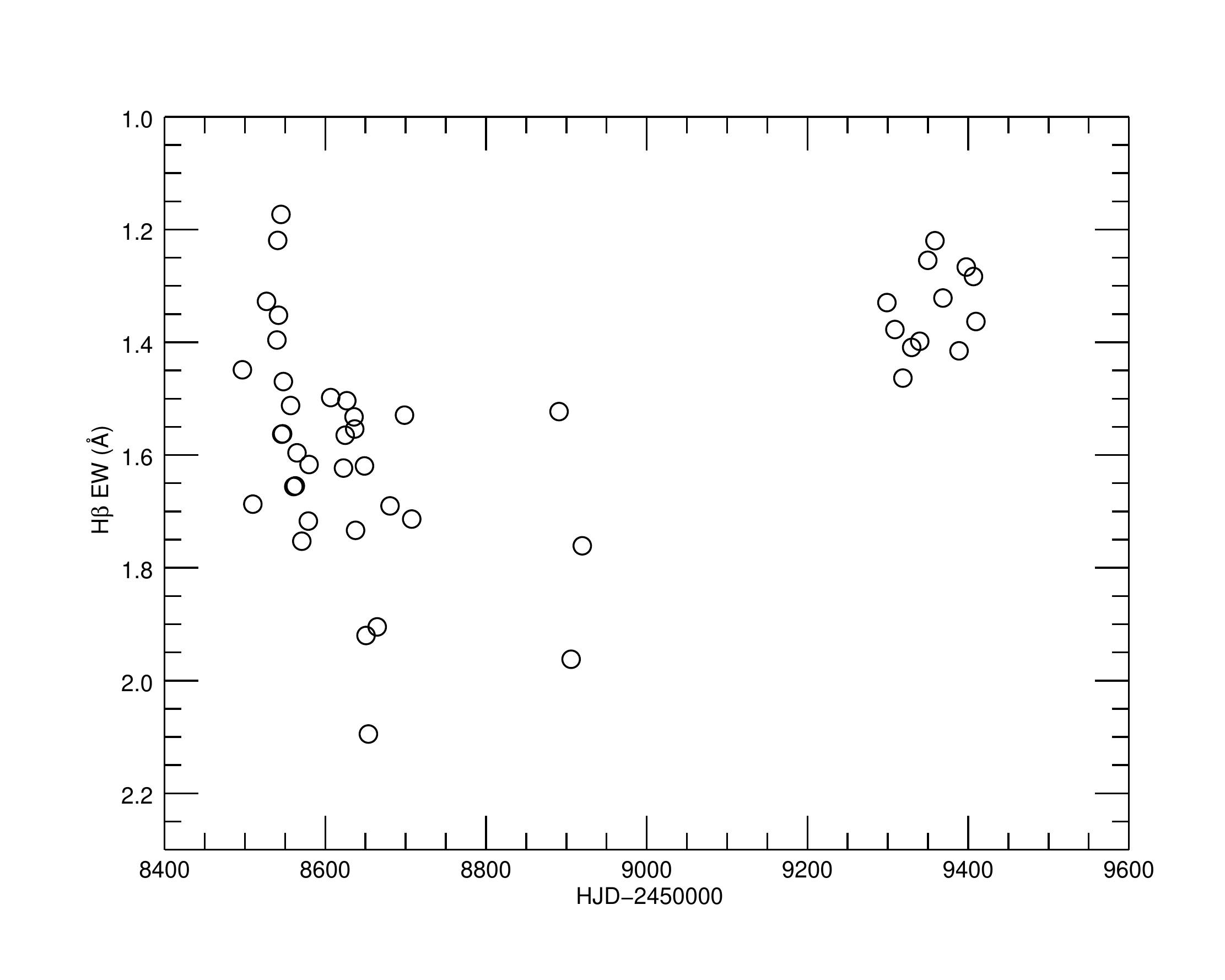}{0.5\textwidth}{(a) H$\beta$ profiles}
        \fig{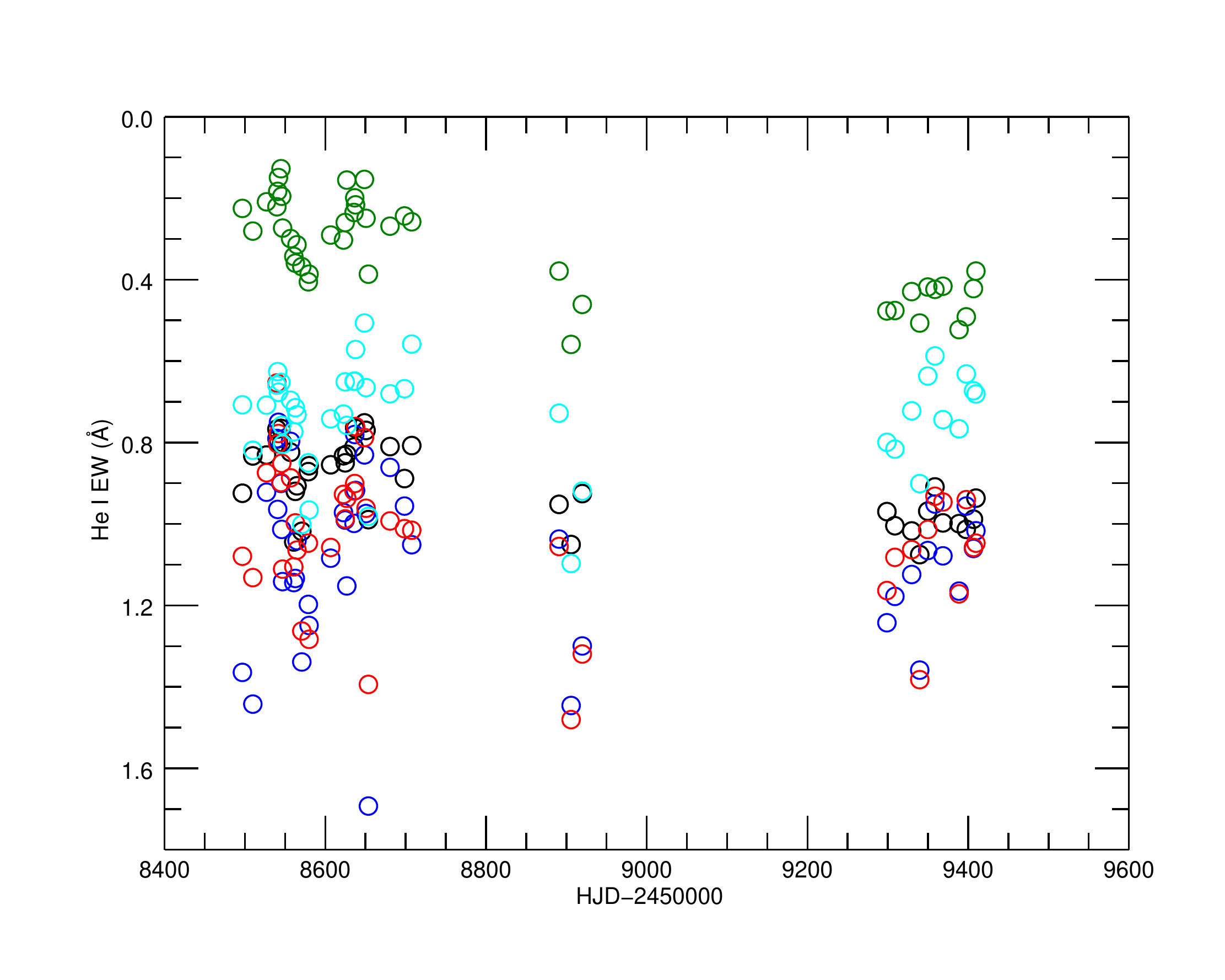}{0.5\textwidth}{(b) \ion{He}{1} profiles}
          }
\gridline{\fig{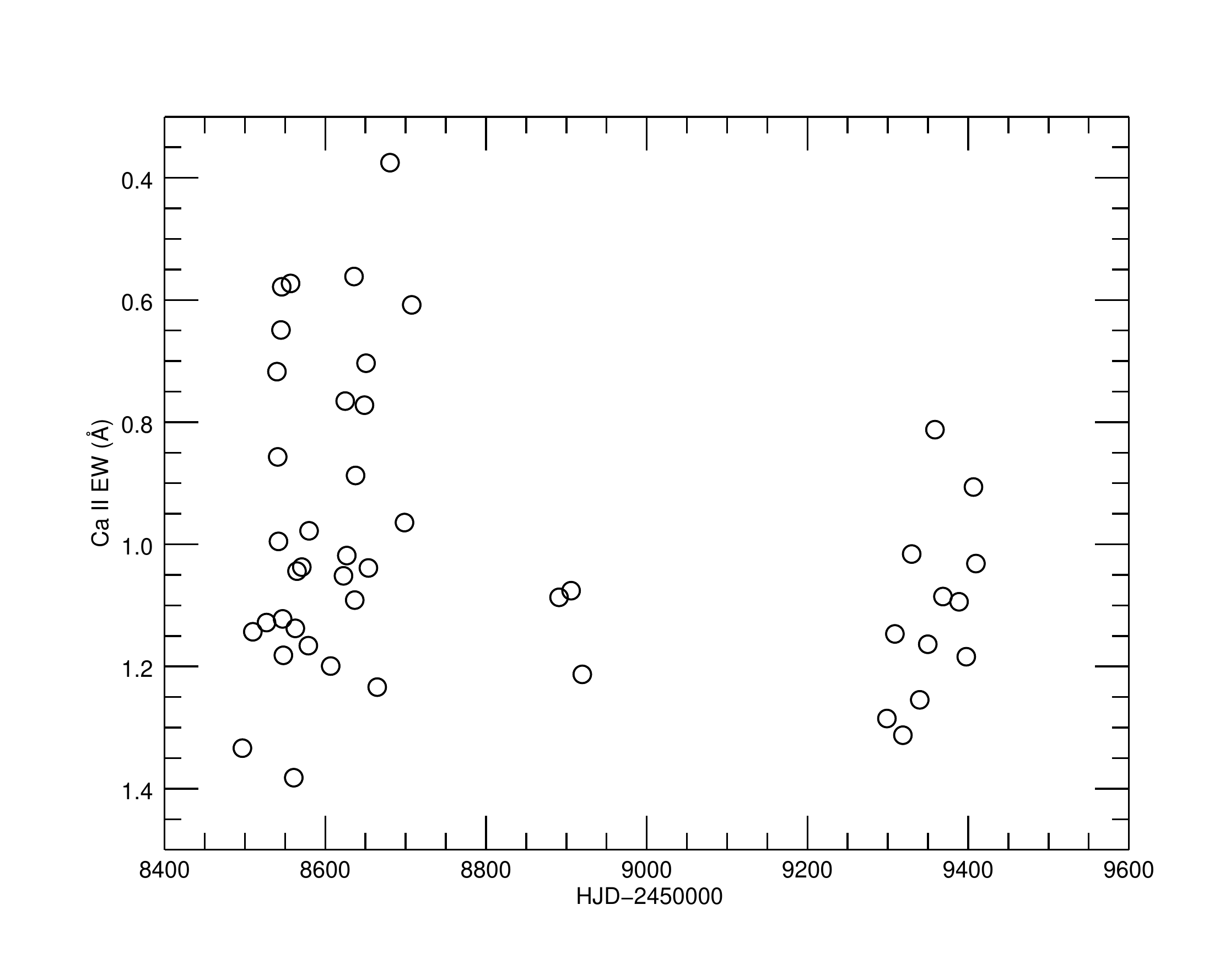}{0.5\textwidth}{(c) \ion{Ca}{2} $\lambda8542$ profiles}
        }
 \caption{The time plot of $W_\lambda$ for H$\beta$ (panel a), \ion{He}{1} (panel b), and \ion{Ca}{2} $\lambda8542$ (panel c) for HD 137387. Measured $W_\lambda$ values for \ion{He}{1} $\lambda4921$ (black), $\lambda5015$ (green), $\lambda5875$ (blue), $\lambda6678$ (red), and $\lambda7065$ (cyan) are shown in panel (b). }
\label{fig:EW_HD137387}
\end{figure*}

\placefigure{fig:EW_HD137387}

\newpage
\null\vspace{5 cm}

\section{Notes on Individual Stars}\label{sec:notes}
\restartappendixnumbering

\begin{itemize}[leftmargin=*]
    \item[] {\emph{HD 113120 (LS Mus)}. \citet{Hohle2010} applied a model fits to derive stellar parameters and estimated that the Be star has a mass of $M_\mathrm{Be} = 8.5 \pm 0.5$ \Mnor. If so, then the sdO star mass is $1.30 \pm0.8$ \Mnor, i.e., just below the Chandrasekhar limit. \citet{Zorec2016} reported that the rapidly rotating Be star has $V\sin{i} = 339 \pm 30$ km s$^{-1}$, $T_{\rm eff} = 22.8 \pm 0.7$ kK, $\log{g} = 3.69 \pm 0.44 $ (cm s$^{-2}$), and $\log{L_\mathrm{Be}} = 3.91 \pm 0.04$ \Lnor. The subdwarf companion of this star was first detected by \citet{Wang2018} through a cross-correlation technique using FUV spectra from IUE and later confirmed by \citet{Wang2021} from HST/STIS observations. Based upon the FUV spectroscopic analysis, \citet{Wang2021} reported that the sdO star has estimated $T_{\rm eff} = 45$ kK, $f_\mathrm{sdO}/f_\mathrm{Be} = 4.1 \pm 0.9\%$ in the FUV, $R_\mathrm{sdO} = 0.30 \pm 0.10$ \Rnor, and $\log{L_\mathrm{sdO}} = 2.52$ \Lnor. Through a spectral energy distribution (SED) fitting to the observations, they suggest that the Be star has a size of $R_\mathrm{Be} = 4.11 \pm 0.79$ \Rnor. \citet{Brandt2021} cross-calibrated Hipparcos and Gaia EDR3 surveys and identified this star as an astrometric accelerating star with a faint companion. }
    \item[] {\emph{HD 137387 ($\kappa$ Aps)}. \citet{Levenhagen2006} derived an age estimation of $\log{Age} (yr) = 7.29$ based on models. \citet{Zorec2016} reported that the Be component has $V\sin{i} = 250 \pm 21$ km s$^{-1}$, $T_{\rm eff} = 24.0 \pm 2.6$ kK, $\log{g} = 4.01 \pm 0.23 $ (cm s$^{-2}$), and $\log{L_\mathrm{Be}} = 4.05 \pm 0.09$ \Lnor. \citet{Wang2021} confirmed the detection of the hot subdwarf companion of this system and reported that the sdO star has $T_{\rm eff} = 40$ kK, $f_\mathrm{sdO}/f_\mathrm{Be} = 3.2 \pm 0.3\%$ in the FUV, $R_\mathrm{sdO} = 0.44 \pm 0.06$ \Rnor, and $\log{L_\mathrm{sdO}} = 2.65$ \Lnor. }
    \item[] {\emph{HD 152478 (V846 Ara)}. \citet{Levenhagen2006} derived an age of $\log{Age} (yr) = 7.34$. \citet{Hohle2010} derived a mass of $M_\mathrm{Be} = 6.3 \pm 0.4$ \Mnor. The Be component has estimates of $V\sin{i} = 295 \pm 18$ km s$^{-1}$, $T_{\rm eff} = 19.8 \pm 0.6$ kK, $\log{g} = 3.74 \pm 0.19 $ (cm s$^{-2}$), and $\log{L_\mathrm{Be}} = 3.09 \pm 0.02$\ \Lnor\ from \citet{Zorec2016}. \citet{Wang2021} reported that the sdO companion has $T_{\rm eff} = 42$ kK, $f_\mathrm{sdO}/f_\mathrm{Be} = 4.9 \pm 0.3\%$ in the FUV, $R_\mathrm{sdO} = 0.27 \pm 0.04$ \Rnor, and $L_\mathrm{sdO} = 2.31$ \Lnor. }
    \item[] {\emph{HD 157042 ($\iota$ Ara)}. \citet{Arcos2018} derived the stellar parameters for the Be component of $T_{\rm eff} = 22 \pm 0.22$ kK, $\log{g} = 3.90 \pm 0.04 $ (cm s$^{-2}$), $R = 5.17 \pm 0.1$ \Rnor, $V\sin{i} = 280 \pm 6$ km s$^{-1}$ from the Be Stars Observation Survey (BeSOS). \citet{Wang2021} reported the determination of atmospheric properties of the sdO companion of $T_{\rm eff} = 33.8$ kK, $f_\mathrm{sdO}/f_\mathrm{Be} = 2.6 \pm 0.3\%$ in the FUV, $R_\mathrm{sdO} = 0.61 \pm 0.09$ \Rnor, and $L_\mathrm{sdO} = 2.64$ \Lnor.  }
    \item[] {\emph{HD 157832 (V750 Ara)}. This star is identified as a $\gamma$\,Cas analog star, which displays photometric variations in TESS light curves \citep{Naze2020}. \citet{Naze2022} confirm this star's hard X-ray emission features from recent space observations. A marginal signature of an sdO companion star was detected from IUE FUV spectroscopy by \citet{Wang2018}, but such a feature was not found in the HST spectra by \citet{Wang2021}. }
\end{itemize}

\bibliography{ms.bib}{}

\begin{thebibliography}{}
\expandafter\ifx\csname natexlab\endcsname\relax\def\natexlab#1{#1}\fi
\providecommand{\url}[1]{\href{#1}{#1}}
\providecommand{\dodoi}[1]{doi:~\href{http://doi.org/#1}{\nolinkurl{#1}}}
\providecommand{\doeprint}[1]{\href{http://ascl.net/#1}{\nolinkurl{http://ascl.net/#1}}}
\providecommand{\doarXiv}[1]{\href{https://arxiv.org/abs/#1}{\nolinkurl{https://arxiv.org/abs/#1}}}

\bibitem[{{Abt} \& {Levy}(1978)}]{Abt1978}
{Abt}, H.~A., \& {Levy}, S.~G. 1978, \apjs, 36, 241, \dodoi{10.1086/190498}

\bibitem[{{Aguilera-Dena} {et~al.}(2022){Aguilera-Dena}, {M{\"u}ller},
  {Antoniadis}, {Langer}, {Dessart}, {Vigna-G{\'o}mez}, \&
  {Yoon}}]{Aguilera-Dena2022}
{Aguilera-Dena}, D.~R., {M{\"u}ller}, B., {Antoniadis}, J., {et~al.} 2022,
  arXiv e-prints, arXiv:2204.00025.
\newblock \doarXiv{2204.00025}

\bibitem[{{Arcos} {et~al.}(2018){Arcos}, {Kanaan}, {Ch{\'a}vez}, {Vanzi},
  {Araya}, \& {Cur{\'e}}}]{Arcos2018}
{Arcos}, C., {Kanaan}, S., {Ch{\'a}vez}, J., {et~al.} 2018, \mnras, 474, 5287,
  \dodoi{10.1093/mnras/stx3075}

\bibitem[{{Bagnuolo} {et~al.}(1994){Bagnuolo}, {Gies}, {Hahula}, {Wiemker}, \&
  {Wiggs}}]{Bagnuolo1994}
{Bagnuolo}, William~G., J., {Gies}, D.~R., {Hahula}, M.~E., {Wiemker}, R., \&
  {Wiggs}, M.~S. 1994, \apj, 423, 446, \dodoi{10.1086/173822}

\bibitem[{{Bjorkman} {et~al.}(2002){Bjorkman}, {Miroshnichenko}, {McDavid}, \&
  {Pogrosheva}}]{Bjorkman2002}
{Bjorkman}, K.~S., {Miroshnichenko}, A.~S., {McDavid}, D., \& {Pogrosheva},
  T.~M. 2002, \apj, 573, 812, \dodoi{10.1086/340751}

\bibitem[{{Brandt}(2021)}]{Brandt2021}
{Brandt}, T.~D. 2021, \apjs, 254, 42, \dodoi{10.3847/1538-4365/abf93c}

\bibitem[{{Carrier} {et~al.}(2002){Carrier}, {Burki}, \&
  {Burnet}}]{Carrier2002}
{Carrier}, F., {Burki}, G., \& {Burnet}, M. 2002, \aap, 385, 488,
  \dodoi{10.1051/0004-6361:20020174}

\bibitem[{{Chojnowski} {et~al.}(2018){Chojnowski}, {Labadie-Bartz}, {Rivinius},
  {Gies}, {Panoglou}, {Borges Fernandes}, {Wisniewski}, {Whelan}, {Mennickent},
  {McMillan}, {Dembicky}, {Gray}, {Rudyk}, {Stringfellow}, {Lester},
  {Hasselquist}, {Zharikov}, {Levenhagen}, {Souza}, {Leister}, {Stassun},
  {Siverd}, \& {Majewski}}]{Chojnowski2018}
{Chojnowski}, S.~D., {Labadie-Bartz}, J., {Rivinius}, T., {et~al.} 2018, \apj,
  865, 76, \dodoi{10.3847/1538-4357/aad964}

\bibitem[{{de Mink} {et~al.}(2013){de Mink}, {Langer}, {Izzard}, {Sana}, \& {de
  Koter}}]{deMink2013}
{de Mink}, S.~E., {Langer}, N., {Izzard}, R.~G., {Sana}, H., \& {de Koter}, A.
  2013, \apj, 764, 166, \dodoi{10.1088/0004-637X/764/2/166}

\bibitem[{{Ekstr{\"o}m} {et~al.}(2008){Ekstr{\"o}m}, {Meynet}, {Maeder}, \&
  {Barblan}}]{Ekstrom2008}
{Ekstr{\"o}m}, S., {Meynet}, G., {Maeder}, A., \& {Barblan}, F. 2008, \aap,
  478, 467, \dodoi{10.1051/0004-6361:20078095}

\bibitem[{{Ge} {et~al.}(2020){Ge}, {Webbink}, {Chen}, \& {Han}}]{Ge2020}
{Ge}, H., {Webbink}, R.~F., {Chen}, X., \& {Han}, Z. 2020, \apj, 899, 132,
  \dodoi{10.3847/1538-4357/aba7b7}

\bibitem[{{Gies} {et~al.}(1998){Gies}, {Bagnuolo}, {Ferrara}, {Kaye},
  {Thaller}, {Penny}, \& {Peters}}]{Gies1998}
{Gies}, D.~R., {Bagnuolo}, William~G., J., {Ferrara}, E.~C., {et~al.} 1998,
  \apj, 493, 440, \dodoi{10.1086/305113}

\bibitem[{{Gies} \& {Wang}(2020)}]{Gies2020}
{Gies}, D.~R., \& {Wang}, L. 2020, \apjl, 898, L44,
  \dodoi{10.3847/2041-8213/aba51c}

\bibitem[{{G{\"o}tberg} {et~al.}(2019){G{\"o}tberg}, {de Mink}, {Groh},
  {Leitherer}, \& {Norman}}]{Gotberg2019}
{G{\"o}tberg}, Y., {de Mink}, S.~E., {Groh}, J.~H., {Leitherer}, C., \&
  {Norman}, C. 2019, \aap, 629, A134, \dodoi{10.1051/0004-6361/201834525}

\bibitem[{{Granada} {et~al.}(2013){Granada}, {Ekstr{\"o}m}, {Georgy},
  {Krti{\v{c}}ka}, {Owocki}, {Meynet}, \& {Maeder}}]{Granada2013}
{Granada}, A., {Ekstr{\"o}m}, S., {Georgy}, C., {et~al.} 2013, \aap, 553, A25,
  \dodoi{10.1051/0004-6361/201220559}

\bibitem[{{Grundstrom}(2007)}]{Grundstrom2007}
{Grundstrom}, E.~D. 2007, PhD thesis, Georgia State University.
\newblock \url{https://scholarworks.gsu.edu/phy_astr_diss/19/}

\bibitem[{{Hanuschik} {et~al.}(1996){Hanuschik}, {Hummel}, {Sutorius},
  {Dietle}, \& {Thimm}}]{Hanuschik1996}
{Hanuschik}, R.~W., {Hummel}, W., {Sutorius}, E., {Dietle}, O., \& {Thimm}, G.
  1996, \aaps, 116, 309.
\newblock \url{https://articles.adsabs.harvard.edu/pdf/1996A\%26AS..116..309H}

\bibitem[{{Harmanec} {et~al.}(2020){Harmanec}, {Lipt{\'a}k}, {Koubsk{\'y}},
  {Bo{\v{z}}i{\'c}}, {Labadie-Bartz}, {{\v{S}}lechta}, {Yang}, \&
  {Harmanec}}]{Harmanec2020}
{Harmanec}, P., {Lipt{\'a}k}, J., {Koubsk{\'y}}, P., {et~al.} 2020, \aap, 639,
  A32, \dodoi{10.1051/0004-6361/202037964}

\bibitem[{{Harmanec} {et~al.}(2000){Harmanec}, {Habuda}, {{\v{S}}tefl},
  {Hadrava}, {Kor{\v{c}}{\'a}kov{\'a}}, {Koubsk{\'y}}, {Krti{\v{c}}ka},
  {Kub{\'a}t}, {{\v{S}}koda}, {{\v{S}}lechta}, \& {Wolf}}]{Harmanec2000}
{Harmanec}, P., {Habuda}, P., {{\v{S}}tefl}, S., {et~al.} 2000, \aap, 364, L85.
\newblock \doarXiv{astro-ph/0011516}

\bibitem[{{Harmanec} {et~al.}(2002){Harmanec}, {Bozi{\'c}}, {Percy}, {Yang},
  {Ruzdjak}, {Sudar}, {Wolf}, {Iliev}, {Huang}, {Buil}, \&
  {Eenens}}]{Harmanec2002}
{Harmanec}, P., {Bozi{\'c}}, H., {Percy}, J.~R., {et~al.} 2002, \aap, 387, 580,
  \dodoi{10.1051/0004-6361:20020453}

\bibitem[{{Harmanec} {et~al.}(2022){Harmanec}, {Bo{\v{z}}i{\'c}},
  {Koubsk{\'y}}, {Yang}, {Ru{\v{z}}djak}, {Sudar}, {{\v{S}}lechta}, {Wolf},
  {Kor{\v{c}}{\'a}kov{\'a}}, {Zasche}, {Opli{\v{s}}tilov{\'a}}, {Vr{\v{s}}nak},
  {Ak}, {Eenens}, {Baki{\c{s}}}, {Baki{\c{s}}}, {Otero}, {Chini}, {Demsky},
  {Barlow}, {Svoboda}, {Jon{\'a}k}, {Vitovsk{\'y}}, \&
  {Harmanec}}]{Harmanec2022}
{Harmanec}, P., {Bo{\v{z}}i{\'c}}, H., {Koubsk{\'y}}, P., {et~al.} 2022, \aap,
  666, A136, \dodoi{10.1051/0004-6361/202244006}

\bibitem[{{Hinkle} {et~al.}(2003){Hinkle}, {Wallace}, \&
  {Livingston}}]{Hinkle2003}
{Hinkle}, K.~H., {Wallace}, L., \& {Livingston}, W. 2003, in American
  Astronomical Society Meeting Abstracts, Vol. 203, American Astronomical
  Society Meeting Abstracts, 38.03

\bibitem[{{Hohle} {et~al.}(2010){Hohle}, {Neuh{\"a}user}, \&
  {Schutz}}]{Hohle2010}
{Hohle}, M.~M., {Neuh{\"a}user}, R., \& {Schutz}, B.~F. 2010, Astronomische
  Nachrichten, 331, 349, \dodoi{10.1002/asna.200911355}

\bibitem[{{Iglesias-Marzoa} {et~al.}(2015){Iglesias-Marzoa},
  {L{\'o}pez-Morales}, \& {Jes{\'u}s Ar{\'e}valo
  Morales}}]{Iglesias-Marzoa2015}
{Iglesias-Marzoa}, R., {L{\'o}pez-Morales}, M., \& {Jes{\'u}s Ar{\'e}valo
  Morales}, M. 2015, \pasp, 127, 567, \dodoi{10.1086/682056}

\bibitem[{{Jilinski} {et~al.}(2010){Jilinski}, {Ortega}, {Drake}, \& {de la
  Reza}}]{Jilinski2010}
{Jilinski}, E., {Ortega}, V.~G., {Drake}, N.~A., \& {de la Reza}, R. 2010,
  \apj, 721, 469, \dodoi{10.1088/0004-637X/721/1/469}

\bibitem[{{Klement} {et~al.}(2022{\natexlab{a}}){Klement}, {Baade}, {Rivinius},
  {Gies}, {Wang}, {Labadie-Bartz}, {Ticiani Dos Santos}, {Monnier}, {Carciofi},
  {M{\'e}rand}, {Anugu}, {Schaefer}, {Le Bouquin}, {Davies}, {Ennis},
  {Gardner}, {Kraus}, {Setterholm}, \& {Labdon}}]{Klement2022b}
{Klement}, R., {Baade}, D., {Rivinius}, T., {et~al.} 2022{\natexlab{a}}, arXiv
  e-prints, arXiv:2210.03090.
\newblock \doarXiv{2210.03090}

\bibitem[{{Klement} {et~al.}(2022{\natexlab{b}}){Klement}, {Schaefer}, {Gies},
  {Wang}, {Baade}, {Rivinius}, {Gallenne}, {Carciofi}, {Monnier}, {M{\'e}rand},
  {Anugu}, {Kraus}, {Davies}, {Lanthermann}, {Gardner}, {Wysocki}, {Ennis},
  {Labdon}, {Setterholm}, \& {Le Bouquin}}]{Klement2022a}
{Klement}, R., {Schaefer}, G.~H., {Gies}, D.~R., {et~al.} 2022{\natexlab{b}},
  \apj, 926, 213, \dodoi{10.3847/1538-4357/ac4266}

\bibitem[{{Kolbas} {et~al.}(2015){Kolbas}, {Pavlovski}, {Southworth}, {Lee},
  {Lee}, {Lee}, {Kim}, {Kim}, {Smalley}, \& {Tkachenko}}]{Kolbas2015}
{Kolbas}, V., {Pavlovski}, K., {Southworth}, J., {et~al.} 2015, \mnras, 451,
  4150, \dodoi{10.1093/mnras/stv1261}

\bibitem[{{Koubsk{\'y}} {et~al.}(2012){Koubsk{\'y}}, {Kotkov{\'a}}, {Votruba},
  {{\v{S}}lechta}, \& {Dvo{\v{r}}{\'a}kov{\'a}}}]{Koubsky2012}
{Koubsk{\'y}}, P., {Kotkov{\'a}}, L., {Votruba}, V., {{\v{S}}lechta}, M., \&
  {Dvo{\v{r}}{\'a}kov{\'a}}, {\v{S}}. 2012, \aap, 545, A121,
  \dodoi{10.1051/0004-6361/201219679}

\bibitem[{{Koubsk{\'y}} {et~al.}(2014){Koubsk{\'y}}, {Kotkov{\'a}}, {Kraus},
  {Yang}, {{\v{S}}lechta}, {Harmanec}, {Wolf}, {Votruba}, {Kub{\'a}t},
  {Kub{\'a}tov{\'a}}, {Niemczura}, \& {{\v{S}}koda}}]{Koubsky2014}
{Koubsk{\'y}}, P., {Kotkov{\'a}}, L., {Kraus}, M., {et~al.} 2014, \aap, 567,
  A57, \dodoi{10.1051/0004-6361/201424022}

\bibitem[{{Lanz} \& {Hubeny}(2003)}]{Lanz2003}
{Lanz}, T., \& {Hubeny}, I. 2003, \apjs, 146, 417, \dodoi{10.1086/374373}

\bibitem[{{Lanz} \& {Hubeny}(2007)}]{Lanz2007}
---. 2007, \apjs, 169, 83, \dodoi{10.1086/511270}

\bibitem[{{Laplace} {et~al.}(2021){Laplace}, {Justham}, {Renzo}, {G{\"o}tberg},
  {Farmer}, {Vartanyan}, \& {de Mink}}]{Laplace2021}
{Laplace}, E., {Justham}, S., {Renzo}, M., {et~al.} 2021, \aap, 656, A58,
  \dodoi{10.1051/0004-6361/202140506}

\bibitem[{{Levenhagen} \& {Leister}(2006)}]{Levenhagen2006}
{Levenhagen}, R.~S., \& {Leister}, N.~V. 2006, \mnras, 371, 252,
  \dodoi{10.1111/j.1365-2966.2006.10655.x}

\bibitem[{{Lopes de Oliveira} \& {Motch}(2011)}]{Lopes2011}
{Lopes de Oliveira}, R., \& {Motch}, C. 2011, \apjl, 731, L6,
  \dodoi{10.1088/2041-8205/731/1/L6}

\bibitem[{{Meilland} {et~al.}(2007){Meilland}, {Stee}, {Vannier}, {Millour},
  {Domiciano de Souza}, {Malbet}, {Martayan}, {Paresce}, {Petrov}, {Richichi},
  \& {Spang}}]{Meilland2007}
{Meilland}, A., {Stee}, P., {Vannier}, M., {et~al.} 2007, \aap, 464, 59,
  \dodoi{10.1051/0004-6361:20064848}

\bibitem[{{Meynet} {et~al.}(2007){Meynet}, {Ekstr{\"o}m}, {Maeder}, \&
  {Barblan}}]{Meynet2007}
{Meynet}, G., {Ekstr{\"o}m}, S., {Maeder}, A., \& {Barblan}, F. 2007, in
  Astronomical Society of the Pacific Conference Series, Vol. 361, Active
  OB-Stars: Laboratories for Stellar and Circumstellar Physics, ed. A.~T.
  {Okazaki}, S.~P. {Owocki}, \& S.~{Stefl}, 325.
\newblock \doarXiv{astro-ph/0601339}

\bibitem[{{Mourard} {et~al.}(2015){Mourard}, {Monnier}, {Meilland}, {Gies},
  {Millour}, {Benisty}, {Che}, {Grundstrom}, {Ligi}, {Schaefer}, {Baron},
  {Kraus}, {Zhao}, {Pedretti}, {Berio}, {Clausse}, {Nardetto}, {Perraut},
  {Spang}, {Stee}, {Tallon-Bosc}, {McAlister}, {ten Brummelaar}, {Ridgway},
  {Sturmann}, {Sturmann}, {Turner}, \& {Farrington}}]{Mourard2015}
{Mourard}, D., {Monnier}, J.~D., {Meilland}, A., {et~al.} 2015, \aap, 577, A51,
  \dodoi{10.1051/0004-6361/201425141}

\bibitem[{{Naz{\'e}} {et~al.}(2022){Naz{\'e}}, {Rauw}, {Czesla}, {Smith}, \&
  {Robrade}}]{Naze2022}
{Naz{\'e}}, Y., {Rauw}, G., {Czesla}, S., {Smith}, M.~A., \& {Robrade}, J.
  2022, \mnras, 510, 2286, \dodoi{10.1093/mnras/stab3378}

\bibitem[{{Naz{\'e}} {et~al.}(2020){Naz{\'e}}, {Rauw}, \&
  {Pigulski}}]{Naze2020}
{Naz{\'e}}, Y., {Rauw}, G., \& {Pigulski}, A. 2020, \mnras, 498, 3171,
  \dodoi{10.1093/mnras/staa2553}

\bibitem[{{Nemravov{\'a}} {et~al.}(2012){Nemravov{\'a}}, {Harmanec},
  {Koubsk{\'y}}, {Miroshnichenko}, {Yang}, {{\v{S}}lechta}, {Buil},
  {Kor{\v{c}}{\'a}kov{\'a}}, \& {Votruba}}]{Nemravova2012}
{Nemravov{\'a}}, J., {Harmanec}, P., {Koubsk{\'y}}, P., {et~al.} 2012, \aap,
  537, A59, \dodoi{10.1051/0004-6361/201117922}

\bibitem[{{Panoglou} {et~al.}(2018){Panoglou}, {Faes}, {Carciofi}, {Okazaki},
  {Baade}, {Rivinius}, \& {Borges Fernandes}}]{Panoglou2018}
{Panoglou}, D., {Faes}, D.~M., {Carciofi}, A.~C., {et~al.} 2018, \mnras, 473,
  3039, \dodoi{10.1093/mnras/stx2497}

\bibitem[{{Paredes} {et~al.}(2021){Paredes}, {Henry}, {Quinn}, {Gies},
  {Hinojosa-Go{\~n}i}, {James}, {Jao}, \& {White}}]{Paredes2021}
{Paredes}, L.~A., {Henry}, T.~J., {Quinn}, S.~N., {et~al.} 2021, \aj, 162, 176,
  \dodoi{10.3847/1538-3881/ac082a}

\bibitem[{{Peters} {et~al.}(2008){Peters}, {Gies}, {Grundstrom}, \&
  {McSwain}}]{Peters2008}
{Peters}, G.~J., {Gies}, D.~R., {Grundstrom}, E.~D., \& {McSwain}, M.~V. 2008,
  \apj, 686, 1280, \dodoi{10.1086/591145}

\bibitem[{{Peters} {et~al.}(2013){Peters}, {Pewett}, {Gies}, {Touhami}, \&
  {Grundstrom}}]{Peters2013}
{Peters}, G.~J., {Pewett}, T.~D., {Gies}, D.~R., {Touhami}, Y.~N., \&
  {Grundstrom}, E.~D. 2013, \apj, 765, 2, \dodoi{10.1088/0004-637X/765/1/2}

\bibitem[{{Peters} {et~al.}(2016){Peters}, {Wang}, {Gies}, \&
  {Grundstrom}}]{Peters2016}
{Peters}, G.~J., {Wang}, L., {Gies}, D.~R., \& {Grundstrom}, E.~D. 2016, \apj,
  828, 47, \dodoi{10.3847/0004-637X/828/1/47}

\bibitem[{{Poeckert}(1981)}]{Poeckert1981}
{Poeckert}, R. 1981, \pasp, 93, 297, \dodoi{10.1086/130828}

\bibitem[{{Pols} {et~al.}(1991){Pols}, {Cote}, {Waters}, \& {Heise}}]{Pols1991}
{Pols}, O.~R., {Cote}, J., {Waters}, L.~B.~F.~M., \& {Heise}, J. 1991, \aap,
  241, 419.
\newblock \url{https://articles.adsabs.harvard.edu/pdf/1991A\%26A...241..419P}

\bibitem[{{Porter} \& {Rivinius}(2003)}]{Porter2003}
{Porter}, J.~M., \& {Rivinius}, T. 2003, \pasp, 115, 1153,
  \dodoi{10.1086/378307}

\bibitem[{{Pr{\v{s}}a} {et~al.}(2016){Pr{\v{s}}a}, {Harmanec}, {Torres},
  {Mamajek}, {Asplund}, {Capitaine}, {Christensen-Dalsgaard}, {Depagne},
  {Haberreiter}, {Hekker}, {Hilton}, {Kopp}, {Kostov}, {Kurtz}, {Laskar},
  {Mason}, {Milone}, {Montgomery}, {Richards}, {Schmutz}, {Schou}, \&
  {Stewart}}]{Prsa2016}
{Pr{\v{s}}a}, A., {Harmanec}, P., {Torres}, G., {et~al.} 2016, \aj, 152, 41,
  \dodoi{10.3847/0004-6256/152/2/41}

\bibitem[{{Quirrenbach} {et~al.}(1997){Quirrenbach}, {Bjorkman}, {Bjorkman},
  {Hummel}, {Buscher}, {Armstrong}, {Mozurkewich}, {Elias}, \&
  {Babler}}]{Quirrenbach1997}
{Quirrenbach}, A., {Bjorkman}, K.~S., {Bjorkman}, J.~E., {et~al.} 1997, \apj,
  479, 477, \dodoi{10.1086/303854}

\bibitem[{{Reig}(2011)}]{Reig2011}
{Reig}, P. 2011, \apss, 332, 1, \dodoi{10.1007/s10509-010-0575-8}

\bibitem[{{Rivinius} {et~al.}(2013){Rivinius}, {Carciofi}, \&
  {Martayan}}]{Rivinius2013}
{Rivinius}, T., {Carciofi}, A.~C., \& {Martayan}, C. 2013, \aapr, 21, 69,
  \dodoi{10.1007/s00159-013-0069-0}

\bibitem[{{Rivinius} {et~al.}(2012){Rivinius}, {Vanzi}, {Chacon}, {Leyton},
  {Helminiak.}, {Baffico}, {{\v{S}}tefl}, {Baade}, {Avila}, \&
  {Guirao}}]{Rivinius2012}
{Rivinius}, T., {Vanzi}, L., {Chacon}, J., {et~al.} 2012, in Astronomical
  Society of the Pacific Conference Series, Vol. 464, Circumstellar Dynamics at
  High Resolution, ed. A.~C. {Carciofi} \& T.~{Rivinius} (San Francisco, CA:
  ASP), 75.
\newblock \doarXiv{1209.5275}

\bibitem[{{Ru{\v{z}}djak} {et~al.}(2009){Ru{\v{z}}djak}, {Bo{\v{z}}i{\'c}},
  {Harmanec}, {Fi{\v{r}}t}, {Chadima}, {Bjorkman}, {Gies}, {Kaye},
  {Koubsk{\'y}}, {McDavid}, {Richardson}, {Sudar}, {{\v{S}}lechta}, {Wolf}, \&
  {Yang}}]{Ruzdjak2009}
{Ru{\v{z}}djak}, D., {Bo{\v{z}}i{\'c}}, H., {Harmanec}, P., {et~al.} 2009,
  \aap, 506, 1319, \dodoi{10.1051/0004-6361/200810526}

\bibitem[{{Schneider} {et~al.}(2021){Schneider}, {Podsiadlowski}, \&
  {M{\"u}ller}}]{Schneider2021}
{Schneider}, F.~R.~N., {Podsiadlowski}, P., \& {M{\"u}ller}, B. 2021, \aap,
  645, A5, \dodoi{10.1051/0004-6361/202039219}

\bibitem[{{Shafter} {et~al.}(1986){Shafter}, {Szkody}, \&
  {Thorstensen}}]{Shafter1986}
{Shafter}, A.~W., {Szkody}, P., \& {Thorstensen}, J.~R. 1986, \apj, 308, 765,
  \dodoi{10.1086/164549}

\bibitem[{{Shao} \& {Li}(2014)}]{Shao2014}
{Shao}, Y., \& {Li}, X.-D. 2014, \apj, 796, 37,
  \dodoi{10.1088/0004-637X/796/1/37}

\bibitem[{{Shao} \& {Li}(2021)}]{Shao2021}
---. 2021, \apj, 908, 67, \dodoi{10.3847/1538-4357/abd2b4}

\bibitem[{{Shenar} {et~al.}(2020){Shenar}, {Bodensteiner}, {Abdul-Masih},
  {Fabry}, {Mahy}, {Marchant}, {Banyard}, {Bowman}, {Dsilva}, {Hawcroft},
  {Reggiani}, \& {Sana}}]{Shenar2020}
{Shenar}, T., {Bodensteiner}, J., {Abdul-Masih}, M., {et~al.} 2020, \aap, 639,
  L6, \dodoi{10.1051/0004-6361/202038275}

\bibitem[{{Slettebak}(1982)}]{Slettebak1982}
{Slettebak}, A. 1982, \apjs, 50, 55, \dodoi{10.1086/190820}

\bibitem[{{Stancliffe} \& {Eldridge}(2009)}]{Stancliffe2009}
{Stancliffe}, R.~J., \& {Eldridge}, J.~J. 2009, \mnras, 396, 1699,
  \dodoi{10.1111/j.1365-2966.2009.14849.x}

\bibitem[{{Temmink} {et~al.}(2022){Temmink}, {Pols}, {Justham}, {Istrate}, \&
  {Toonen}}]{Temmink2022}
{Temmink}, K.~D., {Pols}, O.~R., {Justham}, S., {Istrate}, A.~G., \& {Toonen},
  S. 2022, arXiv e-prints, arXiv:2209.12707.
\newblock \doarXiv{2209.12707}

\bibitem[{{Tetzlaff} {et~al.}(2011){Tetzlaff}, {Neuh{\"a}user}, \&
  {Hohle}}]{Tetzlaff2011}
{Tetzlaff}, N., {Neuh{\"a}user}, R., \& {Hohle}, M.~M. 2011, \mnras, 410, 190,
  \dodoi{10.1111/j.1365-2966.2010.17434.x}

\bibitem[{{Thaller} {et~al.}(1995){Thaller}, {Bagnuolo}, {Gies}, \&
  {Penny}}]{Thaller1995}
{Thaller}, M.~L., {Bagnuolo}, William~G., J., {Gies}, D.~R., \& {Penny}, L.~R.
  1995, \apj, 448, 878, \dodoi{10.1086/176016}

\bibitem[{{Tokovinin} {et~al.}(2013){Tokovinin}, {Fischer}, {Bonati},
  {Giguere}, {Moore}, {Schwab}, {Spronck}, \& {Szymkowiak}}]{Tokovinin2013}
{Tokovinin}, A., {Fischer}, D.~A., {Bonati}, M., {et~al.} 2013, \pasp, 125,
  1336, \dodoi{10.1086/674012}

\bibitem[{{van Bever} \& {Vanbeveren}(1997)}]{vanBever1997}
{van Bever}, J., \& {Vanbeveren}, D. 1997, \aap, 322, 116.
\newblock \url{https://articles.adsabs.harvard.edu/pdf/1997A\%26A...322..116V}

\bibitem[{{Wang} {et~al.}(2017){Wang}, {Gies}, \& {Peters}}]{Wang2017}
{Wang}, L., {Gies}, D.~R., \& {Peters}, G.~J. 2017, \apj, 843, 60,
  \dodoi{10.3847/1538-4357/aa740a}

\bibitem[{{Wang} {et~al.}(2018){Wang}, {Gies}, \& {Peters}}]{Wang2018}
---. 2018, \apj, 853, 156, \dodoi{10.3847/1538-4357/aaa4b8}

\bibitem[{{Wang} {et~al.}(2021){Wang}, {Gies}, {Peters}, {G{\"o}tberg},
  {Chojnowski}, {Lester}, \& {Howell}}]{Wang2021}
{Wang}, L., {Gies}, D.~R., {Peters}, G.~J., {et~al.} 2021, \aj, 161, 248,
  \dodoi{10.3847/1538-3881/abf144}

\bibitem[{{Zorec} {et~al.}(2016){Zorec}, {Fr{\'e}mat}, {Domiciano de Souza},
  {Royer}, {Cidale}, {Hubert}, {Semaan}, {Martayan}, {Cochetti}, {Arias},
  {Aidelman}, \& {Stee}}]{Zorec2016}
{Zorec}, J., {Fr{\'e}mat}, Y., {Domiciano de Souza}, A., {et~al.} 2016, \aap,
  595, A132, \dodoi{10.1051/0004-6361/201628760}

\end{thebibliography}
\bibliographystyle{aasjournal}



\end{CJK*}
\end{document}